\documentclass[12pt,a4paper]{article}
\pdfoutput=1
\usepackage{amssymb,amsmath,amsthm,amsfonts}
\usepackage{mathtools}
\usepackage{fullpage}
\usepackage{natbib}
\usepackage[affil-it]{authblk}

\usepackage[left=2cm, right=2cm, top=2cm, bottom=2cm]{geometry}
\usepackage{rotating}
\usepackage{graphicx} 
\usepackage{float}
\usepackage{array}
\usepackage{subcaption}
\usepackage{caption}
\usepackage{multirow}
\usepackage{csquotes}
\usepackage[usenames,dvipsnames]{xcolor}
\usepackage{tcolorbox}
\usepackage{enumerate}
\usepackage{adjustbox}
\usepackage{kpfonts}
\usepackage{booktabs}
\usepackage{subfiles}
\usepackage[colorlinks]{hyperref}
\AtBeginDocument{
\hypersetup{
citecolor=MidnightBlue,
linkcolor=Brown,   
urlcolor=MidnightBlue}
}
\usepackage{setspace}

\usepackage{tikz}
\usetikzlibrary{shapes.geometric, arrows}

\tikzstyle{full_node} = [circle, draw, minimum size=0.75cm, text centered]
\tikzstyle{blank_node} = [circle, draw, minimum size=0.75cm, text centered]
\tikzstyle{post_node} = [circle, draw, minimum size=0.75cm, text centered, fill=lightgray]

\newtheorem{assumption}{Assumption}

\newcommand\cites[1]{\citeauthor{#1}'s\ (\citeyear{#1})}

\title{Social~contagion~and~asset~prices: Reddit's~self-organised~bull~runs\footnote{
{\textbf{Acknowledgements}: We are grateful to participants of the Alan Turing Institute interest group on economic data science, the American Finance Association annual conference, the Harvard Macroeconomics and MIT finance PhD workshops, the Man Group and Capital Fund Management academic seminars. We thank Rick Van der Ploeg, Andrei Shleifer, Xavier Gabaix, Tim de Silva, Adrien Verdelhan, Xiaowen Dong, J. Doyne Farmer, Jean-Philippe Bouchaud, Paul Tetlock, Laura Veldkamp, David Hirshleifer, Steve Bond, Kieran Marray, Francis DiTraglia, Kevin Sheppard, John Pougu\'{e}-Biyong, Ilan Strauss, Jangho Yang, Fran\c{c}ois Lafond, Matthias Winkler, Jos\'{e} Moran, Farshad Ravasan, Pedro Bordalo, Cars Hommes, Stefan Zohren, Mirta Galesic, and Renaud Lambiotte for their insightful comments. We thank Baillie Gifford and the Institute for New Economic Thinking at the Oxford Martin School for funding our work at the University of Oxford. 
}}}

\author[1]{Valentina Semenova\thanks{valentina.semenova@maths.ox.ac.uk}}
\author[1]{Julian Winkler\thanks{julian.winkler@economics.ox.ac.uk}}

\affil[1]{University of Oxford}
\date{}
\begin{document}

\maketitle
\vspace{-1em}

\begin{abstract}

    Can unstructured text data from social media help explain the drivers of large asset price fluctuations? This paper investigates how social forces affect asset prices, by using machine learning tools to extract beliefs and positions of `hype' traders active on Reddit's WallStreetBets (WSB) forum. Our stylized model shows that peer effects help explain return predictability and reversals, as well as bubble dynamics. We empirically document that sentiments expressed by WSB users about assets' future performances (bullish or bearish) are in part due to sentiments of their peers and past asset returns. The paper directly estimates the effect of WSB activity on asset prices. We document: that retail trader demand follows WSB discussions through using Trade and Quote (TAQ) data, the predictability of prices from retail trader discourse, the amplified market impact of idiosyncratic investor sentiment from viral content online, and the greater exposure of hype investors to bubbles in the markets. 
    


    

\end{abstract}

\vfill

\newpage



\section{Introduction}
\label{sec:introduction}
    
    In investigating the stock market crash of May 28, 1962, the Securities and Exchange Commission (SEC) found that: `investor ``psychology'' being what it is, the increasing decline in one or several issues can easily spread to others. Once the process becomes generally operative, the stage is set for a serious market break' (\citeyear{sec1963special}). The SEC concluded that large institutions acted as a balancing force during the collapse. The report pointed at retail traders as the \textit{key players} behind the panic. Over half a century later, we are again confronted with the consequences of investors' social behaviours. As online discussants on Reddit's `WallStreetBets' (WSB) forum drove up the price of GameStop shares in January, 2021, retail investors regained a spotlight on the (virtual) trading floor. A key difference between 1962 and today is the internet, which offers both a coordination platform on an unprecedented scale, and a new datasource on investor narratives, interactions and psychology. 
    
    This paper sets out to reconcile observed behaviours on social media with economic theory by examining the beliefs and positions of individuals active on WSB. We propose several mechanisms for how social and psychological forces -- information assimilation from peers and extrapolation -- can affect asset prices and result in bubbles and volatility. The paper validates our assumptions of the importance of social forces using the WSB dataset, which is concurrently used to explore several other aspects of investor behaviour, such as their reaction to market surprises and the passthrough of beliefs to asset demand. In a final empirical exercise, we explore whether there is evidence of the fact that social investing has impacted the markets. We show that: retail trading behaviours follow WSB discussions, the interplay of asset prices and social forces is consistent with our model, the heavy-tailed nature of online discussions can lead to the amplification of idiosyncratic individual demand shocks, and that WSB discussions are tied to bubble-like dynamics in the markets.

    Our approach addresses several challenges in the literature. Current research often relies on investor survey data for information on beliefs, and filings (such as 13F filings) to study holdings. These are reported at a fairly low frequency (quarterly or semi-annually), typically cover high net worth individuals or institutional investors (and only their largest asset holdings) and do not explicitly match holdings to beliefs. Fortunately, new, unstructured datasets and ML tools offer a potential solution. Through applying large language models to the WSB dataset, we are able to examine small retail investors and observe their reactions to information at a more granular timescale. We are able to track positions and beliefs of the same individuals over time, as well as the information that they are exposed to from peers. In combination with novel econometric techniques, this rich data allows us to produce estimates for retail investors' reactions to information shared by peers as well as large market moves, and the extent to which WSB discourse has moved asset prices at a daily or weekly frequency. It complements the existing literature on social investing by studying a relatively unexplored dataset (compared to StockTwits or SeekingAlpha) which arguably constitutes the rise to prevalence of a new type of retail trader - the `hype' trader. We directly explore peer effects and their consequences through extending existing models and through our empirical analyses, which incorporate recent techniques for identifying retail trades through Trades and Quotes (TAQ) and methods developed for data with heavy tails (granularity). 

    We begin with an empirical analysis of the data. First, we examine the extent to which individuals trade based on their stated beliefs about an asset's predicted performance. We manually extract positions from screenshots publicised on WSB and sentiments using a large language model which has been fine-tuned on financial data. Expressing a positive sentiment about an asset on WSB raises the probability of a long investment in the same asset in the future by over six times. The effect is not symmetric - expressing a negative sentiment raises the probability of a short investment only 2.5 times. Neutral sentiments appear to be highly predictive of long positions. The strong, statistically significant link between the sentiment of a WSB user and their subsequent positions demonstrates the credibility of the discourse on the forum, in terms of compelling users to trade along their stated interests.
    
    We perform a preliminary analysis of how beliefs relate to asset prices. We observe that higher asset returns are associated with high current sentiment, but lower past sentiments. On a day when users on WSB express sentiments that are two times more bullish than bearish about an asset, we would expect to see an excess positive return of approximately 0.2 percentage points (pp) in the asset. However, on the next day, we would expect excess returns to decline by 0.05 pps, constituting a price reversal. We track the link between returns and sentiments in the twenty days surrounding activity on WSB: we find a statistically significant, positive relationship between current sentiments and returns up to five days in the past, and a negative relationship with returns up to four days in the future. Investors on WSB appear to follow trends in prices, after which returns revert.
    
    Our stylized model sheds light on the relationship between investors influencing each other online and returns. In a dynamic setting, we expect hype investors to drive prices up contemporaneously but depress future returns since they are willing to hold the asset purely due to social pressure. The model demonstrates how heterogeneous, idiosyncratic sentiments can survive aggregation and impact asset prices due to viral online content. In a final modeling exercise, we show that social investing can have pervasive effects through demonstrating potential impacts in models for bubbles. 

    The second part of our paper presents an empirical exercise estimating the degree of social contagion and price extrapolation among users of WSB -- two fundamental components of our model. Our main goal is to quantify the extent to which expressed sentiments are influenced by so-called `peer effects'. To accomplish this, we test how peer sentiment impacts investor decision-making in two ways. First, we select individuals who express sentiments about an asset multiple times and observe how peers discuss the same asset in-between. We use historic peer sentiment as an Instrumental Variable (IV), which mitigates the common shock problem. This approach is inspired by the peer effects in the classroom literature, which gauges future student performance based on entry exams \citep{duflo2011peer}. Second, we leverage the WSB network of interactions to identify the information to which an individual investor has been exposed. The network links an older submission about an asset to a new submission if the author of the new submission comments on the older submission. To estimate a parameter for social contagion, we regress the sentiment expressed by the new submission on the average sentiment of older, linked submissions. We use the timing of our IVs to control for common shocks and instrument the sentiment of linked submissions to control for an author's endogenous choice to comment.

    In both approaches, exogenous variation in average peer sentiment is a statistically significant predictor for the change in author sentiment. This finding suggests that retail investors experience investment complementarities and adapt their strategies based on those of their peers. The IV results reveal that when the odds of peers expressing bullish over bearish sentiments double, the odds of a given user expressing bullish over bearish sentiment increase by an average of 14\%. 

    Our empirical exercise sheds light on several other phenomena. First, we confirm that retail investors on WSB tend to be trend-followers. Specifically, a log-return of 0.1 on a given day increases the probability of a user posting a bullish over bearish post by twenty percent. However, among hype investors, peer effects appear to play \textit{a greater rol}e in sentiment formation than extrapolation. Second, we explore how retail investor sentiments respond to market surprises (defined as log-returns that are two standard deviations above or below a stock's monthly average), as well as reinforcement between peer sentiments and market performance. Interestingly, the effect is negative and significant for negative market surprises, however, positive market surprises appear to have no impact. This finding suggests that downside panic can spread quickly among investors.

    In the final section, we synthesize our model and empirical observations of WSB users to measure their impact on returns and trading patterns. First, we consider whether changes in retail investor trading patterns can be explained by variation in WSB discussions. We extract intra-day trade data in the most popular stocks on WSB from the TAQ dataset and identify retail trades as those that are traded at the sub-penny increment \citep{boehmer2021tracking}. Our approach demonstrates that changes in interest in certain assets over others on WSB can explain changes in the fraction of retail trade volume observed in the market. Contrary to intuition, the effect appears more pronounced for stocks with large market caps, which are not conventionally considered `meme' stocks. 
    
    In order to study price impacts, we forecast variation in sentiments among WSB users unrelated to current price changes, leveraging the strong temporal persistence of sentiment due to the peer effect channel. Our estimates are both statistically and economically significant in predicting changes in weekly average log-returns, providing evidence for a relationship between social dynamics, as proxied by WSB conversations, and financial markets. 

    An important question that remains unanswered is whether viral content can destabilize markets. To tackle this, we leverage the framework of \citet{gabaix2020granular,gabaix2021search} for Granular Instrumental Variables (GIVs) to estimate the impact of the idiosyncratic demand shocks, in the presence of viral content. To construct our GIV, we compare the average sentiment of all submissions to the popularity-weighted sentiment of submissions. Our analysis reveals a statistically significant relationship between our GIV and future returns. Specifically, a doubling in the odds of a very popular submission expressing bullish over bearish sentiments results in an average increase in returns of approximately 1\% the following week. 

    In a final empirical exercise, we tie WSB activity to bubble-like dynamics in the markets. We extend a model for index bubbles from \cite{greenwood2019bubbles} to individual stocks and show that there is a statistically significant difference in activity levels on WSB in stocks that exhibit bubble-like dynamics, with those that experience run-ups in price but no subsequent downturn. The findings underscore the importance of understanding the timing of retail investor trading patterns, and individual losses and gains resulting from social investing \citep{pearson2021chinese}.

    Economists have long deliberated to what extent social dynamics and human psychology play a role in economic decision-making \citep{shiller1984stock,black1986noise}. An outstanding question is, to what extent do social behaviours impact financial markets and how has social media changed the investment landscape? We shed light on this question, using data from WSB, and hope to demonstrate concurrently the opportunities for using unstructured text data and ML tools for finance. The paper argues that the rise to prominence of the WSB forum constitutes a change in the financial climate, where socially-driven `hype' investors now play a prominent role in the markets. 
    
    \paragraph{Related literature} This paper contributes to several rapidly-evolving domains within the finance literature. The first, and perhaps most relevant, empirically documents online investor dynamics. We add to this literature through tracking an additional phenomenon, namely peer effects, and evaluating the impact on the markets. This paper draws on empirical techniques in other economic domains, such as papers studying peer effects in the classroom as well as novel ML approaches for extracting signal from text data. A closely-related strand of literature proposes novel frameworks to better understand the effect of the spread of information and psychological biases on financial markets; we use these models to inform our empirical approach. We extend two frameworks: one with information complementarities and another with extrapolative bubble dynamics to demonstrate the relevance of social investing in these settings. Finally, we use recent techniques in the finance literature to evaluate market impact: this paper leverages TAQ data to track shifting retail investor preferences, and a granular instrumental variable approach to demonstrate the market effects of heavy-tails in popularity on social media. 

    Investor discussion forums emerged as a potential way to study investor behaviour when messages shared among investors were documented to have predictive power for the market \citep{antweiler2004all,sabherwal2011internet,chen2014wisdom,azar2016wisdom, agrawal2018momentum}. Since then, forums have been used to uncover various aspects of investor behaviour, such as: the impact of financial peer effects on susceptibility to the disposition effect \citep{heimer2016peer}, the existence of echo chambers \citep{cookson2022echo}, disagreements due to differing information sets \citep{cookson2020don}, how impression management can lead to the propagation of noise \citep{chen2022listening}, how social network investor centrality can lead to different market response to earnings announcements \citep{hirshleifer2021social}, to name a few. \cite{han2014investor} take the opposite approach of inferring the information transmission network from trade decisions, and evaluate the role of investor network centrality on returns. This paper would be remiss not to mention several interesting papers focusing on our forum of interest -- WallStreetBets: \cite{hu2021rise} classify and study the interactions of three types of investors -- fanatic, rational and naïve; \cite{bradley2021place} study the informational content of due diligence posts for prediction; \cite{mancini2022self} and several other works focus on the role of consensus-formation in driving the GameStop short-squeeze, rather than broader forum dynamics. We study a different question to those presented above, and propose an empirical methodology to more precisely identify how information shared by peers contributes to sentiment formation, as well as the impact of market surprise and reinforcement. The contribution rests on: i) a novel way to analyze text data, leveraging large language models to extract sentiment \citep{araci2019finbert}, ii) our identification strategy, which uses temporal variation in peer composition, to study peer effects directly, and iii) the use of Trade and Quotes (TAQ) data, a granular instrumental variable approach and a proposed methodology for identifying bubble-like dynamics to track market impact and changes in retail investor preferences \citep{boehmer2021tracking, gabaix2020granular,greenwood2019bubbles}. 

    Other related work investigates the diffusion of micro-finance decisions in a social network \citep{banerjee2013diffusion}, the effect of peers on risk taking \citep{lahno2015peer}, the effect of social networks on saving \citep{breza2019social}, and the role of `social learning' versus `social utility' in financial decision-making \citep{bursztyn2014understanding}. By studying a broader set of investors in a natural experiment, our research question is similar to \cite{pool2015people}, who demonstrate that socially connected fund managers appear to hold similar stocks. Several studies in the peer effects literature leverage naturally occurring variation in peers for their identification strategy. An area which pioneered many of these techniques investigates peer effects in the classroom (see \cite{epple2011peer,sacerdote2011peer} for a general overview, and \cite{duflo2011peer} for a prominent example). Social networks are also an active area of study (see \cite{bramoulle2020peer} for a recent review). The present paper highlights how to transfer well-established techniques from the empirical peer effects in the classroom and networks literatures, as well as the recently-proposed granular instrumental variable approach (designed specifically to tackle confounding issues in financial markets \citep{gabaix2020granular,gabaix2021search}), to social media data.
    
    The economic interest in asset mispricing has a long history, with examples dating back to Tulipmania in the Netherlands in the 17\textsuperscript{th} century \citep{garber1989tulipmania}. Since then, numerous frameworks have been proposed to explain the gradual increase and sudden drop in financial assets. These frameworks include the spread of information \citep{veldkamp2006media}, strategic complementarities \citep{hellwig2009knowing,zenou2016key}, and psychological models, such as diagnostic expectations \citep{bordalo2021diagnostic} and extrapolation \citep{glaeser2017extrapolative} -- with \cite{hirshleifer2015behavioral} offering a review, while also promoting a move to `social finance', which we strive to contribute to. Several noteworthy works propose models for the impact on observable network ties on asset price fluctuation (for prominent examples see \cite{pedersen2022game}, \cite{golub2010naive}), however they are less relevant for this paper, as we do not focus on the impact of specific social connections, but rather of the \textit{average informational content} shared by peers. We justify our approach using strategic information complementarities and extrapolation. We also demonstrate how our model estimates can be used in conjunction with existing models for bubbles to understand the role that social dynamics play \citep{barberis2018extrapolation,hirshleifer2020presidential}. Our goal of extending existing models to the current setting is to: i) explain a setting in which investors share their strategies, ii) justify the negative relationship between investor sentiment and returns, iii) explain channels through which investor information sharing online can impact markets -- such as through heavy tails in the popularity of certain content or through the formation of asset bubbles. The modeling contribution should be considered primarily within its empirical context rather than as a standalone study. Several recent papers empirically investigate bubbles, with \cite{greenwood2019bubbles} proposing a methodology to study price run-ups in sectors and with \cite{pearson2021chinese} considering brokerage-level data to understand the Chinese warrant bubble. We adapt the framework of \cite{greenwood2019bubbles} to our setting and demonstrate the increased activity of hype investors in assets that exhibit bubble-like dynamics.

    \paragraph{Road map} We present our results in five sections. The following section comprehensively describes the data source and relevant variables. Section \ref{sec:model} presents a model for price dynamics in the presence of information sharing among investors. Section \ref{sec:social_dynamics} presents empirical evidence for our proposed investor dynamics. Section \ref{sec:market_impact} empirically evaluates the effect of retail investors on financial markets. Section \ref{sec:conclusion} concludes.

\section{What is \textit{WallStreetBets}?}
\label{sec:data_method}
        
    Reddit, launched in 2005, is a  social news aggregation, web content rating, and discussion website. It was ranked as the 8\textsuperscript{th} most visited site globally in November 2022,\footnote{\url{https://www.statista.com/statistics/1201880/most-visited-websites-worldwide/}} with over 430 million anonymous users by the end of 2019.\footnote{\url{https://redditblog.com/2019/12/04/reddits-2019-year-in-review/}} The website's contents are self-organised by subject into smaller sub-forums, `subreddits', which discuss a unique, central topic.
    
    \paragraph{Structure of WSB} Within subreddits, users publish titled posts (called `submissions'), typically accompanied with a body of text or a link to an external website. These submissions can be commented and `upvoted' or `downvoted' by other users. A ranking algorithm raises the visibility of a submission with the amount of upvotes it receives, but lowers it with age. Therefore, the first submissions that visitors see are i) highly upvoted, and ii) recent, with the precise algorithm considered private intellectual property and discussed further in Appendix \ref{app:extended_description}.\footnote{\url{https://www.reddit.com/r/help/comments/7l7686/order_of_posts/}} Comments on a submission, visible to anyone, are subject to a similar scoring system, and can, themselves, be commented on.
    
    \paragraph{Features} The WSB subreddit was created on January 31, 2012, and reached one million followers in March 2020.\footnote{\url{https://subredditstats.com/r/wallstreetbets}} As per a Google survey from 2016, the majority of WSB users are `young, male, students that are inexperienced investors utilizing real money (not paper trading); most users have four figures in their trading account'.\footnote{\url{https://andriymulyar.com/blog/how-a-subreddit-made-millions-from-covid19}} Individuals on the forum discuss and express their sentiments about stock-related news. In addition to market discussions, there is ample evidence of users pursuing the investment strategies encouraged in WSB conversations. Users post screenshots of their investment gains and losses, which subreddit moderators are encouraged to verify -- a dynamic reminiscent of \cites{shiller2005irrational} description of an asset bubble. The discussions are whimsical, but mostly investment-focused. 
    
    \paragraph{Available data} We downloaded WSB data using the PushShift API.\footnote{\url{https://pushshift.io/}} PushShift records all comment and submission data at the time of creation. The full dataset consists of two parts. The first is a total of 1.4 million submissions, with their authors, titles, text and timestamps. The second is comprised of 16.5 million comments, with their authors, text, timestamp, and the identifier of the parent comment or submission. Submission and comment numbers have grown exponentially since 2015 -- Figure \ref{fig:monthly_post_hist} in Appendix \ref{app:extended_description} displays the forum's exponential growth. 
    
    Our dataset spans January, 2012 to July, 2020. Importantly, it does not include the events of the 2021 GameStop (GME) short squeeze. The decision to focus on this timeframe is intentional: before the GME short squeeze, WSB received less attention from institutional investors, as well as less bot-activity. As such, our sample tracks retail investor discussions more precisely, without systematic external influence. The short squeeze constituted a regime shift in the dynamics of WSB -- our goal is to study the conditions and behaviours that preceded the hype, rather than the eventual outcome thereof. Furthermore, ample research has emerged focusing exclusively on the GameStop short squeeze (discussed within our literature review), whereas our goal is to characterise investor behaviour, rather than examine a single event. 

    \paragraph{Identifying assets} The following sections predominantly rely on submissions for text data, since they are substantially richer than individual comments. Comments are used to trace interactions between discussants. In order to understand how users discuss specific assets, we extract mentions of \textit{tickers} from the WSB submissions' text data. A ticker is a short combination of letters, used to identify an asset on trading platforms. For example, `AAPL' refers to shares in Apple, Inc. Appendix \ref{app:most_frequent_tickers} documents how tickers are extracted from submissions. Table \ref{tab:most_frequent_ticker} in Appendix \ref{app:most_frequent_tickers} displays the twenty tickers that feature most prominently in WSB conversations up to July, 2020. These are typically stocks in technology firms, such as AMD or FB. A handful of Exchange Traded Funds (ETFs) are also present, notably the S\&P 500 (SPY) and a leveraged gold ETF (JNUG). 
    
    A small fraction of the 4,650 tickers we extract dominate the discourse on WSB: 90\% of tickers are mentioned fewer than 31 times, and more than 60\% are mentioned fewer than five times. Appendix \ref{app:most_frequent_tickers} documents the heavy-tailed nature of ticker discussions. In total, we are left with 111,765 submissions that mention one, unique ticker and were posted before July 1\textsuperscript{st}, 2020. These submissions have 1.9 million comments in total.
    
    \paragraph{Sentiment model} In addition to extracting tickers, we gauge whether submissions express an expectation for an asset's future price to rise, the \textit{bullish} case, to fall, the \textit{bearish} case, or to remain unpredictable/stable, the \textit{neutral} case. We identify sentiment using a supervised-learning approach, with a hand-labeled dataset of almost five thousand submissions for training, validation and testing \citep{araci2019finbert}. The sentiment model outputs a probability for each sentiment category, achieving 70\% accuracy in categorising the manually labeled test set. Appendix \ref{app:bert} discusses details of this Natural Language Processing (NLP) model and the distribution of labels.

    \paragraph{Key sentiment variable}
    The sentiment classifier assigns three probability scores to each submission about a ticker: the probability of a submission being bullish, $P(\phi = +1)$, bearish, $P(\phi = -1)$, neutral, $P(\phi = 0)$. The probabilities sum to one. At the time $t$ when an author $i$ posts about asset $j$, we use the probability scores above to calculate a continuous sentiment score between $(-\infty, \infty)$:
    \begin{align}
    \Phi_{i,j,t} = \frac{1}{2} \log \left( \frac{\text{P}(\phi_{i,j,t} = +1)}{\text{P}(\phi_{i,j,t} = -1)} \right). \label{eq:sentiment_characteristic}
    \end{align}
    Submissions labeled as bullish ($\text{P}(\phi = +1) = 1$), or bearish ($\text{P}(\phi = +1) = 1$), are set to $\text{P}(\phi = +1) = 0.98$, or $\text{P}(\phi = -1) = 0.98$, to retrieve a finite value for the log-odds. We also extract three categorical variables (bullish, bearish, neutral) which are encoded with a one if the label received the highest probability from our classifier: the categorical variable $\phi_{i,j,t}^{+1}$ will be equal to one is author $i$'s post about asset $j$ at time $t$ is categorised as bullish; $\phi_{i,j,t}^{0}$ and $\phi_{i,j,t}^{-1}$ will be zero. We leverage these variables to investigate investor sentiment throughout the paper. Appendix \ref{app:bert} shows the distribution of our key sentiment variable $\Phi_{i,j,t}$.

    \paragraph{Additional Information} We provide a detailed data appendix -- Appendix \ref{app:data_appendix}  -- which discusses: i) the growth of the forum over time, ii) the way content is presented to users, followership ties on Reddit and why our identification assumptions hold, iii) the details of the ticker and sentiment extraction methodologies, and iv) the prevalence of hype traders. 
    
    \subsection{Isn't all of this just talk?}
    \label{subsec:investment_screenshots}

\begin{table}[ht!] 
\begin{center}
  \caption{Follow-through on WSB Advice} 
  \label{tab:screenshots} 

\begin{tabular}{@{\extracolsep{5pt}}lcc} 
\\[-1.8ex]\hline 
\hline \\[-1.8ex] 
 & \multicolumn{2}{c}{\textit{Dependent variable:} Position in Asset $j$ of Author $i$} \\ 
\cline{2-3} 
\\[-1.8ex] &  \multicolumn{2}{c}{$B_{i,j}$ - categorical}\\
& (1) & (2)\\
\hline \\[-1.8ex]
 $\Phi_{i,j}$ &   1.50 (0.20) ***& \\
 $\phi_{i,j}^{-1}$  &  & -0.97 (0.29) ***\\
 $\phi_{i,j}^{0}$ &  & 0.66 (0.21) ***\\
 $\phi_{i,j}^{+1}$ &  & 1.84 (0.27) ***\\
 \hline \\[-1.8ex]
Observations  & 278 & 278\\ 
Pseudo-R$^{2}$ & 0.13 & 0.17 \\ 
\hline 
\hline 
\end{tabular} 
\end{center}

\footnotesize{ \textit{Notes}: This table presents estimated log-odds coefficients for two logit models for the relationship between the sentiment expressed by user $i$ about asset $j$ and the subsequent long/short position the user reports (see Eq. \ref{eq:positions_logit}). Sentiment estimates are presented in two ways: (1) the continuous log-odds of the author expressing positive over negative sentiment $\Phi_{i,j}$, and (2) as a categorical variable where $\phi_{i,j}^{-1}$ corresponds to the expression of negative sentiment, $\phi_{i,j}^0$ - neutral, $\phi_{i,j}^{+1}$ - positive.

*** Significant at 1\% level ** Significant at 5\% level * Significant at 10\% level}

\end{table}

    Why should we care about the sentiments people express about assets online? Anecdotally, the GameStop short squeeze demonstrated that the online discussions on the WSB forum have impact on assets. However, this does not constitute evidence of the fact that people follow through on the investment strategies they discuss online systematically. 
    
    To address the concern that WSB sentiment data has limited impact on investment decisions, we utilize screenshots that users post of their investment positions to test whether they follow through on their expressed sentiments. We extract approximately 9,000 images from WSB -- we focus only on image-related URLs (such as ones with the domain name `imgur', an image-hosting site) mentioned in posts of authors who had previously posted about a ticker. We hand-annotate a third of the images. Specifically, we manually annotate the image, if it is a  position screenshot, with i) the tickers in the screenshot and ii) the positions (long or short) the author displays. The position taken by author $i$ in asset $j$, $B_{i,j}$, are annotated as +1 if the author is long in the asset, and -1 if the author is short.

    We note that the sample of screenshots is biased. Authors on WSB are socially incentivized to share extreme losses or gains. We, therefore, observe relatively few positions, as compared to sentiments. The positions data is also skewed towards long positions, which is consistent with the skew towards bullish sentiments on the forum. However, despite these shortcomings, the positions provide sufficient variety in investment strategies to test whether people trade on their expressed sentiments on WSB. 
    
    We match the ticker screenshot to a submission posted before/simultaneously with that screenshot by the same author and about the same ticker. We regress the most recently expressed sentiment by author $i$ about asset $j$ (our key sentiment variable $\Phi_{i,j}$) on the log-odds of the position $B_{i,j}$ extracted from their screenshot being long versus short:
    \begin{align}
     \log\left(\frac{\text{P}(B_{i,j} = +1)}{\text{P}(B_{i,j} = -1)} \right) = \lambda ^s \Phi_{i,j} + u^p_{i,j,t},
        \label{eq:positions_logit}
    \end{align}
    where $\lambda^s$ measures the pass-through rate of sentiment into eventual investment positions. In an alternative formulation, we represent the past sentiment as three categorical variables: $\phi_{i,j}^{-1}$, $\phi_{i,j}^0$, $\phi_{i,j}^{+1}$, which take on a value of one if the author's sentiment is labeled short, neutral or long, respectively (and a value of zero otherwise).
    
    \paragraph{Results} Table \ref{tab:screenshots} presents the coefficients of past sentiments regressed on future positions, estimated using a logistic regression. We observe that an author's sentiment is highly correlated to their subsequent holdings of the stock. Let us consider the results in column (2) -- an author creating a bullish post about an asset raises the probability of a long versus short investment by over six times. Furthermore, this specification of the logistic regression predicts an author's position with over 75\% accuracy. The positions data gives us confidence that investors do trade based on their discussions and expressed sentiments. 
    
    \subsection{Predicting stock returns with WSB sentiments}
    \label{subsec:wsb_characteristics}

    Is WSB activity linked stock market returns? We run a set of simple exercises to motivate our future analysis. We average the sentiment characteristics in Eq. \ref{eq:sentiment_characteristic} by stock $j$ and trading day $t$, denoting these mean sentiments by $\Bar{\Phi}_{j,t}$. We merge these daily sentiment observations with US common stock returns reported by CRSP (detailed in Appendix \ref{app:market_data}), and transform the reported returns into log returns. 

    We first consider a regression of log returns on mean daily sentiment,
    \begin{align}
        r_{j,t} = \lambda_{1,T} \Bar{\Phi}_{j,t+T} + \eta^r_t + v^r_{j,t}, \label{eq:temporal_coefficient}
    \end{align}
    where $v^r_{j,t}$ is a residual, $\eta^r_t$ is a daily fixed effect, and $T$ denotes a lag varying from -10 to 15 days. The OLS estimates for coefficients $\lambda_{1,T}$ describe how WSB sentiments are temporally related to stock returns.
    
    Subsequently, we regress daily log returns on current mean sentiments, as well as previous day sentiments:
    \begin{align}
        r_{j,t} =  \lambda_1 \Bar{\Phi}_{j,t} + \lambda_2 \Bar{\Phi}_{j,t-1} + \eta^r_t + v^r_{j,t}, \label{eq:return_regression}
    \end{align}
    where $\eta^r_t$ is a daily fixed effect, $v^r_{j,t}$ an error term, and $\lambda_1, \lambda_2$ our coefficients of interest. This specification gives a sense for the dynamic properties of WSB sentiments -- leading up to a trade day, plus their response on that day.

\begin{figure}[ht!]
        \centering
        \includegraphics{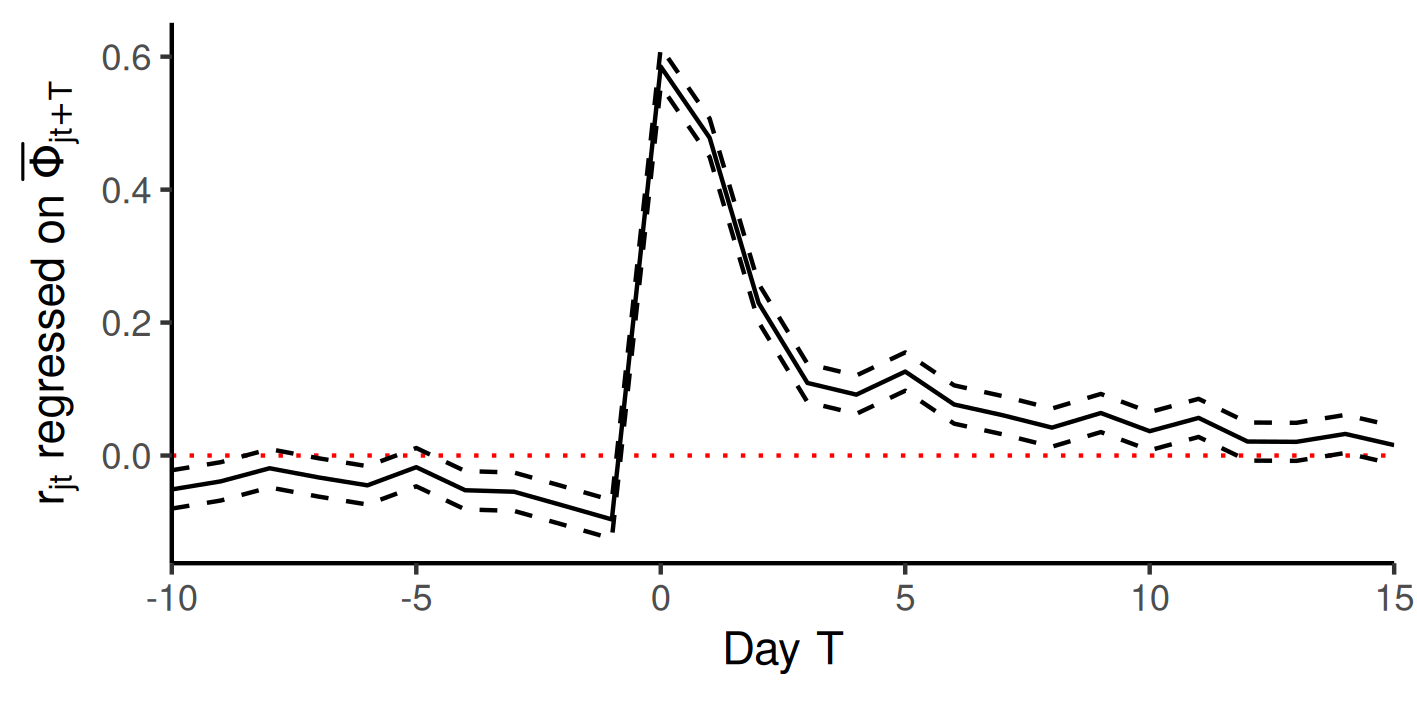}
        \caption{\footnotesize{\textbf{Returns correlate negatively with past WSB sentiments, and positively with current and future WSB sentiments;} Daily stock log returns are regressed on daily average sentiments expressed in submissions on that stock on WSB, where sentiments are lagged by -10 to 15 days -- the estimated relationship is described in Eq. \ref{eq:temporal_coefficient}. A lag of -5 implies that sentiments precede the returns observation by five days. The point estimate for the correlation is plotted by the solid black line, with the 99\% confidence interval in dashed black lines. A correlation of zero is highlighted by the dotted red line. All variables are demeaned by their daily average. Return correlation with past WSB sentiments is negative, but strongly positive with current and future sentiments.}}
        \label{fig:historical_return_correlation}
        \vspace{-0.5em}
    \end{figure}

    \paragraph{Results} Figure \ref{fig:historical_return_correlation} plots the OLS estimates for $\lambda_{1,T}$ in Eq. \ref{eq:temporal_coefficient} as a function of lag $T$. Generally, past sentiments appear negatively related with current returns, although the effect is small, and not highly significant beyond four lags. Current sentiments are strongly correlated with current and past returns, and this effect is significant for up to five days in the past, before dissipating. This implies that a large return in an asset today will have a persistent impact on investor sentiment for five days into the future. Investor sentiments do not anticipate future returns, but rather follow the trend. 

\begin{table}[ht!] 
\begin{center}
  \caption{Stock returns versus WSB characteristics} 
  \label{tab:characteristic_regression} 
\begin{tabular}{@{\extracolsep{5pt}}lcc} 
\\[-1.8ex]\hline 
\hline \\[-1.8ex] 
 & \multicolumn{2}{c}{\textit{Dependent variable:}} \\ 
\cline{2-3} 
\\[-1.8ex] & \multicolumn{2}{c}{$r_{j,t}$} \\ 
\\[-1.8ex] & (1) & (2)\\ 
\hline \\[-1.8ex] 
 $\bar{\Phi}_{j,t}$ & 0.60$^{***}$ (0.04) &  \\ 
  $\bar{\Phi}_{j,t-1}$ & $-$0.16$^{***}$ (0.02) & $-$0.07$^{***}$ (0.02) \\ 
  $r_{j,t-1}$ &  & $-$0.06$^{***}$ (0.004) \\ 
  $\bar{\Phi}_{j,t-1} \times r_{j,t-1}$ &  & 0.01 (0.01) \\ 
 \hline \\[-1.8ex] 
Day FE & Yes & Yes \\ 
Observations & 8,287,639 & 8,287,639 \\ 
R$^{2}$ & 0.0004 & 0.003 \\ 
\hline 
\hline
\end{tabular} 
\vspace{-0.7em}
\end{center}

\footnotesize{ \textit{Notes}: This table presents the OLS estimates for the relationship between stock log returns, $r_{j,t}$ and average expressed sentiment on WSB, $\Bar{\Phi}_{j,t}$ and $\Bar{\Phi}_{j,t-1}$. $t$ represents time in days. The regression highlights the existence of a positive relationship between current sentiments and current returns, and a negative one between current returns and past sentiments. The negative relationship persists when controlling for previous day returns $r_{j,t-1}$ in Column (2). Accompanying standard errors, displayed in brackets, are clustered at the stock level, and calculated in the manner of \citet{mackinnon1985some}.

*** Significant at 1\% level ** Significant at 5\% level * Significant at 10\% level}

\end{table}

    Table \ref{tab:characteristic_regression} reports OLS estimates for coefficients from Eq. \ref{eq:return_regression} in Column 1. Returns again relate positively to contemporaneous sentiment, and negatively with previous day sentiments. Both of these are statistically significant at the 1\% level. However, the implied effects are relatively small; on an average day, returns are 0.1 log points lower if sentiments expressed on the previous day are twice more likely to be bullish than bearish. 
    
    In Column 2, we estimate Eq. \ref{eq:return_regression} with the interaction between lagged, average sentiment and lagged returns, to capture a non-linearity for sentiments in stocks that garner exceedingly high amounts of attention. The slight negative relationship between current and past returns could potentially confound the effect of past sentiment, as seen in the smaller coefficient for lagged sentiments. However, there is no clear evidence that the interaction between sentiments and outsized returns produce a significant effect on subsequent returns.

    \subsection{Motivation}    
    If Table \ref{tab:screenshots} is to be believed, the sentiments expressed on WSB induce trading activity. However, the negative relationship between current returns and past sentiment reported in Table \ref{tab:characteristic_regression} is puzzling, and would suggest that the authors of those submissions erred in their assessments. We argue that the correlations in Table \ref{tab:characteristic_regression} are not a manifestation of (erroneous) information spreading on WSB. Rather, the trading patterns of these retail investors are responsible for reversals in prices as they seeks to find and follow risky strategies. The following section builds a hypothesis on the emergence of bubble-like dynamics as a function of social contagion in investor strategies, whereby return expectations are based, in part, on experiences of peers.

\section{Social dynamics and asset prices}
\label{sec:model}

What motivates investors to share trading advice online, and how does such behaviour affect asset prices? Some seasoned traders might argue that one can only have an investment edge while others are unaware of your strategy. We rationalize observed online information sharing through an asset demand model by incorporating information complementarities in investment decisions. This gives rise to \textit{social contagion} in asset demand; investors buy the asset because others do as well, irrespective of their personal beliefs. The model explores the patterns of returns when this behaviour is present. 

Our proposed mechanism is motivated by a growing literature on strategic information complementarities \citep{hellwig2009knowing,zenou2016key}, as well as studies of diagnostic expectations \citep{bordalo2021diagnostic} -- we choose to focus on information-sharing, rather than network ties since WSB does not have explicit friendship ties, making the information framework more appropriate for our setting. Investors trade on the momentum of the stock price, against a supply of shares provided by noise traders.  Our inclusion of a social component subsequently induces persistence in asset demand over time, which leads to reversal in future returns. The goal of this section is to present what role a social component might play in asset returns and to frame our empirical analysis. 

\paragraph{General setup}
We analyse the price of one asset traded by $N$ investors, indexed by $i$. Each investor derives CARA utility from consuming $c$, $U_i(c_i) = -\text{exp}(-\gamma c_i)$, where $\gamma$ is the constant absolute rate of risk aversion -- the model setup is consistent with various behavioural models, models for bubble formation, and is justified by empirical observations on the relationship between investor sentiment and volatility \citep{bordalo2021diagnostic,barberis2018extrapolation}. We do not include any discounting in their decision-making, but assume they evaluate an asset according to a log-normal distributed value $v = \log(V)$ with expectation $\mathbb{E}_i(v)$ and variance $\sigma^2_i(v)$. In the static model, investor $i$ purchases $\phi_i$ shares at the current market log-price $p$ to optimise the mean-variance objective function \begin{align}
\label{eq:mean_variance_tradeoff}
    \mathcal{L}(\phi_i) &= [\mathbb{E}_i(v) - p]\phi_i - \frac{\gamma}{2}\sigma^2_i(v)\phi_i^2, \\
    \Rightarrow \phi_i^\ast &= \frac{\mathbb{E}_i(v) - p}{\gamma \sigma^2_i(v)}, \label{eq:optimal_demand}
\end{align}
where an asterisk denotes the value that maximises objective $\mathcal{L}$. In this way, we distinguish between beliefs about value $\mathbb{E}_i(v)$ from investor $i$'s decision to buy amount $\phi_i$. Eq. \ref{eq:optimal_demand} yields a familiar expression for asset demand in equilibrium: namely a ratio of expected net returns over their variance.

Assuming that asset supply originates from noise traders, $S$, as in \citet{bordalo2021diagnostic}, common variance $\sigma^2 = \sigma^2_i(v)$ and averaging expected values $\mathbb{E}(v) = {1}/{N} \sum \mathbb{E}_i(v)$, we can re-arrange Eq. \ref{eq:optimal_demand} to yield the following expression for the market-clearing price $p$:
\begin{align}
    p = \mathbb{E}(v) - \frac{S}{N} \gamma \sigma^2. \label{eq:simple_equilibrium}
\end{align}
Eq. \ref{eq:simple_equilibrium} accounts for the price level by investor's average expected value of the asset, in addition to their ability to absorb the exogenous level of assets supplied. This ability depends on the depth of the investor pool -- reflected by the number of investors $N$ -- as well as their risk appetite $\gamma \sigma^2$. In this simple market, the price increases with expected value, and decreases with supply. Appendix \ref{app:model_complementarity} further links this model framework to complementarities in asset demand.

\paragraph{Granular shocks}

An investor's demand may have some idiosyncratic preference which is not captured by common factors. For example, an investor may place particular confidence in products he enjoys using, or admire the corporate strategy of certain company leaders. Such sentiments would manifest as idiosyncratic, heterogeneous investor demand, where $e_i$ is the idiosyncratic component of $i$' asset demand. Under the assumption that these shocks have a finite variance and mean zero, they should average out to zero by the Central Limit Theorem. However, consider a scenario where some investors have different levels of importance $s_i$ for aggregate demand:
\begin{align}
    S &= \sum_{i=1}^N s_i \phi_i^\ast, \\
    \Rightarrow p &= \sum_{i=1}^N s_i \mathbb{E}_i(v) - \gamma \sigma^2 S + \sum_{i=1}^N s_i e_{i},
\label{eq:granular_social1}
\end{align}
where weights $\sum_i s_i = 1$ for demand to equal supply. This could be the case for several reasons: investors could have different amount of capital or, of greater interest to this paper, some investors may have \textit{more sway} in forming public opinion than others. The importance of key players in a social context has been explored in several economic settings -- see \cite{zenou2016key} for a thorough overview. Typically, the most central nodes in a social network have the ability to quickly diffuse information and, therefore, have a high influence on others. 
    
We justify this weighting scheme by the fact that certain users on WSB have a disproportionate effect in shaping the broader discourse. Indeed, WSB is structured to promote viral content, and we would expected a consensus to be formed by key players -- or, rather, around key submissions. If the distribution of importance does not have a finite variance -- i.e. it is `heavy-tailed' -- then the idiosyncratic shocks would not average out to zero. We use this framework to evaluate the impact of viral content from WSB on market returns in section \ref{sec:giv}.

\subsection{Equilibrium price dynamics with peer effects} 

To study the joint dynamics of an asset's price and demand by social investors, we treat aggregate asset demand $\phi = \sum_i \phi_i$ and log price $p$ as state variables for a dynamic system, indexed by time $t$. In doing so, we assume that cumulative demand $\phi_{t}$ reflects a difference between individual valuations of the asset and the price. We distinguish between two independent components of individual valuations: the private signal of individuals $g(b_{i,t})$ and the signal individuals draw from observations of peers $f(\phi_{i,t})$. Aggregate asset demand and price are
\begin{align}
    \phi_t &= \frac{\mathbb{E}_t[g(b_{i,t})] + \mathbb{E}_t[f( \phi_{i,t})] - p_t}{\gamma \sigma^2}, \label{eq:demand_system} \\
    p_t &= \mathbb{E}_t[g(b_{i,t})] + \mathbb{E}_t[f( \phi_{i,t})] - \frac{S_t}{N}\gamma \sigma^2. \label{eq:price_system}
\end{align}
Cumulative demand is, therefore, the difference between individual valuations and the market-clearing price, normalized by risk-aversion. The market-clearing price, on the other hand, is the difference between individual valuations and the rate of asset supply, normalized by the number of investors $N$ and their risk appetite $\gamma\sigma^2$, similar to Eq. \ref{eq:simple_equilibrium}. 

A focus of this paper is the relationship between valuations $g(b_{i,t})$ and the social component $f(\phi_{i,t})$. Studies in behavioural finance suggest different expectation formation mechanisms that ultimately deviate from rational expectations \citep{barberis2018extrapolation,bordalo2021diagnostic}. We combine two such features in Assumptions \ref{assumption:complementarity} and \ref{assumption:mech_extrapolation} to propose a testable structure for $\mathbb{E}_t[f(\phi_{i,t})]$ and $\mathbb{E}_t[g(b_{i,t})]$.

\paragraph{Persistent demand} The mechanism by which past demand enters current asset demand is by the complementarity in investor payoffs. Investor $i$'s payoff to holding the asset is assumed to increase linearly in average asset demand by others.

\begin{assumption} [Persistent Demand]
\label{assumption:complementarity}  
   Social investor $i$'s expectation of future returns is linearly increasing in average asset demand by others: $f(\phi_{i,t}) = \alpha \phi_{t-1},$ where $\phi_{t-1} = 1/N \sum_i \phi_{i,t-1}$ is average asset demand.
    
\end{assumption}

\paragraph{Mechanical Extrapolation} We assume that investors partially trade on the momentum of the asset's price, which \citet{bordalo2021diagnostic} term `mechanical extrapolation'. The functional form of $\mathbb{E}_t[g(b_i)]$ is specified in Assumption \ref{assumption:mech_extrapolation}.

\begin{assumption}[Mechanical Extrapolation]
\label{assumption:mech_extrapolation}
    
   The average investor projects past price increases into the future using the updating rule:
   \begin{align}
       \mathbb{E}_t[g(b_{i,t})] = p_t + \beta(p_t - p_{t-1}), \label{eq:mech_extrapolation}
   \end{align}
   where $\beta$ captures a fixed degree of price extrapolation.
    
\end{assumption}

The validity of these assumptions is discussed in Appendix \ref{app:model}.

\paragraph{System for price and demand} Combining Assumptions \ref{assumption:complementarity} and \ref{assumption:mech_extrapolation} into Eqs. \ref{eq:demand_system}-\ref{eq:price_system} yields demands and returns $r_t = p_t - p_{t-1}$:
\begin{align}
    \phi_t &= \frac{\alpha \phi_{t-1} + \beta r_t}{\gamma \sigma^2}, 
    \label{eq:model_sentiment_return1}
\end{align}
\begin{align}
    r_t &= -\frac{\alpha}{\beta} \phi_{t-1} + \frac{S_t\gamma \sigma^2}{\beta N}.
\end{align}
In this scenario, asset demand and returns are determined simultaneously. The first mechanism is through market clearing, where demand has to adjust to supply. The second is the adjustment of the expected value for the asset to the realised return through $\beta$ and the social signal through $\alpha$. 

As a result, returns are accounted for by current and past asset demand:
\begin{align}
\label{eq:return_against_demand}
    r_t = \frac{\gamma\sigma^2 }{\beta} \phi_t - \frac{\alpha}{\beta} \phi_{t-1},
\end{align}
since supply must equal demand at time $t$. This equation uncovers several important mechanisms at play. Returns are related positively to current demand $\phi_{t}$ through market clearing -- supply $S$ is exogenous and must meet current demand. Higher demand drives up returns. Returns are, however, negatively related to past demand $\phi_{t-1}$ through the expected value of an asset -- a \textit{valuation mechanism}. If there is a positive social signal, investors still value an asset highly, even in the presence of low returns. 

To explain the basic intuition, we consider the following scenarios: i) one where the asset has a positive return $r_t$ and no social signal $\phi_{t-1}$, and ii) one where investors observe a positive social signal $\phi_{t-1}$. In scenario (i), demand is driven by the extrapolation component alone -- investors believe that returns will continue to increase based on the current trend. In scenario (ii) on the other hand, investors do not require a large return to demand the asset -- positive past sentiment drives current demand $\phi_t$. Under exogenous supply, a strong positive signal from peers means that the extrapolated return is \textit{less} important in justifying a higher price. The underlying reasoning relies on the fact that the system is in equilibrium. Therefore, both returns and sentiments have adjusted to reflect a new steady-state, where sentiments are at a certain level $\phi_t$. 

Finally, we observe that the ratios of the coefficients, $\gamma/\beta$ and $\alpha/\beta$, play an important role. $\beta$ effectively anchors the demand of investors in reality -- a greater value of $\beta$ implies that social signals carry less weight, and investors focus on price trends to forecast and expect asset values to grow at some constant rate. As $\beta$ decreases, returns are determined more by social forces -- hype from peers, rather than past performance, now justifies returns and demand. In our data, we observe that $\beta$ is roughly five times $\alpha$ -- individuals weight the sentiments of peers, however, returns are necessary to justify their investment strategies. $\sigma^2$ serves to taper the impact of sentiment, since investors are less certain in their signal and demand less of the asset. The model also explains persistent fluctuations in asset demand, which is discussed in Appendix \ref{app:model}.

\subsection{Bubbles with peer effects}

As noted in our introduction and in the excellent overview of \cite{hirshleifer2015behavioral}, `social investing' may play a role in a variety of price dynamics, beyond the mechanisms highlighted above. Our model considers a specific equilibrium scenario where information sharing by peers and subsequent price dynamics are explained by strategic information complementarities. Quantifying peer dynamics is, however, equally relevant for other models. Consider, for example, the model for bubbles with extrapolation by \cite{barberis2018extrapolation}. We argue that in this scenario, sentiments expressed by peers may also play a role. \cite{barberis2018extrapolation} propose a model where extrapolators determine their demand based on some weight, $w_i$, placed on `fundamental trader' valuation of an asset, while weight $(1-w_i)$ is placed on an extrapolation component, which we denote as $M_t/(\gamma\sigma^2)$ (originally $X_t/(\gamma\sigma^2)$). Consistently with our model, traders maximize a CARA utility function defined over next period’s wealth. The extrapolation signal, $M_t$, is determined by discounting past asset returns. We propose that this formulation can be modified to incorporate the effect of information shared by peers. The extension allows us to justify bubble dynamics in the absence of a change in fundamental news about the asset, and presents a potential model for a setting with news which alter the fundamental value of an asset, as well as `animal spirits' among investors. We present the model extension in Appendix \ref{app:model}, and simulate price dynamics using our parameter estimates within the next section.

\subsection{Model predictions}

A simple linear regression exercise in Table \ref{tab:characteristic_regression} provides some evidence of the validity of the model proposed by Eq. \ref{eq:return_against_demand}. We summarise our asset demand model with social contagion by four further predictions, which we seek to validate in our WSB data.

\paragraph{Prediction 1: A mechanism for peer effects in asset demand} \textit{Given that asset demands by social investors are complementary, a marginal increase(decrease) in peer outlook about an asset will raise(lower) the future outlook of an investor about the asset.}

We dedicate Section \ref{sec:social_dynamics} to investigating strategic complementarities among investors on WSB. Besides testing for the direct effect of peers on investor sentiment (Prediction 1), we also use the opportunity to test our assumption for mechanical extrapolation: A uniform, marginal increase(decrease) in an asset's returns will raise (lower) the future outlook of an investor about the asset. It will also indirectly increase(decrease) the outlook of an investor through increasing(decreasing) the outlook of peers.

\paragraph{Prediction 2: Return predictability} \textit{An increase in asset demand explained by past investor sentiment increases the asset's price.}

According to Eq. \ref{eq:return_against_demand}, there is a positive contemporaneous correlation between returns and WSB sentiments, but a negative correlation with lagged sentiments. However, one issue with the positive correlation is the challenge of identification. The data typically reflects an equilibrium outcome where sentiments and returns are positive, and vice versa. The main challenge is to identify exogenous variations in current sentiment relative to current returns. If current sentiments do not depend on future returns, we can estimate the impact of WSB sentiments on returns using sentiment scores predicted from preceding discussions. Stock-specific characteristics will also drive persistent heterogeneity in the expressed sentiments of WSB users. We tackle the question of whether social dynamics can account for some return predictability in Section \ref{sec:market_impact}.

\paragraph{Prediction 3: Peer-driven bubble dynamics} \textit{Peer effects are a mechanism behind asset bubbles.}

We use parameter estimates from Section \ref{sec:social_dynamics} to demonstrate how peer effects can amplify bubbles in the presence of a change to the fundamental value of an asset, and also drive bubble-like dynamics in the presence of a social shock. We later empirically validate the presence of bubble-like dynamics in assets discussed on WSB. 

\paragraph{Prediction 4: Granular social forces} \textit{Idiosyncratic demand shocks do not average out in the presence of heavy-tailed attention and impact asset prices.}

Eq. \ref{eq:granular_social1} predicts that, in the presence of heavy-tailed attention to certain investors, heterogeneity in investor sentiments will not average out, and will instead have an impact on asset prices. An emphasis on viral content compels WSB discussants to follow specific, popular strategies, which are predicted to have an outsized impact on asset price returns. We investigate the role of these granular idiosyncratic sentiments in Section \ref{sec:market_impact}.

\section{Social dynamics in WSB}
\label{sec:social_dynamics}
    
 This section provides empirical evidence for the existence of two mechanisms underlying asset demand -- namely peer effects and extrapolation -- among investors on WSB. Section \ref{sec:model} proposes the framework. We seek to test whether these complementarities manifest in the sentiments expressed about the future outlook of an asset among investors on WSB.

    
    \paragraph{Testable prediction} Prediction 1 in Section \ref{sec:model} establishes the behaviours we expect to see within the WSB community. In this section, we argue that user sentiment data observed on WSB are consistent with our model: investors are influenced by peer sentiments, and extrapolate past returns. WSB, as a platform, is a venue for `social investors' to realise their strategic information complementarities.

    \paragraph{Estimating equation} The target independent variable of interest for studying hype investor sentiment is the log-odds of bullish over bearish sentiment,
    \begin{align}
        \Phi_{i,t} &= g(b_{i,t}) + f(\Bar{\phi}_{-i,(t-1,t)}) + \varepsilon_{i,t},
        \label{eq:continuous_ind_var}
    \end{align}
    derived from our utility framework in Appendix \ref{app:utility_framework}. One key addition is the time subscript, $t$. An author chooses a bullish over bearish strategy depending on: i) a signal $b_{i,t}$, and ii) the observed sentiments of peers, $\Bar{\phi}_{-i,(t-1,t)}$.

    \subsection{Empirical strategy: consensus formation among investors}
    \label{subsec:consensus}
    
    We use two approaches to estimate Eq. \ref{eq:continuous_ind_var}: i) the \textit{Frequent Posters} approach, and ii) the \textit{Commenter Network} approach. Both leverage different features of our data. For the \textit{Frequent Posters} approach, we leverage the fact that certain users post multiple submissions about the same asset (hence, \textit{frequent}). For the \textit{Commenter Network} approach, we use instances in which users comment on others' submissions to more precisely gauge the transmission of sentiments about the same asset.
    
    For the \textit{Frequent Posters} approach, we observe that 8,173 authors create at least two submissions about the same ticker. We quantify peer influence by identifying the impact of other authors who write submissions about the same asset \textit{between} an individual's two submissions. We use an IV of previous, expressed peer sentiments to control for exogenous shocks (see Figure \ref{fig:frequent_posters_dag} for an illustration). Our approach allows us to control for the author's sentiment prior to exposure to his peers, in addition to market moves.
    
    We argue that the peer sentiments that an individual is exposed to have random, temporal variation: the posts that an individual is exposed to on WSB depend on what other anonymous, disconnected users have posted on the forum shortly before the author logs on, and what topic has recently gained popularity (see Section \ref{sec:data_method} and Appendix \ref{app:extended_description} for a detailed description). Users are `disconnected' in the sense that Reddit does not have friendship/follower ties \textit{within} specific forums -- followership ties on Reddit more broadly and the fact that they do not impact our approach are discussed in Appendix \ref{app:extended_description}. Individuals cannot, therefore, filter exposure to certain sentiments over others. We argue that this randomised exposure of users to different opinions (similar in spirit to random assignment of individuals to groups, such as in \cite{weidmann2021team}) allows us to estimate direct peer effects.
    
    The \textit{Commenter Network} approach considers a submission-to-submission network, with an earlier submission exerting peer influence on a future submission if the author of the later submission commented on the earlier one. The submission-to-submission network helps identify peers an author interacts with more precisely. Here, we also control for market moves, and employ a set of IVs to address endogeneity concerns. As our IVs, we measure: i) sentiments of submissions to which the influencing submission is connected (the `friends of friends' -- detailed in Figure \ref{fig:sentiment_projection}), and ii) the historic sentiment of neighbours. The underlying argument rests on the premise that neighbours of network distance two exert an influence on user sentiments through peer effects (consistently with \cite{bond201261}). A user's endogenous choice to comment on certain posts over others would therefore not account for users one step removed.
    
    \subsubsection{Identifying peer influence: Frequent Posters}
    \label{subsec:mult_post_2sls}
    
    Within WSB, we observe author $i$ initially express a sentiment about an asset $j$, $\Phi_{i,j,(t-1)}$ (the continuous log-odds of a post expressing bullish over bearish sentiment, as per Eqs. \ref{eq:sentiment_characteristic}\&\ref{eq:continuous_ind_var}), and, subsequently, write a new submission about the same asset at a later time, with an updated sentiment $\Phi_{i,j,t}$ (where time $t$ is in event time). In the time between these posts, the author may observe submissions by others on the same asset expressing average sentiment $\Bar{\Phi}_{-i,j,(t-1,t)}$, in addition to outside information related to the asset. Our goal is to identify the effect that expressed peer sentiments have on changing author $i$'s sentiment.
    
    \begin{figure}[ht]
        \centering
        \includegraphics[width = 0.9\textwidth]{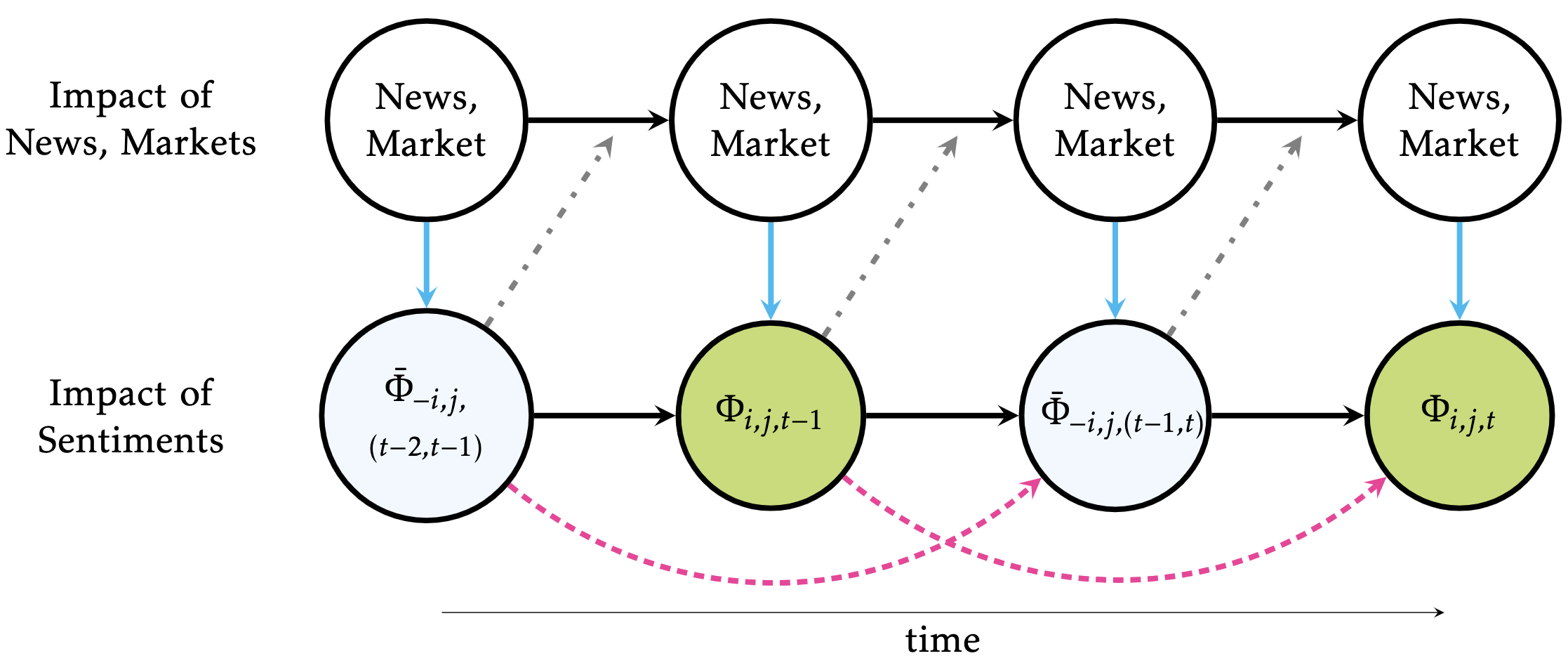}
        \caption{\footnotesize{\textbf{\textit{Frequent Posters} Directed Acyclic Graph (DAG);} we trace the flow of information within our system. Arrows represent the impact that information from one source has on the next source. Light blue nodes $\Bar{\Phi}_{-i,j,(t-2,t-1)}$ and $\Bar{\Phi}_{-i,j,(t-1,t)}$ represent peer sentiment; green nodes $\Phi_{i,j,t-1}$ and $\Phi_{i,j,t}$ represent the sentiment of investor $i$ - our target variable. Time $t$ is expressed in event time. The magenta, dashed line represents the impact that historical sentiment expressed about an asset has on the author's own future opinion. In our first stage, we estimate $\Bar{\Phi}_{-i,j,(t-1,t)}$ by $\Bar{\Phi}_{-i,j,(t-2,t-1)}$, controlling for the market move at the time of the peer's initial post while estimating the coefficients. In this way, we are able to isolate the impact that peer sentiment $\Bar{\Phi}_{-i,j,(t-1,t)}$ has on individual $i$ at time $t$, $\Phi_{i,j,t}$ .}}
         \label{fig:frequent_posters_dag}
    \vspace{-1em}
    \end{figure}
    
    \paragraph{Reduced form} We first estimate the effect of average peer sentiment between an author's two submissions with the following linear model: 
    \begin{align}
    \label{eq:2sls_stage2}
    \Phi_{i,j,t} = \kappa \Bar{\Phi}_{-i,j,(t-1,t)} + X_{i,j,t} \beta + \epsilon_{i,j,t},
    \end{align}
    where the vector of control variables, $X_{i,j,t}$, is composed of stock-specific fixed effects, author $i$'s past sentiment, and stock log returns, both on day $t$ and the average of the five days preceding $t$, and the variance in log returns on the five days prior to day $t$; $\beta$ is a vector of corresponding coefficients. Even though peers appear randomly on the forum in this formulation (as discussed earlier in this section), an exogenous shock in the period $(t-1,t)$ may affect the views of both peers and the author in question simultaneously. For this reason, the OLS estimates do not enable us to precisely estimate peer influence.
    
    \paragraph{Instrumenting peer sentiment I} To tackle this issue, we use the historical views of peers as an IV for their views expressed within $(t-1,t)$. Our choice of IV is founded in psychology: \cite{ross1975perseverance} find that `once formed, impressions are remarkably persevering and unresponsive to new input', with later studies, such as \cite{anderson1980perseverance}, supporting these findings. We reason about our choice of IV through the Directed Acyclic Graph (DAG) shown in Figure \ref{fig:frequent_posters_dag}. We consider that historic news and market moves are fully reflected in the news and market information available within the following timestep. Information shared by peers is also fully incorporated from one timestep to the next; however, dotted pink lines indicate the persistence of individual author sentiments (the persistence of individual formed impressions).

    Leveraging the structure of our DAG, we estimate investor $k$'s sentiment (a peer of investor $i$) about asset $j$, $\Phi_{k,j,t}$, based on the sentiment they expressed previously, $\Phi_{k,j,t-1}$, and control for asset returns at the time of their original post, $r_{j,t-1}$: 
    \begin{align}
    \label{eq:2sls_stage1}
    \Phi_{k,j,t} = \kappa^0_1 \Phi_{k,j,t-1} + \kappa^0_2 r_{j,t-1} + \epsilon^0_{k,j,t},
    \end{align}
    where $\epsilon^0_{k,j,t}$ is an idiosyncratic error. The coefficient $\kappa^0_1$ estimates the true effect of an individual's historical sentiment. Consistently with our DAG, controlling for $r_{j,t-1}$ allows us to accurately estimate $\kappa^0_1$, while controlling for confounders. Eq. \ref{eq:2sls_stage1} is estimated using a sample containing submissions by all authors who post multiple times. The F-statistic for this first stage estimate, presented in Panel B.1 of Table \ref{tab:reg_consensus_main}, suggests that this is a strong instrument. Our choice of IV gives a good approximation for author sentiment, while allowing us to control for common shocks affecting the sentiments of peers and investor $i$ in the period $(t-1,t)$. We use the predicted outlook of peers between an author's posts, $\Hat{\Phi}_{-i,j,(t-1,t)}$, to estimate peer effects as our Second Stage regression, while keeping all other controls the same -- historic peer sentiment is used for prediction. Appendix \ref{app:consensus} provides further details on our variable construction and method. Appendix \ref{app:market_data} describes the construction of market variables, and their matching to WSB data.
    
    \paragraph{Credible estimation} We check whether our estimation strategy is credible, with respect to the three challenges presented by \cite{zenou2016key} and \citet{athey2017state} in estimating peer effects. The first lies is in distinguishing peer effects from contextual effects -- the tendency of perspectives to vary with some observable characteristics of the group, rather than individuals influencing each other. Controls for asset price movements and ticker specific characteristics -- the main sources of exogenous variation -- address this concern. Second, the random, anonymous nature of WSB, as well as controlling for ticker-specific fixed effects, address the possibility for correlated effects. The specification with the IV addresses the common shock problem. A more rigorous, statistical analysis of our identification strategy is included with the results.  
        
    \subsubsection{Identifying peer influence -- Commenter Network}
    \label{subsec:network_2sls}
    
\begin{figure}[ht!]
    \begin{center}

        \begin{subfigure}[t]{.4\textwidth}
            \includegraphics[width=\textwidth]{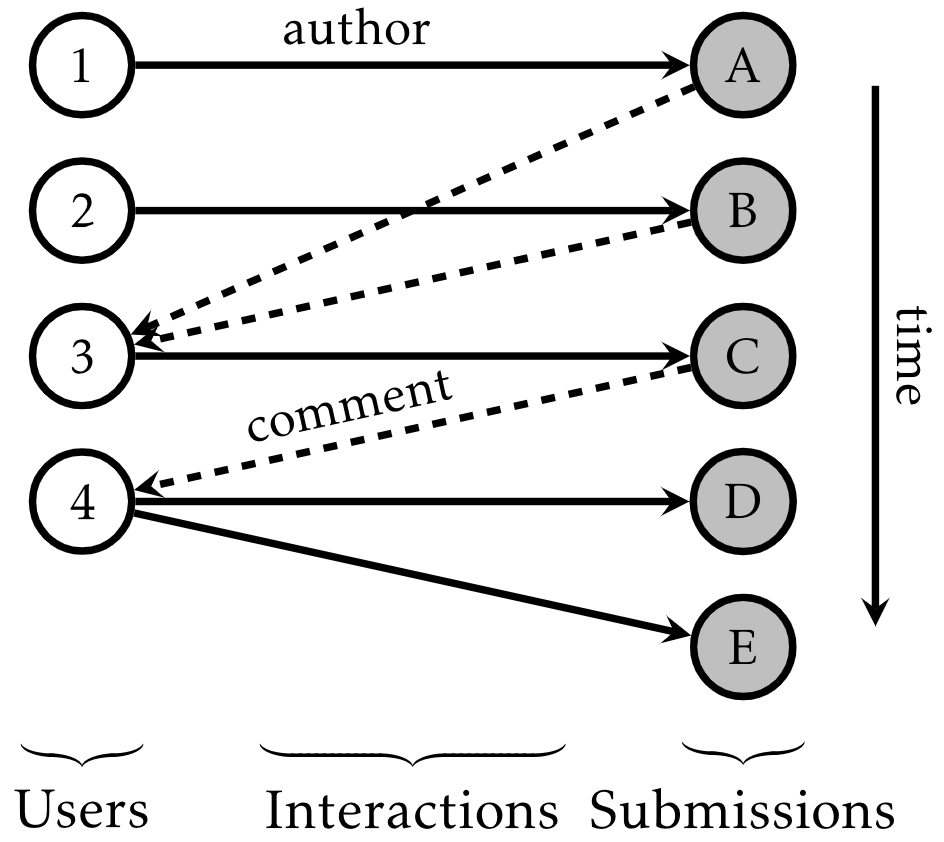}
            \subcaption{Bipartite network between authors and submissions}
            \label{fig:sentiment_diagram}
        \end{subfigure}
        \hspace{2cm}
        \begin{subfigure}[t]{.35\textwidth}
            \includegraphics[width = \textwidth]{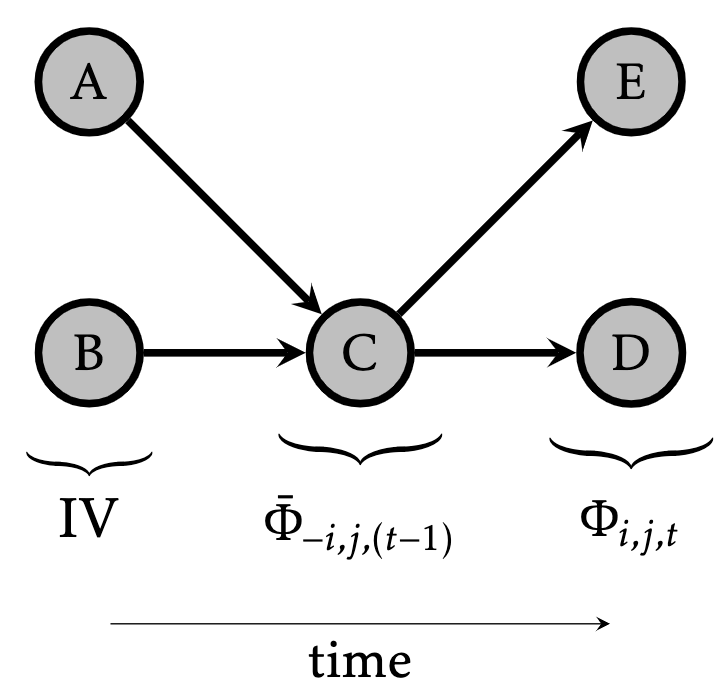}
            \subcaption{Submission-to-submission projection of network in Figure \ref{fig:sentiment_diagram}}
            \label{fig:sentiment_projection}
        \end{subfigure}  
        \bigskip
        \begin{subfigure}[t]{.49\textwidth}
            \includegraphics[width=\textwidth]{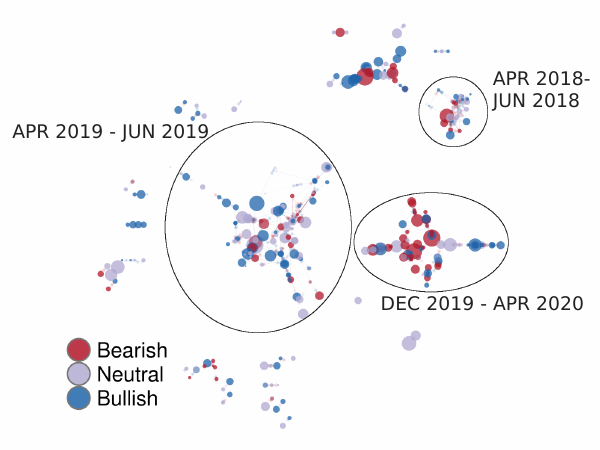}
            \subcaption{Submission-to-submission network: DIS}
            \label{fig:sentiment_network_dis}
        \end{subfigure}  
        \hfill
        \begin{subfigure}[t]{.49\textwidth}
            \includegraphics[width=\textwidth]{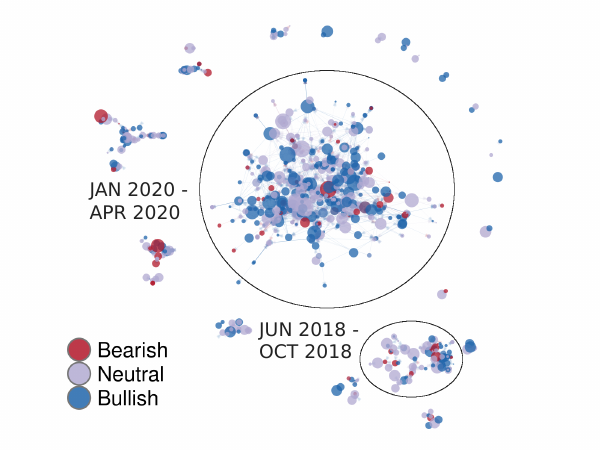}
            \subcaption{Submission-to-submission network: MSFT}
            \label{fig:sentiment_network_msft}
        \end{subfigure} 
        \vspace{-0.8em}
        \caption{\footnotesize{\textbf{User networks in WSB conversations}; WSB data is summarised as a bipartite graph, illustrated in Figure \ref{fig:sentiment_diagram}, where users (left) are linked to submissions (right) when they author the submission (solid edge) or comment on the submission (dashed edge). The resulting projection of submissions, in Figure \ref{fig:sentiment_projection}, tracks the propagation of sentiments $\Phi$. The submission-to-submission networks for two stocks in Figures \ref{fig:sentiment_network_dis} and \ref{fig:sentiment_network_msft} reveal that individuals post more submissions that are bullish(bearish) at times when the price of an asset increases(decreases) dramatically, with some visual evidence that similar sentiments tend to cluster.}}
        \label{fig:sentiment_network}
        \end{center}
    \end{figure}

    WSB allows us to trace the interactions of users through a commenting network, even though there are no user friendship ties. We exploit a submission-to-submission interaction network for each asset, tracking which submissions in the past influence future submissions based on authors' commenting histories. This method offers a more precise way to identify a user's peers by observing which individuals, and submissions, an author explicitly interacts with. Figures \ref{fig:sentiment_diagram} and \ref{fig:sentiment_projection} illustrate the approach.
    
    Two examples of submission-to-submission networks in our data are displayed in Figures \ref{fig:sentiment_network_dis} and \ref{fig:sentiment_network_msft}. Distinct temporal clusters emerge, as a certain asset gains and loses prominence on WSB. Some discussions appear fragmented: the \textit{DIS} discussion in Figure \ref{fig:sentiment_network_dis}, for example, contains several smaller clusters, with perceptible differences in overall sentiments. Others, such as the \textit{MSFT} discussion in Figure \ref{fig:sentiment_network_msft}, contain a giant component where investors with different sentiments interact.
    
    Our network approach uses a similar Reduced Form and Second Stage to the \textit{Frequent Posters} approach in Eq. \ref{eq:2sls_stage2}. We modify our control for an author's past sentiment about the stock to account for authors who post for the first time: a dummy variable encodes whether the author's most recent previous post is bearish, neutral, bullish or missing.
    
    \paragraph{Instrumenting peer sentiment II} We use an IV approach to estimate peer influence. As the First Stage, we estimate the sentiments of neighbours to estimate an author's view. As indicated in Figure \ref{fig:sentiment_projection}, the sentiments in submissions \textbf{A}, \textbf{B} can be used to predict that of submission \textbf{C}. The \textit{predicted} sentiment of \textbf{C} can then, in turn, be used to predict the sentiments of \textbf{D} and \textbf{E}. This choice of IV is well-established in the networks literature \citep{zenou2016key,patacchini2016social,bifulco2011effect}, and helps control for the exogenous choice to comment on certain submissions and not others. We also include the neighbour's own historical sentiment, as a set of categorical variables, as the second IV (similarly to the \textit{Frequent Posters} approach). Our Eq. \ref{eq:2sls_stage1}, therefore includes a set of author controls $X_{i,j,t}^0$:
    \begin{align*}
    \Phi_{k,j,t} = \kappa^0 \Phi_{k,j,t-1} + X_{i,j,t}^0 \beta^0 +\epsilon^0_{k,j,t},
    \end{align*}
    where the superscript denotes the estimation of the First Stage. In the results in Table \ref{tab:reg_consensus_main}, we display the estimate for our main IV - neighbours of neighbours in the commenting network; the additional IV of author historical sentiment is displayed in Appendix \ref{app:consensus}.  
    
    \paragraph{Timing of observations} We use the timings of events to mitigate the common shock problem for both our IVs: the neighbour's historical sentiment and the `friends or friends' submissions. For the latter, we calculate the time period of influence for a given post, which ends when the last comment is made on a submission. This effectively marks the point when a particular submission ceases to be of interest to the WSB community. We filter for instances where the period of influence for a submission used as an IV for another submission ends before the new submission we are modeling is created. In practice, if submission \textbf{C} in Figure \ref{fig:sentiment_projection} occurs on July 1st at 2:31PM, the final comments on posts \textbf{A} and \textbf{B} must occur before, in order to ensure that our IV is not affected by a common shock. We also include an author's own, historical sentiment as an IV only if his previous submission occurs at least two business days before the current one. 
        
    The \textit{Commenter Network} offers certain upsides, but also certain shortcomings, as compared to the \textit{Frequent Posters} approach. The network method more precisely identifies the channels of influence between authors. However, the allocation of peers is no longer random, since the network structure is governed by a \textit{choice} to comment.
    
    
    \subsection{Results: peer effects and extrapolation}
    \label{sec:consensus_results}
    
    In this section, we present the Reduced Form, Second Stage, and First Stage regression estimates for both the \textit{Frequent Posters} and \textit{Commenter Network} approaches. The Reduced Form and Second Stage estimates, across both model specifications, show that peer sentiments directly impact an individual's sentiment about an asset, with user sentiments conforming to those of their peers.

    \begin{table}[ht!]
\begin{center}
\caption{Peer influence in WSB sentiments}
\label{tab:reg_consensus_main}
\resizebox{\textwidth}{!}{
\begin{tabular}{lcc}
\toprule
&    Frequent Posters  &  Network \\ 
& (1) & (2) \\
\cline{2-3} \\[-1.8ex]

\multicolumn{3}{l}{\textbf{Panel A}: Reduced Form -- peer influence estimated using \textit{observed} average sentiment of peers} \\\\[-1.8ex]
&  \multicolumn{2}{c}{\textit{Dependent Variable}: Investor Sentiment $(\Phi_{i,j,t})$} \\ \cline{2-3} \\[-1.8ex]

    Average peer sentiment, & & \\
    $\Bar{\Phi}_{-i,j,(t-1)}$ \textit{(observed)} & 0.055 (0.011) ***  &  0.041 (0.009) *** \\ 
    $r_{j,t}$ & 0.020 (0.003) ***  & 0.022 (0.004) *** \\ \\[-2ex]
    Author \& asset controls ($X_{i,j,t}$) & Yes  &  Yes \\
    
    Number of obs. & 14,376 & 24,902\\
    F-statistic & 67 & 582 \\

\toprule \\[-1.8ex] 

\multicolumn{3}{l}{\textbf{Panel B.1}: Second Stage -- peer influence estimated using \textit{predicted} average sentiment of peers} \\\\[-1.8ex]
&  \multicolumn{2}{c}{\textit{Dependent Variable}: Investor Sentiment $(\Phi_{i,j,t})$} \\ \cline{2-3} \\[-1.8ex]
    
    Average peer sentiment, && \\
    $\hat{\Phi}_{-i,j,(t-1)}$ \textit{(predicted)} & 0.036 (0.010) ***  & 0.022 (0.009) ** \\
    $r_{j,t}$ & 0.025 (0.005) ***  & 0.023 (0.006) *** \\ \\[-2ex]
    Author \& asset controls ($X_{i,j,t}$) & Yes  &  Yes \\
    
    Number of obs. & 11,075 & 16,514 \\
    J-statistic & NA & 0.43 \\ 
    F-statistic & 73 & 1,207 \\

\toprule \\[-1.8ex] 

\multicolumn{3}{l}{\textbf{Panel B.2}: First Stage -- estimating peers' sentiments}\\
&  \multicolumn{2}{c}{\textit{Dependent Variable}: Sentiment of Peers} \\ \cline{2-3} \\[-1.8ex]
    Historical Sentiment of Peers & 0.31 (0.01) *** &   \\
    Sentiment of Neighbours' Neighbours &  &  0.14 (0.01) *** \\ \\[-2ex]
    Author controls ($X_{i,j,t}^0$) & No  &  Yes \\
    Controls for returns ($r_{j,t-1}$) & Yes  &  No \\
    Number of obs. & 19,370 & 24,013\\
    F-statistic & 1,105 & 118 \\

\bottomrule
\end{tabular}
}
\end{center}
\vspace{-0.6em}
\footnotesize{ \textit{Notes}: this table presents the First Stage, Second Stage and Reduced Form OLS estimates for peer influence on WSB. In column (1), the First Stage is estimated using the initial sentiment expressed by an author about an asset to estimate his sentiment in the following post. In column (2), the First Stage is estimated using the sentiment of previous submissions that an author commented on, regarding the same asset. The Second Stage is estimated using the average predicted sentiment of peers. Ticker-level dummies, asset return and volatility controls, and the intercept are included in the Second Stage and Reduced Form estimates, but not shown here; additional author-specific IVs in the network approach are also included but not shown -- the complete estimates are presented in Appendix \ref{app:consensus}. Robust standard errors, clustered at the ticker level for Panels A and B.1, are presented in parentheses. Observations with incomplete data are dropped. \\
*** Significant at 1\% level
** Significant at 5\% level
* Significant at 10\% level}

\end{table}

    Table \ref{tab:reg_consensus_main} presents the normalized coefficients, with Panel A presenting OLS estimates for $\kappa$, from Eq. \ref{eq:2sls_stage2}, using observed variation in peer sentiments, and Panel B.1 using predicted variation in peer sentiments -- independent variables are normalized with respect to their mean and standard deviation (explained further in Appendix \ref{app:consensus}). We relegate estimated coefficients for control variables as well as the non-normalized coefficient estimates to Appendix \ref{app:consensus}. The \textit{Frequent Posters} approach indicates that peer effects are approximately 1.5 times \textit{more important} in individual sentiment formation, as compared to extrapolation. Our non-normalized coefficient estimates in Table \ref{tab:reg_consensus_non_norm} of 0.19 on predicted peer sentiments means that doubling in the odds of peers expressing bullish over bearish sentiments increases the odds of a given submission to be bullish, over bearish, by 14.1\%. In all cases, the robust standard errors, clustered at the ticker level, produce estimates statistically significant at the 1\% level. The \textit{Commenter Network} approach yields a similar result.

    The estimated coefficients in columns (1) and (2) of Panel B.1 suggest that an exogenous increase in average peer outlook appears to increase an investor's own future view about an asset. These findings demonstrate that the data are consistent with Prediction 1. As a result, we conclude that the data supports a model where strategic complementarities govern the investment decisions of retail traders sampled on WSB. The results in Table \ref{tab:reg_consensus_main} also support our model with extrapolation. We observe that recent returns are highly predictive of expressed sentiments across all specifications. 
    
    \paragraph{Support for identification} One potential concern is that individuals who post multiple times about the same asset, or those who comment on others' submissions, may differ from the rest of the population on the forum. If this were the case, our findings would not allow us to draw valid conclusions about the overall population of investors. We provide evidence that sentiments expressed by our samples are similarly distributed to those of the overall user population in Appendix \ref{app:consensus}. 
    
    A second concern is whether our proposed independent variables -- asset price movements, ticker fixed effects and author historical sentiments -- are effective controls for unobserved ticker characteristics. If our controls in the \textit{Frequent Posters} formulation are valid, then a randomly selected cohort of individuals who post on the same ticker \textit{before} the author's first post, should have no effect on the sentiments expressed in dependent submissions. Similarly, if our controls are useful in the \textit{Commenter Network} formulation, a random rewiring of the network should yield no effect. The results are detailed in Appendix \ref{app:consensus}: no statistically significant correlation emerges from the randomly selected cohorts. This provides further evidence that unobserved factors influencing within-ticker variation in both peer composition and author sentiment are not confounding.

    
     A final concern with our \textit{Commenter Network} approach is overidentifying restrictions. A J-statistic of 0.43, and a corresponding p-value of 51\%, leads us to believe that our additional instruments are exogenous (see Appendix \ref{app:consensus} for further details). We explore further dynamics observed on WSB, such as whether there is contagion in asset interest among investors online \citep{banerjee1993economics,shiller2017narrative}, in Appendix \ref{app:consensus}.

    \subsection{Further insights}
    \label{subsec:consensus_further_insights}
    WSB data provide additional opportunities to test investor responses to a market surprise, and the reinforcement mechanism between peers and asset prices. We consider the sentiments expressed by investors $i$ about asset $j$ at time $t$, $\Phi_{i,j,t}$, as our dependent variable and use the controls from Eq. \ref{eq:2sls_stage2} to test for two additional effects: market surprise and reinforcement. 
    
    We define two types of surprises: i) a positive surprise if asset $j$ experiences a return which is two standard deviations higher than the 30-day historical average for the stock on day $t$ or on the day before, and ii) a negative surprise if asset $j$ experiences a return which is two standard deviations lower than the average for the stock on day $t$ or on the day before. We compute the average and standard deviation for stock $j$ using data of the thirty trading days before $t$. We also interact returns and predicted peer sentiment to see the extent to which peer effects are reinforced by returns. We use the \textit{predicted} peer sentiment from our \textit{Frequent Posters} approach in our regressions to control for sentiments that respond to current price changes. 

    \begin{table}[ht!]
\centering
\caption{Additional effects: surprise and reinforcement}
\begin{tabular}{l@{\hskip 0.3in}l@{\hskip 0.4in}c@{\hskip 0.4in}c}
\toprule \toprule
 && \multicolumn{2}{c}{\textit{Dependent Variable: $\Phi_{i,j,t}$}}\\ [1ex]
 && (1) & (2) \\ 
\toprule
\multirow{7}{*}{\rotatebox{90}{\textit{Independent}}}
\multirow{7}{*}{\rotatebox{90}{\textit{Variables}}}
& $\Hat{\Phi}_{-i,j,(t-1,t)}$ &   0.037 (0.009) *** &    0.036 (0.012) *** \\
& $r_{j,t}$                   &   0.019 (0.006) ***  &   0.017 (0.005) *** \\ \\[-2.2ex]
& Positive Surprise             &      -0.018 (0.031) &  \\ \\[-2.2ex]
& Negative Surprise           &      -0.114 (0.038) *** &   \\ \\[-2.2ex]
& $(r_{j,t} \times \Hat{\Phi}_{-i,j,(t-1,t)})^+$        &   &     0.074 (0.025) ***  \\ 
& $(r_{j,t} \times \Hat{\Phi}_{-i,j,(t-1,t)})^-$        &   &     0.070 (0.108)  \\ \\[-2.2ex]
& Author \& asset controls ($X_{i,j,t}$) & Yes & Yes \\
\toprule
& No. Observations: & 11,073 & 11,116 \\
& $R^2$: & 0.08 & 0.08\\
& $R^2_{adj}$: & 0.06 & 0.06\\
\bottomrule
\end{tabular} \\
\bigskip
\vspace{-.7em}
\begin{tabular}{cccc}
\multicolumn{4}{l}{%
  \begin{minipage}{17cm}%
    \footnotesize{ \textit{Notes}: The dependent variable -- individual investor sentiment about an asset, scaled continuously between $(-\infty, \infty)$ -- is estimated using the variables in Eq. \ref{eq:2sls_stage2} and additional variables, using OLS. The additional variables in column (1) are categorical variables for positive and negative market surprises at time $t$ in asset $j$; in column (2) the additional variables are a cross term between asset $j$'s returns and the estimated sentiments of peers: $(r_{j,t}$ X $\Hat{\Phi}_{-i,j,(t-1,t)})^+$ is the product if the predicted sentiment of peers is positive and returns are also positive, and zero otherwise; $(r_{j,t}$ X $\Hat{\Phi}_{-i,j,(t-1,t)})^-$ is product if the predicted sentiment of peers is negative and returns are also negative, and zero otherwise. $(r_{j,t}$ X $\Hat{\Phi}_{-i,j,(t-1,t)})^+$ captures the extent to which positive peer predictions correspond to observed market moves; the reverse is true for $(r_{j,t}$ X $\Hat{\Phi}_{-i,j,(t-1,t)})^-$. Peer sentiment $\Hat{\Phi}_{-i,j,(t-1,t)}$ is estimated using the \textit{Frequent Posters} approach to control for confounders. Robust standard errors, clustered at the ticker level, are presented in parentheses. Observations with incomplete data are dropped.\\
*** Significant at 1\% level
** Significant at 5\% level
* Significant at 10\% level}
  \end{minipage}}
\end{tabular}
\vspace{-0.5em}
\label{tab:surprise_reinforcement}
\end{table}

    \paragraph{Surprise} Table \ref{tab:surprise_reinforcement} presents the results from our exploration of surprise and reinforcement. Column (1) contains the OLS estimates when including positive and negative categorical variables for market surprise. A negative market surprise appears to significantly affect investor sentiments. The result is not symmetric -- a positive surprise does not appear to convince investors of the upside potential of a stock. This observation suggests that downside panic spreads quickly within the investor population. This effect is in addition to the large impact returns have on sentiment. 
    
    \paragraph{Reinforcement} Column (2) considers the effect from market reinforcement of peer sentiments by including the cross term between returns and the predicted sentiments of peers. The cross terms are separated depending on whether the predicted peer sentiment $\Hat{\Phi}_{-i,j,(t-1,t)}$ is positive or negative: $(r_{j,t} \times \Hat{\Phi}_{-i,j,(t-1,t)})^+$ is the \textit{bullish} interaction when $\Hat{\Phi}_{-i,j,(t-1,t)}$ and returns are both positive and zero otherwise, whereas the bearish interaction $(r_{j,t} \times \Hat{\Phi}_{-i,j,(t-1,t)})^-$ is positive if predicted sentiment and returns take negative values, zero otherwise. Therefore, a large value for the bullish interaction corresponds to peers forecasting positive returns in asset $j$ and the asset $j$ simultaneously experiencing positive returns on the day of author $i$'s submission. 
    
    In Table \ref{tab:surprise_reinforcement}, the bullish interaction is highly significant. WSB users are spurred by peers predicting positive returns and subsequently observing the asset outperform in the market, possibly suggesting some `irrational exuberance' \citep{shiller2005irrational}. The reverse is not true for bearish reinforcement. 

    \subsection{Asset prices in a bubble with peer effects}

        \begin{figure}[ht!]
            \begin{subfigure}{0.45\textwidth}
                \centering
                \includegraphics[width=\linewidth]{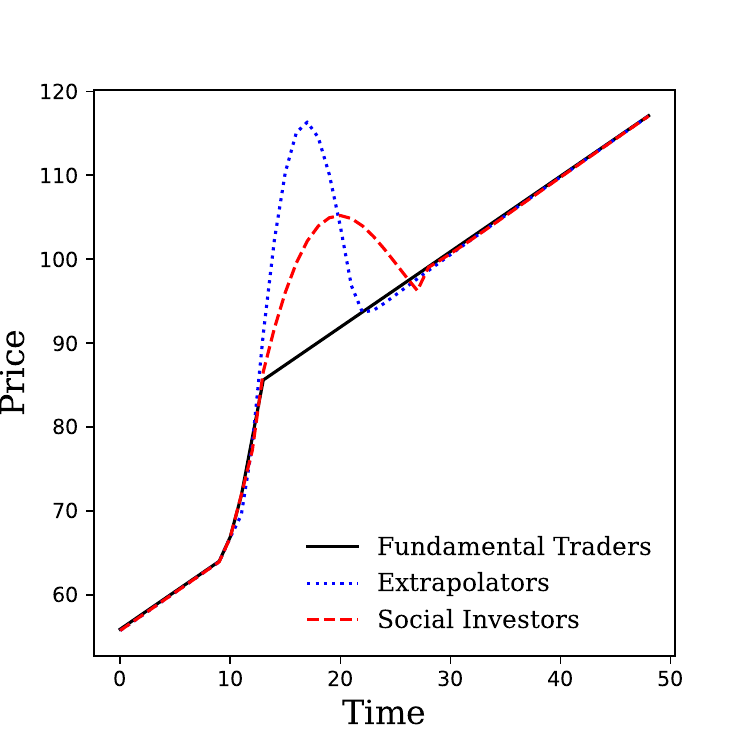}
                \caption{Bubbles with Extrapolation, Sentiments} 
                \label{fig:barberis_bubble1}
            \end{subfigure}
            \hfill
            \begin{subfigure}{0.45\textwidth}
                \centering \includegraphics[width=\linewidth]{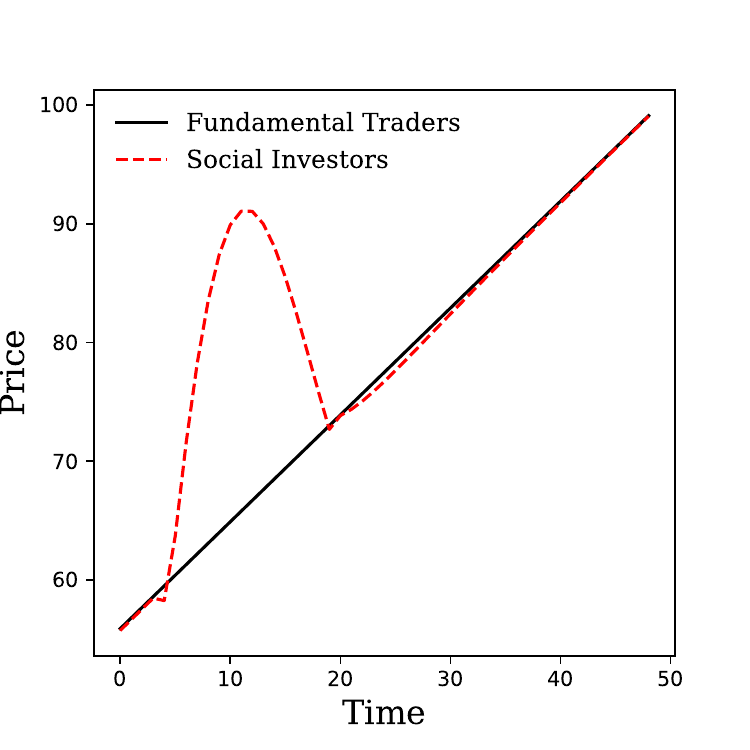}
                \caption{Sentiment-Driven Bubbles} 
                \label{fig:barberis_bubble2}
            \end{subfigure}
            \caption{\footnotesize{\textbf{Bubbles with Social Investors}; We simulate our modified version of the model for bubbles with extrapolation \citep{barberis2018extrapolation}. We choose initial parameters similar to those from Figure (1) in \cite{barberis2018extrapolation}: $w_i= 0.1$, $\sigma^2=3$, $\gamma = 0.1$, $\theta_E = 0.1$; extrapolators / social investors make up 70\% of investors, the remainder are fundamental traders; the quantity of the asset is set to 1. In Figure \ref{fig:barberis_bubble1}, we choose that the fundamental value of the asset remains unchanged except for periods 11-14 when information is revealed resulting in an increase in the future dividend of the asset by 2,4,6,6 (respectively). In Figure \ref{fig:barberis_bubble2}, we impose a comparable sentiment shocks of 6 in periods 5,6,7.}}
            \label{fig:barberis_bubbles_main}
    \end{figure} 

    Our empirical methodology validates the assumptions behind our proposed model with information complementarities. We simultaneously propose that peer effects may amplify bubble dynamics in the markets. To validate this, we leverage parameter estimates from our empirical exercise and utilize them in our extension of a model for bubbles with extrapolation -- a modification of \citep{barberis2018extrapolation}, which we analyze to better understand asset fluctuations at the shorter time horizon (days rather than quarters). From Table \ref{tab:reg_consensus_main}, Column (1), we observe that individuals place a relative weight of 0.6 on the sentiments of their peers and 0.4 on recent market returns, which we propose as estimates for $\alpha^*$ and $\beta^*$ respectively from the model extension in Appendix \ref{app:bubble_model}. We simulate the modified model for bubbles and present the results in Figure \ref{fig:barberis_bubbles_main}. The simulations demonstrate two things. First, in an extrapolative setting, the existence of a  social signal makes the bubble formation process have longer memory; the bubble takes longer to form, and has a less defined peak. Second, the modification allows a mechanism for bubbles to form as a result of a purely social signals, thereby introducing the potential for `animal spirits' among investors to result in bubble-like dynamics.

\section{Has WSB destabilised markets?}
\label{sec:market_impact}

    Prediction 2-4 consider different ways in which strategic complementarities impact asset returns. In Section \ref{sec:social_dynamics}, we focus on identifying the extent to which investors consider sentiments from peers, as well as recent market moves, while updating their own sentiments about future asset returns. The evidence for peer effects in asset demand is robust in two, separate estimation strategies we consider. In this section, we conduct a quantitative analysis of asset demand and returns in order to validate the link between these behaviours and stock returns. 

    We investigate changes in retail trader asset demand through tracking retail investor trades, using the methodology from \cite{boehmer2021tracking}. We observe that changes in asset interest on WSB explain a significant fraction of changes to retail trading behaviour. A key challenge to analysing returns is that sentiments and returns are co-determined in equilibrium. If returns are high, sentiments are also high. Conversely, if sentiments are high, buying pressure will also increase returns contemporaneously. We consider two empirical strategies to investigate the extent to which WSB \textit{caused} returns to exceed a benchmark without social contagion. The first strategy exploits variation in sentiments that we can explain using the history of WSB conversations, investigating Prediction 2. The second strategy exploits the granularity of discussions on WSB to verify Prediction 4. In a final empirical test, we identify bubble-like dynamics in assets and verify that WSB users have a significantly greater interest in assets whose price experiences as a sharp rise, but subsequently implodes, as compared to those that have a sustained increased in price, verifying Prediction 3. 

    \subsection{Evidence of trading}
    Our model in Section \ref{sec:model} predicts that behaviourally-driven changes in investor demand result in changes in asset prices. We test for a link between discussions on WSB and changes to retail investor demand, approximated by the fraction of retail investor trades executed in the market. 
    
    \paragraph{Variable Definition}  Retail trader activity is identified from Trade and Quote (TAQ) data -- a dataset containing all transactions for listed stocks in the United States. We leverage the fact that retail transactions are offered price improvements and, therefore, may execute at a fraction of a penny. To identify retail traders, we first filter trades to those with \textit{exchange code} = `D' in TAQ. In the remaining trades, we identify those that execute at a fraction of a penny as retail transactions. Specifically, let $Z_{j,t} = mod(100* P_{j,t}, 1)$, the fraction of a penny associated with the transaction price in stock $j$ at time $t$. If $Z_{j,t}$ is in the interval (0,0.4) or (0.6,1), the transaction is coded as a retail transaction. We define a metric for retail trade fraction in asset $j$ in week $t_w$ as:
    \begin{align}
        RF_{j,t_w} = Vol_{retail, j, t_w} / Vol_{total, j, t_w},
    \end{align}
    where $Vol_{retail, j, t_w}$ is the sum of the sizes of all trades labeled as retail transactions using the method above across asset $j$ in week $t_w$, and $Vol_{total, j, t_w}$ is the sum of the sizes of all trades in asset $j$ in week $t_w$ from TAQ. 
    Our variable of interest -- the change in retail investor trading fraction -- is defined as:
    \begin{align}
        \Delta RF_{j,t_w} = RF_{j,t_w} - RF_{j,t_w-1}.
    \end{align}
    We design a similar monthly metric, $\Delta RF_{j,t_m}$, to track changes in retail trade volumes on a monthly scale. Importantly, a change in news or asset-level characteristics cannot justify a change to $\Delta RF_{j,t_w}$, since it would affect retail traders and institutional investors. Our metric, therefore, allows us to distinguish between changes to \textit{retail investor preferences} versus overall market shifts. 

    Our goal is to consider whether discussions on WSB explain variation in the fraction of retail investor transactions in the market. We define a metric tracking the prevalence of discussions about asset $j$ in week $t_w$ on the forum versus the overall number of posts about assets on the forum, $DF_{j,t_w}$, defined as the number of posts mentioning asset $j$ in week $t_w$ over the total number of posts mentioning assets in week $t_w$. Our predictor of interest is defined as:
    \begin{align}
        \Delta DF_{j,t_w} = DF_{j,t_w} - DF_{j,t_w-1},
    \end{align}
    the change in the attention allocated to asset $j$ on the forum. 

    \paragraph{Analytic Approach}
    In order to validate our approach, our goal is to demonstrate that changes in discussions among retail investors are accompanied by changes in retail trading volumes. We regress change in weekly (monthly) retail trade fractions on the changes in ticker importance on the forum:
    \begin{align}
        \Delta RF_{j,t_w} = \beta_v \Delta DF_{j,t_w} + \eta_j + \epsilon_{j,t_w},
        \label{eq:taq_regression}
    \end{align}
    where $\eta_j$ are ticker fixed effects, and $\epsilon_{j,t_w}$ is an error term. We repeat the same exercise, except with variables computed at the monthly time scale, $t_m$. 

    \begin{table}[ht!] 
\begin{center}
  \caption{Retail trade volume versus the WSB discussion}
  \label{tab:taq} 
\begin{tabular}{@{\extracolsep{-2pt}}lccc|ccc} 
\\[-1.8ex]\hline 
\hline \\[-1.8ex] 
 & \multicolumn{6}{c}{\textit{Dependent variable: $\Delta RF_{j,t}$}} \\ \\[-1.8ex]
\cline{2-7} 
 \\[-1.8ex] & \multicolumn{3}{c}{Weekly Changes}  & \multicolumn{3}{c}{Monthly Changes} \\
 
 & (1) & (2) & (3) & (4) & (5) & (6) \\ 
 & Pooled & Pooled & Top 25\% & Pooled & Pooled & Top 25\% \\
 &   &  & (market cap) & &  & (market cap) \\ \\[-1.8ex] 
\hline \\[-1.8ex] 
\\[-1.8ex] $\Delta DF_{j,t}$  & 0.098$^{***}$ & 0.098$^{***}$ & 0.167$^{***}$ & 0.261$^{***}$ & 0.253$^{***}$ &  0.505$^{***}$ \\
  & (0.019) & (0.019) & (0.030) & (0.049) & (0.049) & (0.033)\\
 \\[-1.8ex] \hline \\[-1.8ex] 
Ticker FE & No & Yes &  No & No & Yes & No \\
Observations & 3,503 & 3,503  & 1,040 & 776 & 776 & 232 \\ 
R$^{2}_{adj}$ & 0.007 & 0.000 & 0.028 & 0.035 & 0.011 & 0.116 \\ \\[-1.8ex] 
\hline 
\hline
\end{tabular} 
\end{center}
\footnotesize{ \textit{Notes}: this table presents the OLS estimates for the influence of changes in WSB discussion interests on retail trading patterns. Columns (1), (2), (3) present the weekly estimates, while (4), (5), (6) present the monthly ones. Columns (3) and (6) consider the estimates for the top quartile of stocks, by market cap, within our sample. All standard errors are clustered at the ticker level.\\
*** Significant at 1\% level
** Significant at 5\% level
* Significant at 10\% level} \\ 
\vspace{-2em}
\end{table} 
    
    \paragraph{Results} We consider tickers that are popular on WSB are more likely to cause changes in retail trader order flow. For this reason, we look at the trading patterns of the twenty most popular tickers, by year, on WSB between the years 2017 and 2020. Table \ref{tab:taq} presents our main result. We observe that changes in discussion popularity of tickers are statistically significant for explaining changes in retail trading behaviour at the weekly and monthly level. The monthly estimates appear more significant and help explain a greater variation in the dependent variables, as per the $R^2_{adj}$. 
    
    In columns (3) and (6), we repeat the exercise but only consider the stocks that are within the highest quartile by average market cap between the years of 2012-2020 within our sample: AAPL, AMZN, BA, BAC, DIS, FB, GE, GILD, MSFT. The sample does not include `meme' stocks, such as GME, TSLA, PLTR. We observe that the coefficient on $\Delta DF_{j,t_w}$ is more significant in this formulation and $\Delta DF_{j,t_w}$ explains a greater fraction of the variation in changes in retail investor trading activity. A 10\% increase in the prevalence of AAPL discussions on WSB is associated with a 5\% increase in retail trading activity in the market. Including ticker fixed effects in our model specifications for columns (3), (6) does not change the coefficients, however, decreases our $R^2_{adj}$ indicating that the effect is driven by changes in discussions and cannot be explained by stock-level differences.

    Our experiments demonstrate that WSB discussions track changes in retail trading behaviour at the weekly and monthly timescales. These results further justify our modeling approach, highlighting how WSB activity is linked to retail trader demand for assets. 
    
    In additional experiments, we extend the approach to look at retail trade imbalances by separately classifying retail buy and retail sell transactions \citep{boehmer2021tracking}. We regress changes in the logarithm of the volume of retail buy over sell trades  in a certain asset on the logged number of positive posts over the number of negative posts. We find evidence of the fact that weekly changes in average sentiment on WSB explain changes in retail trade imbalance, however, we primarily rely on the study of return predictability in the following section to investigate the relationship between sentiments and markets. 
   
    \subsection{Evidence of price impact}

    
    We are interested in finding variation in current sentiments which are exogenous with respect to current returns. The goal is to detect a positive effect carried by retail investor sentiment, proxied by WSB activity. First, we formulate the linear relationship between returns and current sentiments. We then propose a 2SLS estimation strategy to quantify the impact of social contagion on stock market variables. 

    \paragraph{Independent variable}
    Our independent variable estimated from WSB sentiment data measures sentiment change $\Delta \Phi_{j,t}$, the first-difference of stock $j$'s mean daily sentiment between calendar weeks $t$ and $t-1$. The purpose for this variable is to gauge the stock-specific response to a change in WSB's associated attention and sentiments on a week-by-week basis. Measuring the difference in average sentiments between two periods proxies for the change in asset demand due to changes in the intensity of corresponding sentiments.
    
    \paragraph{Reduced Form} We regress changes in weekly log-returns on changes in weekly sentiments:
    \begin{align}
        \Delta \Bar{r}_{j,t} &= \omega \Delta \Phi_{j,t} + \eta_{t} + \varepsilon_{j,t}, \label{eq:market_impact_return_reduced}
    \end{align}
    where $\omega$ is the coefficient of interest, $\eta_{t}$ denotes week fixed effects, and $\varepsilon_{j,t}$ an idiosyncratic error. We specifically use the first difference in weekly returns to account for stock-specific heterogeneity. The Reduced Form approach does not provide conclusive evidence of a causal relationship between social investor activity and stock market activity. The narratives and sentiments expressed on social media platforms are influenced by real-time news, events, and stock market fluctuations, which can result in reverse causality. In other words, positive sentiments may be expressed during weeks of high returns, regardless of previous sentiment. This makes it challenging to establish a clear link between social investor activity and stock market activity.

    \paragraph{First Stage} We use variation in sentiments that can be explained by past activity on WSB and past stock performance to identify our parameter of interest. We predict sentiment $\Phi_{j,t}$ using past stock price behaviour, as well as past sentiments:
    \begin{align}
        \Phi^+_{j,t} = \log \left( \frac{\text{P}(\phi_{j,t} = +1)}{\text{P}(\phi_{j,t} = 0)}  \right) = \lambda^+_r \Bar{r}_{j,t-1} + \lambda^+_\sigma \sigma^2_{j,t-1} + \lambda^+_1 \Phi^+_{j,t-1} + \lambda^+_2 \Phi^-_{j,t-1} + \eta^+_{t} + \varepsilon^+_{j,t}, \\
        \Phi^-_{j,t} = \log \left( \frac{\text{P}(\phi_{j,t} = -1)}{\text{P}    (\phi_{j,t} = 0)}  \right) = \lambda^-_r \Bar{r}_{j,t-1} + \lambda^-_\sigma \sigma^2_{j,t-1} + \lambda^-_1 \Phi^+_{j,t-1} + \lambda^-_2 \Phi^-_{j,t-1} + \eta^-_{t} + \varepsilon^-_{j,t},
    \end{align}
    where superscripts differentiate between the average log-odds of a submission in week $t$ expressing bullish ($+$) versus negative ($-$) sentiments, over neutral sentiments. Week fixed effects remain in the sentiment models, so that the full estimation strategy rests on within-week variation in all explaining, as well as explained, variables. 


    The approach outlined above relies on coarse aggregates for sentiments: the probabilities here are not estimated on data for individual submission sentiments, as is the case in Section \ref{sec:social_dynamics}. Rather, the probabilities are calculated by averaging the probabilities for \textit{all} submissions in week $t$, discussing ticker $j$, to be bullish ($\text{P}(\phi_{j,t} = +1)$), bearish ($\text{P}(\phi_{j,t} = -1)$), or neutral ($\text{P}(\phi_{j,t} = 0)$). Predicted values for our sentiment measure follow from the fitted sentiment model:
    \begin{align}
        \Delta \widehat{\Phi}_{j,t} = \frac{1}{2} \left( \widehat{\Phi}^+_{j,t} - \widehat{\Phi}^-_{j,t} \right) - \Phi_{j,t-1},
    \end{align}
    where a hat denotes the values fitted from the first stage regressions. 

    \paragraph{Results}
    
    In all our estimates, we restrict ourselves to a sub-sample spanning January 2016 to July 2020. This choice serves to limit the amount of missing data in times when activity on WSB was relatively sparse.
    
    Table \ref{tab:market_impact_FS} helps assess the instruments' strength in predicting sentiments on WSB. The high F-statistics justify that the explanatory variables are not weak instruments. In both columns, we find that lagged weekly mean and variance in returns, combined with lagged sentiments, are significant predictors for the current log-odds in average weekly submissions expressing bullish and bearish sentiments. This is in line with our findings in Section \ref{sec:social_dynamics}.
    
    \begin{table}[!ht]
\begin{center}
  \caption{First Stage estimates for consensus and contagion in WSB} 
  \label{tab:market_impact_FS} 
\begin{tabular}{@{\extracolsep{5pt}}lcc} 
\toprule \\[-1.8ex] 

& \multicolumn{2}{c}{\textit{Dependent variable:}} \\ 
& $\Phi^+_{j,t}$ & $\Phi^-_{j,t}$ \\ 
\cline{2-3} \\[-1.8ex]
    $\Bar{r}_{j,t-1}$ & $-$0.0002 (0.52) & $-$1.53$^{**}$ (0.65) \\ 
    $\sigma^2_{j,t-1}$ & $-$3.78$^{***}$ (0.98) & $-$3.75$^{***}$ (0.68) \\ 
    $\Phi^+_{j,t-1}$ & 0.09$^{***}$ (0.02) & $-$0.06$^{***}$ (0.01) \\ 
    $\Phi^-_{j,t-1}$ & $-$0.03$^{***}$ (0.01) & 0.16$^{***}$ (0.01) \\ 
    \hline \\[-1.8ex] 
    Week FE & Yes & Yes \\
    Number of obs. & 6,711 & 6,711 \\ 
    F-statistic & 17.63 & 49.53 \\ 
    
\bottomrule 
\end{tabular} 
\end{center}
\vspace{-0.6em}
\footnotesize{ \textit{Notes}: the dependent variable in Column (2) is the average log-odds of a given submission in week $t$ on stock $j$ to express bullish over neutral sentiment, and in Column (3) -- bearish over neutral sentiments. Explanatory variables include: the average log-return $\Bar{r}_{j,t-1}$, and the variance in log-returns $\sigma^2_{j,t-1}$. The logit-transformed sentiments are regressed on the lag of the weekly mean and variance of log-returns, as well as the lag in logit-transformed sentiments. Each specification includes week-specific fixed effects. Accompanying standard errors, displayed in brackets, are clustered at the stock level, and calculated in the manner of \citet{mackinnon1985some}.

*** Significant at 1\% level ** Significant at 5\% level * Significant at 10\% level}

\end{table}
    
    \begin{table}[ht!]
\begin{center}
\caption{Market impact of WSB discourse}
\label{tab:market_impact}
\begin{tabular}{lc}
\toprule

\multicolumn{2}{l}{\textbf{Panel A}: Reduced Form relationship between WSB and market activity} \\ \\[-1.8ex] 
 
    & \textit{Dependent variable:} \\
    & $\Delta \Bar{r}_{j,t}$ \\
    \\[-1.8ex] 
 
    $\Delta \Phi_{j,t}$ & 0.002$^{***}$ (0.0003) \\ 
    \hline \\[-1.8ex] 
    Week FE & Yes \\
    Number of obs. & 6,671 \\ 
    F-statistic & 24.32 \\
 
\toprule \\[-1.8ex] 

\multicolumn{2}{l}{\textbf{Panel B}: structural relationship between  WSB and market activity} \\ \\[-1.8ex] 

    $\Delta \widehat{\Phi}_{j,t}$ & 0.004$^{***}$ (0.001) \\ 
    \hline \\[-1.8ex] 
    Week FE & Yes \\
    Number of obs. & 6,671 \\ 
    F-statistic & 12.63 \\ 
    J-statistic & 8.108 \\
    
\bottomrule
\end{tabular}
\end{center}
\vspace{-0.6em}
\footnotesize{ \textit{Notes}: this table presents OLS estimates for stock $j$'s change in average log-return, $\Delta \Bar{r}_{j,t}$, in week $t$. We filter the sample to stocks mentioned in at least 31 distinct submissions on WSB, and exclude any ETFs. Explanatory variables include a measure for sentiment change, $\Delta \Phi{j,t}$, which tracks the change in average sentiments on WSB. Each specification includes week-specific fixed effects. Accompanying standard errors, displayed in brackets, are clustered at the stock level, and calculated in the manner of \citet{mackinnon1985some}. Panel A computes the coefficients using values directly from WSB data, whereas Panel B employs sentiments and stock discussion predicted by past sentiments, stock discussions, as well as returns and return volatility, for which results are in Table \ref{tab:market_impact_FS}. The associated J-statistics are recorded at the bottom of Panel B, which are computed by regressing the residuals from the Second Stage on all variables used for predicted $\Delta \widehat{\Phi}_{j,t}$.

*** Significant at 1\% level ** Significant at 5\% level * Significant at 10\% level}

\end{table}

    Table \ref{tab:market_impact} presents our main results. Panel A regresses changes in average returns against \textit{observed} measures for sentiment changes $\Delta \Phi_{j,t}$. Panel B in Table \ref{tab:market_impact} presents causal evidence for the impact of sentiments among WSB users on stock market variables, using \textit{predicted} sentiments from the model presented in Table \ref{tab:market_impact_FS}. The effect in Panel B from our Reduced Form appears larger than our First Stage - we explain this through the fact that the significance of the first stage is decreased due to our weekly fixed effects, which impact returns and sentiments simultaneously. The estimated average effect is small, which is not surprising given that many of the stocks discussed on WSB have large market capitalisation.
    
    We do not argue that WSB alone affects the markets, but rather that WSB data offers a rich sample of retail investor behaviour. Variation in sentiments explained by the past offers a measure for the intensity by which retail investor asset demand propagates from one week to the next. Even though sentiments reflect current returns, prior beliefs are expected to change prices, thus returns, beyond the market average accounted for by time fixed effects.
    
    \subsection{Evidence of granularity}
    \label{sec:giv}
    

    Thus far, our paper demonstrates that strategic information complementarities can drive persistence in sentiments and oscillations in returns. Our empirical exercise in Section \ref{sec:social_dynamics} also shows that individual sentiments are not \textit{fully} explained by information from peers or recent returns. There is, therefore, unexplained heterogeneity in individual investor sentiments. An outstanding question is whether these heterogeneous opinions can survive aggregation across peers and impact asset returns. In this section, we strive to test Proposition 3 from Section \ref{sec:model}.

    \paragraph{Model framework}

    We remind ourselves of the proposed model framework, which captures the impact of individual, idiosyncratic demand shocks for an asset by investor $i$, $e_i$, on price:
    \begin{align*}
        p &= \sum_{i=1}^N s_i \mathbb{E}_i(v) - \gamma \sigma^2 S + \sum_{i=1}^N s_i e_{i}.
    \end{align*}
    In the absence of `granularity' among investors, the shocks average out to zero. However, when certain investors are weighted differently to others due to differences in capital or popularity, the shocks can have an impact on price, as highlighted above. If there is a granular shock at time $t$, we would therefore expect a change in the log price, which would manifest in a correlation between the granular social shock and returns, as well as increased volatility. 
    
    

    \paragraph{Granularity of social attention}

    We leverage the heavy-tailed structure of WSB discussions for our empirical strategy. The intuition is that certain submissions gather many more followers than others, which we measure using the number of comments they receive. Figure \ref{fig:tail_ticker_mention} in Appendix \ref{app:most_frequent_tickers} displays the heavy-tail in discussions \textit{between} assets -- a handful of tickers are mentioned in thousands of posts, while most assets receive just a small number of mentions. Attention \textit{within} stocks is also dominated by a few, heavily-commented submissions. In our modification to our model, the heavy-tail of attention can result in unexplained variation in sentiments surviving aggregation and impacting returns. We use a Granular Instrumental Variable (GIV) approach to investigate the impact on returns \citep{gabaix2020granular,gabaix2021search,galaasen2020granular}. 
    
    

    \begin{figure}[ht!]
    \begin{subfigure}{.47\textwidth}
        \centering
        \includegraphics[width=\textwidth]{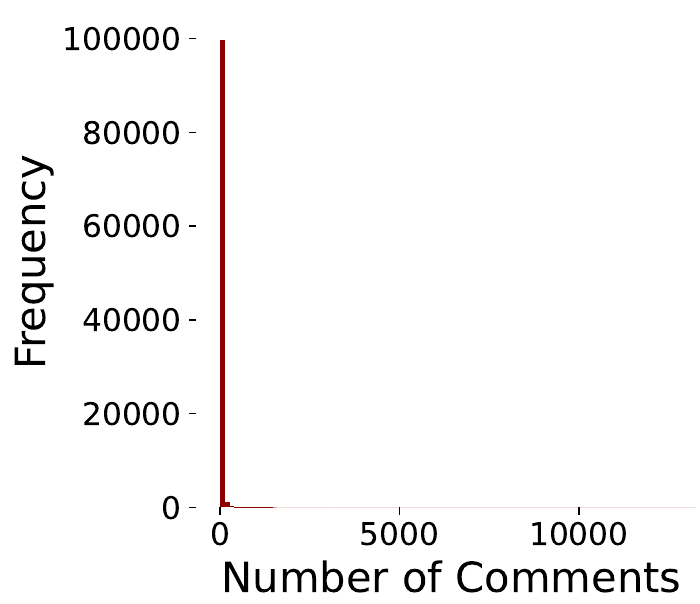}
        \caption{Distribution of comments across all submissions}
        \label{fig:attention_heavy_tail1}
    \end{subfigure}
    \hfill \hfill
    \begin{subfigure}{.47\textwidth}
        \centering
        \includegraphics[width=\textwidth]{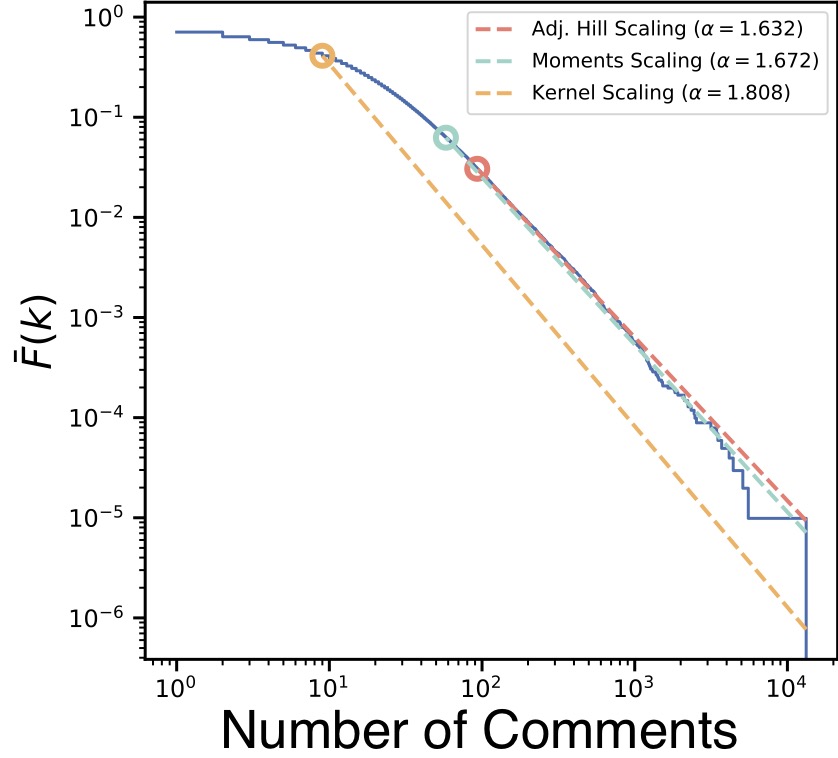}
        \caption{Tail estimate for the distribution of comments}
        \label{fig:attention_heavy_tail_voitalov}
    \end{subfigure}
    \caption{\footnotesize{\textbf{Attention is heavy-tailed}; These graphs show the distribution of comments that submissions that mention a ticker on WSB receive. The left figure plots the distribution. The right plots the tail exponent, estimated using the three different methods outlined in \cite{voitalov2019scale}. The tail exponent is estimated to be less than two across all methods -- this implies that the tail distribution obeys a power law, and is heavy-tailed. }}
    \label{fig:attention_heavy_tail}
    \end{figure}

    We begin by establishing that the \textit{distribution of attention} that information shared by investors online receives is heavy-tailed. We proxy attention by the number of comments that a particular submission receives. Figure \ref{fig:attention_heavy_tail} shows the extreme tail in the distribution of attention -- some submissions appear to receive a large following, while the majority are of little interest. We study the distribution of comments using the approach in \cite{voitalov2019scale}, who propose several methods for estimating the power-exponent of a distribution's tail -- all methods estimate the tail exponent to be less than two, implying that the tail is power-law and heavy-tailed. The heavy-tailed attention online implies that idiosyncratic information contained within the most popular submissions potentially persist after pooling across all investor's opinions, and may have a disproportionate effect on returns. For our identification strategy, we exploit within-ticker-week variation in attention. Therefore, we filter our sample to weeks and tickers where a sufficient number of submissions are made to distinguish between highly popular and less popular sentiments - we choose five submissions for our cutoff. We subsequently test for whether the week in question exhibits granular social attention by fitting a pareto distribution to the popularity of posts within a week and selecting ticker-week combinations where the exponent is less than two. The approach has certain shortcomings as there are challenges to finding the exponent with a small sample of data, however, allows us to estimate whether the activity in a ticker in a given week is a good candidate for granular social dynamics. We ensure our result is robust by imposing several different thresholds and varying our approach between applying and not applying the heavy-tail filter; we observe that all results are consistent with the findings presented below.

    \paragraph{Preliminary analysis}
    As a preliminary analysis, we test for the link between granular social activity and volatility. Table \ref{tab:giv_vol} shows that there is a statistically significant link between the standard deviation of asset $j$'s returns in week $t_w$ and the existence of granularity in social attention in that week (as defined above). The effect indicates that granularity in social discussions are linked to a 2\% increase in asset volatility in a given week, on average.

    \begin{table}[!ht]
\begin{center}
  \caption{Relationship between social granularity and volatility} 
  \label{tab:giv_vol} 
\begin{tabular}{@{\extracolsep{5pt}}lc} 
\toprule \\[-1.8ex] 

& \textit{Dependent variable:} \\ 
& $\sigma^2_{j,t_w}$ \\ 
\cline{1-2} \\[-1.8ex]
    Granularity Indicator & 0.0076$^{***}$ (0.000) \\ 
    $\sigma^2_{j,t_w-1}$ & 0.126$^{***}$ (0.001) \\ 
    $r_{j,t_w}$ & 0.000 (0.000) \\  
    \hline \\[-1.8ex] 
    Week FE & Yes  \\
    Number of obs. & 2,479,664 \\ 
    Adjusted R$^{2}$ &  0.029\\ 
    
\bottomrule 
\end{tabular} 
\end{center}
\vspace{-0.6em}
\footnotesize{ \textit{Notes}: the dependent variable is the variance in the log returns of asset $j$ in week $t_w$, $\sigma^2_{j,t_w}$. Explanatory variables include: the variance in log returns in the previous week $\sigma^2_{j,t_w-1}$, the average log returns in the present week, $r_{j,t_w}$, as well as an indicator for whether we observe social granularity in the week (the indicator equals one  if the ticker receives five posts in a given week and a pareto-fit indicates that the distribution of popularity is heavy-tailed). The sample includes all tickers discussed on WSB since 2016; the sample period begins in 2016 and ends at our WSB cutoff time. Accompanying standard errors, displayed in brackets, are clustered at the stock level.

*** Significant at 1\% level ** Significant at 5\% level * Significant at 10\% level}

\end{table}
    
    The analysis is only a preliminary indication and does not imply causality. Furthermore, we cannot perform the same test for returns, since the direction of idiosyncratic demand (long or short) is important to quantify. Identifying the link between granular social shocks and returns is tricky due to confounding variables -- returns and popularity of online content about an asset may both be driven by news or other market factors. The next step of our empirical approach therefore consists of extracting idiosyncratic social shocks, measured as unexplained idiosyncratic variation in the sentiments of submissions. 

    \paragraph{Estimates of idiosyncratic social shocks}
    
    To extract unexplained variation in sentiments, we regress the sentiment expressed in a given post on the return on day $t$, $r_{j,t}$, the cumulative returns on the week containing day $t$, $r_{j,t_w}$, and average sentiments expressed by peers in the prior week $\Bar{\Phi}_{j,t_w-1}$. Since our analysis includes weekly and daily variables to better extract idiosyncratic social shocks, we distinguish between $t$ time in days, and $t_w$ time in weeks. For a post about asset $j$ made by author $i$ at time $t$, we estimate the idiosyncratic information content of the post as $e_{i,j,t}$ in the following regression:

    \begin{align}
        \Phi_{i,j,t} = \beta_1^A r_{j,t} + \beta_2^A r_{j,t_w} + \beta_3^A X_{t_w} + \beta_4^A \Bar{\Phi}_{j,t_w-1} + e_{i,j,t},
        \label{eq:eq_idiosyncratic_erorr}
    \end{align}
    where $X_{j}$ are asset fixed effects. We add asset fixed effects in order to control for the fact that certain sentiments about assets are persistent on WSB -- for example, the forum's enthusiasm about TSLA. We note that the results remain similar with and without ticker fixed effects.
    
    The strategy follows the reasoning outlined in Section \ref{sec:consensus_results}, Figure \ref{fig:frequent_posters_dag}. We posit that any news that emerges at time $t$ about a company is assimilated by the market and manifests in returns. Any variation in sentiment that is unexplained by market performance at time $t$ and by past discussions is post-specific and is idiosyncratic to news and information more broadly available about that stock at that time.

\begin{figure}[!ht]
        \centering
        \includegraphics[width=0.8\textwidth]{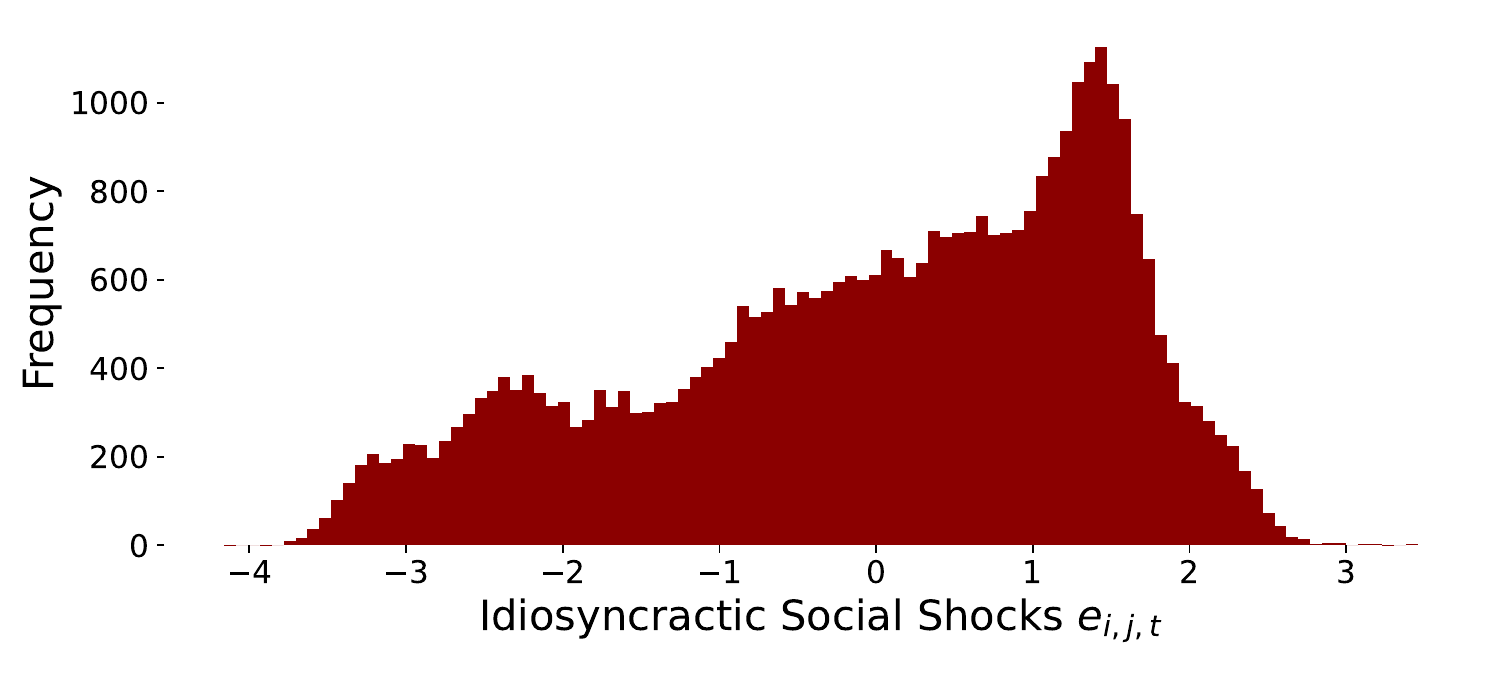}
        \caption{\footnotesize{\textbf{Idiosyncratic social shocks:} this graph plots the distribution of idiosyncratic sentiment heterogeneity, $e_{i,j,t}$, from Eq. \ref{eq:eq_idiosyncratic_erorr}.}}
        \label{fig:giv_residuals}
    \end{figure}

    The object of interest, residual $e_{i,j,t}$, is information shared in the submissions that is orthogonal to asset $j$'s returns at time $t$ or within week $t_w$. We, therefore, would expect $e_{i,j,t}$ to have an impact on the market through social forces, rather than through purely informational content. Figure \ref{fig:giv_residuals} plots the distribution of idiosyncratic social shocks. The distribution is somewhat asymmetric and the modal, unexplained sentiment is bullish, but the left tail of discussions demonstrates the presence of intense bearish discourse.

    \paragraph{Estimating the effect of granular attention}
    
    In order to assess the impact on asset returns of granular social attention, we proceed by analyzing the following relationship:

    \begin{align}
        \centering
        r_{j,t_{w}+1} = \beta_1^B \Bar{e}_{j,t_w} + \beta_2^B r_{j,t_w} + \beta_3^B X_{t_w} + v_{j,t_w},
        \label{eq:giv_simple_form}
    \end{align}
    where $X_{t_w}$ are week fixed effects, $r_{j,t_{w}+1}$ is the cumulative return of stock $j$ in week $t_w+1$, $r_{j,t_w}$ is the cumulative return in asset $j$ in week $t_w$, $\Bar{e}_{j,t_w} = \sum_{i} s^c_{i,j,t_w} e_{i,j,t_w}$ the popularity-weighted average idiosyncratic sentiment shared about stock $j$ in week $t_w$ ($\sum_i s^c_{i,j,t_w} = 1$), and $v_{j,t_w}$ is a stock-week specific error. $s^c_{i,j,t_w}$ is calculated by summing the total number of comments received across all posts in stock $j$ in week $t_w$ -- the comment count for a specific submission is then normalized by the total comment count within week $t_w$; comment count is re-indexed so that the minimum number of comments a post receives is one. We choose to model future returns in order to mitigate any confounding variables. 

    A key identification challenge stems from our goal to identify the impact of social forces and popularity of some content over other content, versus general idiosyncratic sentiment on the forum. We rely on a GIV for our identification strategy. The GIV is defined as the difference between popularity-weighted and equally-weighted social shocks, each aggregated for the stock $j$ at period $t_w$:
    \begin{align}
    \centering
    GIV_{j,t_w} = \sum_{i} s^c_{i,j,t_w} e_{i,j,t_w} - \sum_i \frac{1}{N_{j,t_w}} e_{i,j,t_w},
    \label{eq:giv_social_shocks}
    \end{align}
    where $N_{j,t_w}$ is the total number of posts about stock $j$ in week $t_w$, and $i$ at the authors who are active in the discussion about asset $j$ in week $t_w$. We subsequently replace $\Bar{e}_{j,t_w}$ in Eq. \ref{eq:giv_simple_form} with $\Hat{u}_{j,t_w}$, where $\Hat{u}_{j,t_w}$ is the predicted values from the regression of the GIV on the social shocks $\Bar{e}_{j,t_w}$. The outcome variable $\Hat{u}_{j,t_w}$ is driven by the popularity of certain posts over others, rather than by general idiosyncratic sentiment.

    \paragraph{GIV requirements and threats to identification}
    
    In addition to capturing the impact of social forces, our specification allows us to disentangle asset properties which affect both sentiment and returns simultaneously. The use of the GIV allows us to mitigate the common shocks problem, where certain stock-specific shocks could affect all idiosyncratic sentiments and future returns -- this is similar to \cite{galaasen2020granular} in reasoning. Specifically, stock $j$'s returns in week $t_{w}+1$ may be driven in-part by social forces, but also by asset properties which affect both sentiment and returns, which we are unable to control for while estimating idiosyncratic sentiments in Eq. \ref{eq:eq_idiosyncratic_erorr}. A correlation of these shocks with $\Bar{e}_{i,t_w}$ may result for a biased estimator for $\beta_1^B$. More formally, outcome variable (after imposing controls from Eq. \ref{eq:giv_simple_form}) $y^r_{j,t_{w}+1}$ may be of the form:
    \begin{align}
    \centering
    y^r_{j,t_{w}+1} = \beta_1^{B} \Bar{e}_{j,t_w} + \eta_{j,t_{w}},
    \end{align}
    where $\eta_{j,t_{w}}$ is the `common shock' to asset $j$ in period $t_w$.

    We assume that the idiosyncratic social shock from a post can be expressed as having a stock level component, common to all posts about the stock within that time period, and a post-level component: 
    \begin{align}
    \centering
    e_{i,j,t_w} = \beta^{cs}_{j,t_{w}} \eta_{j,t_{w}} + u_{i,j,t},
    \end{align}
    where $\beta^{cs}_{j,t_{w}}$ is the sensitivity of posts within week $t_w$ to the common shock to stock $j$. 
    
    The identification strategy rests on the assumption that the popularity of the idiosyncratic content of posts $s^c_{i,j,t_w} e_{i,j,t_w}$ is not correlated with stock-week shocks $\eta_{j,t_{w}}$. More formally, we require $E[u_{i,j,t}\eta_{j,t_{w}}] = 0$: the idea is that there are social shocks which make certain content on WSB popular over other content, but that is orthogonal to shocks affecting asset $j$ in week $t_w$. This is not a problem in this setting for several reasons. Firstly, the popularity of a post could potentially be linked to a stock through it's informativeness about the asset's price. However, in the creation of our social idiosyncratic shocks, we extract shocks while controlling for returns on the day and the week of the post. Our social shock time series is, therefore, orthogonal to asset returns at time $t$ and in week $t_w$ and is, therefore, orthogonal to new information available to investors at the time. Furthermore, we find post popularity to be uncorrelated with asset returns on day $t$ and week $t_w$ on which the post is made. As a final precaution, we look at the time period for which a post is active on WSB, where the final time that a post is active is the final comment activity on the post (if the post receives no comments, it is simply the time of the post). We remove posts from our sample that receive commenting activity into the week following the post. On WSB we observe data about relatively unsophisticated retail investors where the sentiments of posts at time $t$ about an asset are systematically linked to \textit{negative} future returns (this holds both when we take a raw average and popularity-weighted average average sentiment). This is additional proof that investors we observe do not have access to information on stock-level shocks. Finally, both our shock and popularity time series is constructed at time $t_w$, while the dependent variable is observed at time $t_{w}+1$, avoiding contemporaneity issues.
    
    \paragraph{Results}

    \begin{table}[ht!] 
\begin{center}
  \caption{Stock returns versus granular social shocks}
  \label{tab:giv1} 
\begin{tabular}{@{\extracolsep{-2pt}}lcc|ccc} 
\\[-1.8ex]\hline 
\hline \\[-1.8ex] 
 & \multicolumn{5}{c}{\textit{Dependent variable: $r_{j,t_w+1}$}} \\ \\[-1.8ex]
\cline{2-6} 
 \\[-1.8ex] & Average & Popularity- & \multicolumn{3}{c}{Instrumented by GIV: $\Hat{u}_{j,t_w}$} \\
 & $\Bar{e}_{j,t_w}$ & weighted $\Bar{e}_{j,t_w}$ && \\
 & (1) & (2) & (3) & (4) & (5) \\
\hline \\[-1.8ex] 
\\[-1.8ex]Granular Social & 0.007 & 0.010$^{***}$ & 0.012$^{***}$ & 0.013$^{***}$ & 0.011$^{**}$\\ 
 Shock & (0.005) & (0.004) & (0.005) & (0.005) & (0.005) \\
\\[-1.8ex] $r_{j,t_w}$  & -0.099$^{***}$ & -0.100$^{***}$ & -0.059$^{***}$ & -0.102$^{***}$ & -0.052$^{*}$ \\
  & (0.037) & (0.037) & (0.012) & (0.037) & (0.028) \\
 \\[-1.8ex] \hline \\[-1.8ex] 
Ticker FE & Yes & Yes &  No & Yes & No \\
Week FE & No & No &  No & No & Yes \\
Controls in Eq. \ref{eq:giv_simple_form} & Yes & Yes & Yes & Yes & Yes \\
Observations & 2,201 & 2,201 & 2,201 & 2,201 & 2,201 \\ 
R$^{2}_{adj}$ & 0.160 & 0.164 & 0.013 & 0.163 & 0.124 \\ \\[-1.8ex] 
\hline 
\hline
\end{tabular} 
\end{center}
\footnotesize{ \textit{Notes}: this table presents the OLS estimates for the influence of idiosyncratic social shocks on WSB. Columns (1) and (2) present the effect of average idiosyncratic shocks and popularity-weighted idiosyncratic shocks, respectively. Columns 3-5 present various specifications of our instrumented idiosyncratic social shocks $\Hat{u}_{j,t_w}$. Robust standard errors, clustered at the ticker level, are shown in parentheses. The F-statistic for the first stage regression is 2,297. Observations with incomplete data are dropped. \\
*** Significant at 1\% level
** Significant at 5\% level
* Significant at 10\% level} \\ 
\vspace{-2em}
\end{table} 

    In order to study the financial consequences of granular social attention, we run the following regression on weekly returns and posts: 
    \begin{align*}
    \centering
        r_{j,t_{w}+1} = \beta_1^B \hat{u}_{j,t_w} + \beta_2^B r_{j,t_w} + \beta_3^B X_{j} + v_{j,t_w},
    \end{align*}
    where $\hat{u}_{j,t_w}$ are the fitted values of idiosyncratic social shocks on our GIV, $r_{j,t_{w}+1}$ is the cumulative weekly return of stock $j$ in week $t_w+1$, $X_{j}$ are ticker fixed effects. The formulation closely follows that of our extended model.  

    Several patterns emerge from our empirical exercise. Firstly, we observe that the data appears to follow the structure proposed in our model -- idiosyncratic social shocks are positively linked to future returns. In column (4), the estimated effect can be summarised as follows: an estimate for $\beta_1^B$ at 0.013, means that the idiosyncratic doubling in the odds of a very popular post expressing bullish over bearish sentiments (while less popular posts do not express an idiosyncratic sentiment) increases returns in the following week by one percent, on average. The effect is small, but persists across specifications. Consistently with our model prediction, the average idiosyncratic noise, in column (1), has no effect. 
    

    \subsection{Evidence of bubbles}

    Can we identify Reddit's bull runs? \cite{greenwood2019bubbles} propose a transparent classification scheme to identify potential bubbles. They determine time windows in which the price indices for various industry market capitalisations grew at a `rapid' rate, constituting a sample of `run-ups'. These run-ups are then separated into those whose price levels remained constant, versus those whose price levels crashed (where the rate of increase and price level of the crash are pre-selected). A run-up followed by a crash constitutes an instance in which the index experienced a `bubble'.
    
    We adapt the method of \cite{greenwood2019bubbles} to identify large run-ups in stock prices, and test whether activity on Reddit during the price run-up is related to an eventual downturn. The two main variables we use to measure social activity are the number of submissions that mention a given stock in the month preceding a price run-up, as well as the average sentiment expressed in that time window.

    The great difficulty in finding price run-ups for individual stocks is that large price swings are more erratic compared to the returns on broad industries considered by \citet{greenwood2019bubbles}. Importantly, run-ups in stock prices can be immediate under small market capitalisation, but slower with large market capitalisation. An arbitrary condition on time and return magnitude thus introduces considerable selection concerns into a sample of run-ups in stock prices. This calls for a more flexible method. 
    
    We tweak the methodology from \citep{greenwood2019bubbles}, but still rely on their theoretical framework: the identification of price run-ups, in relation to corresponding price crashes. The goal of our method is specifically to find instances when a price run-up either precedes or follows a subsequent price decline.

    \begin{figure}[ht!]
        
        \begin{subfigure}{.45\textwidth}
        \centering
            \includegraphics[width=\linewidth]{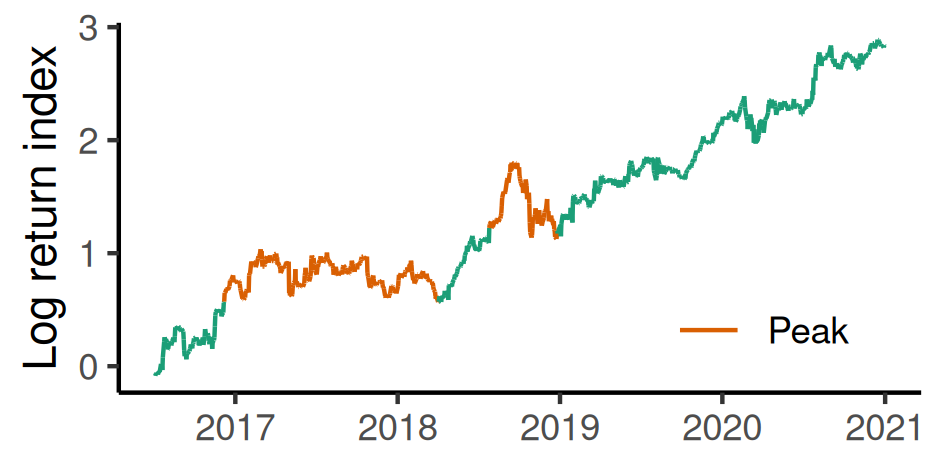}
            \caption{AMD Peaks in returns}
            \label{fig:AMD_peaks}
        \end{subfigure}
        \hfill
        \begin{subfigure}{.45\textwidth}
        \centering
            \includegraphics[width=\linewidth]{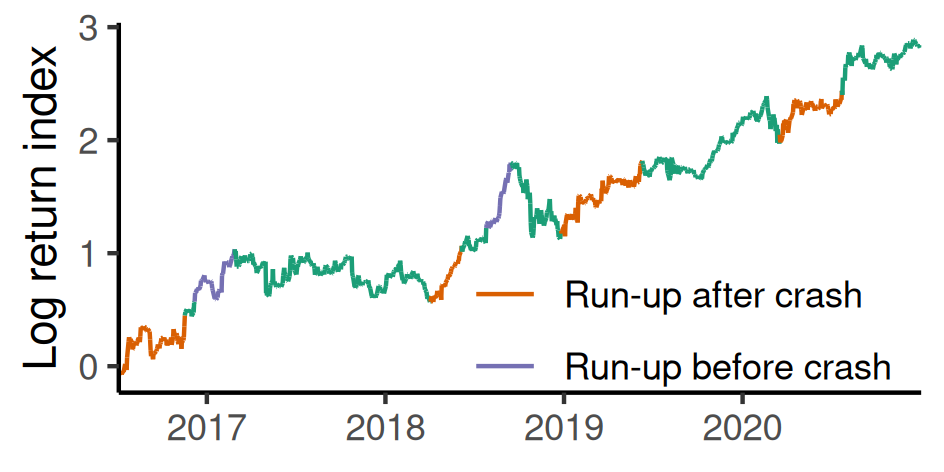}
            \caption{AMD Run-ups relative to crashes}
            \label{fig:AMD_bubbles}
        \end{subfigure}\\\\
        \centering
        \begin{subfigure}{.37\textwidth}
            \includegraphics[width=\linewidth]{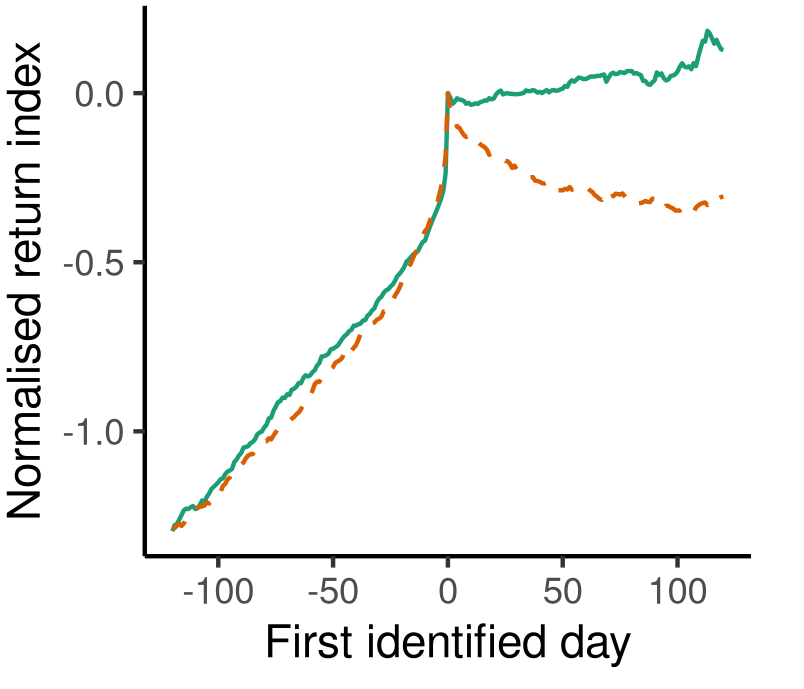}
            \caption{Run-up dynamics}
            \label{fig:bubble_crash}
        \end{subfigure}
        \caption{\textbf{Price run-ups identification}; Figures \ref{fig:AMD_peaks}, \ref{fig:AMD_bubbles} display identified peaks and run-ups in AMD. Figure \ref{fig:bubble_crash} displays run-up dynamics averaged across our sample: returns following a `recovery' run-up, in solid green, remain stable, while they substantially decline after a run-up followed by a crash.}
        \label{fig:detecting_runup}
    \end{figure}

    \paragraph{Identifying price run-ups}
    In order to remain as systematic as possible, we identify `excursions' from local minima in the time series of each stock's log cumulative return index. We define excursions as observations that precede a future minimum, starting from a previous point that matches the cumulative return at that minimum. Specifically, this future minimum is a price level to which the stock does not return at any point following that date. Intuitively, the price of those stocks reverts to a future minimum at the start of the excursion, and thus constitutes a `peak'. Here, we filter excursions with a maximum return from the minimum of over 50\% as peaks. Figure \ref{fig:AMD_peaks} includes two examples of peaks for AMD, one slow run-up starting in late 2016, and an abrupt one in mid 2018.

    In addition to peaks, we can also define troughs as excursions from local maxima. An excursion from a local maximum, as opposed to a minimum, is an observation with a cumulative return lower than the highest value in the stock's history. As such, these function as `inverted' peaks: a series of negative returns following a high price, which ends when the local maximum is recovered. As with peaks, we isolate troughs with a maximum cumulative return over 50\% from the lowest return to the following local maximum.

    In summary, both peaks and troughs constitute price swings of over 50\% in either direction, one with a period of negative returns (troughs) and one of positive returns (peaks). Peaks and troughs differ in the ordering of price run-ups versus crashes: for peaks, the run-up is followed by a crash, whereas the crash is followed by a run-up recovery period for troughs. In this fashion, we can leverage \cites{greenwood2019bubbles} classification of bubbles by studying run-ups that precede crashes (during a peak) and run-ups that follow a crash (during a trough). We illustrate this distinction for the AMD price series in Figure \ref{fig:AMD_bubbles}, where segments in purple correspond to the run-up periods from the peaks highlighted in Figure \ref{fig:AMD_peaks}. Segments in orange represent run-ups that follow crashes.

    We implement this procedure for all stocks traded on the NYSE, NYSE Mkt and NASDAQ, with share codes 10 and 11. In what follows, we restrict the sample of peaks to those after January 1, 2016, and for stocks with an average market capitalisation over one billion USD during the peak. Finally, to harmonise all instances of run-ups with varying lengths of time, we filter out run-ups with an average daily annualised return less than 200\% (corresponding to 0.29\% daily) to match the criterion used by \citet{greenwood2019bubbles}. This leaves us with a sample of 329 run-ups, of which 31 are from a peak -- thus preceding a crash.


    \begin{table}[ht!]
\begin{center}
\caption{WSB features for price runups and crashes}
\label{tab:wsb_bubbles_characteristics}
\begin{tabular}{lcc|cc}
 & Run-up after downturn & Run-up before crash & \textbf{Difference} & \textbf{T test} \\
 & (1) & (2) & (3) & (4) \\
  \\[-1.8ex]
 \cline{2-5} 
 \\[-1.8ex]
 Log Mentions & 0.02 (0.19) & 0.19 (0.63) & 0.17 & 15.27 \\ 
  Sentiment & 0.02 (0.31) & 0.13 (0.52) & 0.10 & 11.41 \\ 
  \end{tabular}
\end{center}
\footnotesize{ \textit{Notes:} this table presents the prevalence of WSB discussions in assets that experience bubble-like dynamics - the mean for each stock sample is shown, with its standard deviation in parentheses. The first column displays that discussions are limited in stocks that are in recovery and do not experience a downturn; however, both mentions and sentiment are higher for stocks that experience a run-up and a subsequent crash, as shown in column (2). }
\end{table}
\vspace{-1.5em}

    \paragraph{Results}
    
    For each type of run-up -- before or after a crash -- we count the number of times a submission is made on WSB in the corresponding time window, mentioning the stock experiencing the run-up. Similarly, we compute the average sentiment expressed in these submissions. Table \ref{tab:wsb_bubbles_characteristics} summarises our findings, along with the corresponding standard deviation. The striking result is that WSB activity, in both the log of mentions as well as the average sentiment, features significantly more prominently in run-ups before crashes, rather than after crashes -- the difference in means is highly significant.

    Are users similarly aware of both types of stocks? One might argue that stocks that experience a crash may have some underlying characteristics which make them of greater interest to the WSB crowd. The share of run-ups preceding crashes during which the corresponding stock is mentioned on WSB, 64.52\%, is close to the same share for run-ups following crashes, at 54.36\%. The samples are thus similarly represented in terms of discussions in WSB. Figure \ref{fig:bubble_crash} demonstrates that the average price trends for both types of run-ups are nearly identical before the cutoff date, then subsequently diverge.

\paragraph{Summary}
Our findings demonstrate a link between bubble-like dynamics in the markets and discussions among investors, and prompt additional important questions for research. Recently, rich datasets have become available to study individual investor portfolios, allowing the study of investor attributes and portfolio choices \citep{balasubramaniam2023owns}, individual extrapolation during asset price bubbles \citep{pearson2021chinese}, and other important characteristics. Combining discussion data and portfolio data offers a promising venue for further research into the profit and loss profiles of hype investors, as well as the behaviour of other market participants faced with hype investor demand. 

\section{Conclusion}
\label{sec:conclusion}

    We contribute to the growing literature on social investing and behavioural finance in several ways. We demonstrate how behavioural finance frameworks can be modified to incorporate peer effects in a setting with strategic information complementary and granularity in social influence, as well as a setting studying asset price bubbles. We justify the assumptions of our modeling framework through a study of peer effects and extrapolation in investor sentiment formation on WSB, and show how our estimates can be used to simulate bubbles. We show that the proposed mechanisms appear to have direct market consequences through impacting retail trader investing patterns, through direct price impacts and through a link to bubble-like market dynamics.
    
    We specifically report empirical estimates for complementarities in asset demand among retail investors, proxied by expressed sentiments. User sentiments are, on average, 14\% more likely to be bullish rather than bearish, if the odds of peers expressing bullish over bearish sentiments double. These results are consistent with the findings of \cite{pool2015people} and \cite{bursztyn2014understanding}. Our observed group of hype traders on WSB appear to weight the sentiments of peers more heavily than extrapolation, when forming their expectations for future price movements. 

    Theoretically, the combination of trend following and peer influence in a standard asset demand context carries certain implications for the behaviour of corresponding prices. Social investors are willing to pay a higher price for an asset they believe others to buy as well. Other factors that determine demand, such as returns when these investors following trends in prices, become less important. This trend following component specifically creates reversals in returns. In the context of bubbles with extrapolation, social influence increases the longevity, but decreases the peak of bubbles, since peer effects imply that individual sentiments have a longer memory. 
    
    In a quantitative exercise, we show that instrumented sentiments from WSB are closely linked to simultaneous returns, as predicted by the model. The instrument allows us to isolate the effect of WSB sentiments on returns, as opposed to the effect of returns on WSB sentiments. We also demonstrate that, in the presence of `granular' heavy-tailed attention, \textit{heterogeneous} investment decisions are not averaged out and can impact returns. Using the method of \citet{gabaix2020granular}, we show that idiosyncratic sentiment heterogeneity among users (which is not reflective of fundamental news) impacts the market due the heavy-tailed nature of the popularity of online content.

    Isn't the WallStreetBets forum a one-off phenomenon? If so, perhaps the behaviours explored here aren't relevant outside of the confines of this study? Even though investor discussion forums have existed for decades, WSB was arguably the first to reach an unprecedented retail following -- the subreddit succeeded at attracting followers not only through lucrative trade ideas, but also through the promise that coordination among smaller retail traders could enable them to oust investment titans. In January, 2021 the forum experienced its first taste of victory, in the form of the GME short squeeze -- beginning a new era of the `hype' trader. The usage of retail investment trading platforms skyrocketed, with \textit{Trading 212} temporarily pausing new account openings in February 2021 due to huge demand, and several other providers struggling to cope with the influx of eager retail investors. Others have developed new features to allow traders to seamlessly execute on the psychological biases explored within this paper: eToro, for example, now offers a \textit{CopyTrader} feature allowing users to precisely mimic the portfolios of others. Other discussion forums similar to WSB are rising to prominence: for example, the new forum \url{r/StockMarketLeakz} is currently one of the top-growing forums on subreddit. Given the recent trends, it is likely that we have only seen the tip of the iceberg, in terms of the impact that hype investors can have on the market. The silver lining is the fact that social media allows us to operate in data-rich landscape, providing opportunities for faster regulatory action and novel research.

\bibliography{bibliography}
\bibliographystyle{agsm}

\normalsize


\appendix
\section*{Appendix}

\section{Data appendix}
\label{app:data_appendix}
\subsection{Extended description of WSB}
\label{app:extended_description}

        \begin{figure}[ht]
                \begin{subfigure}{.45\textwidth}
                \centering
                \frame{\includegraphics[width=\linewidth]{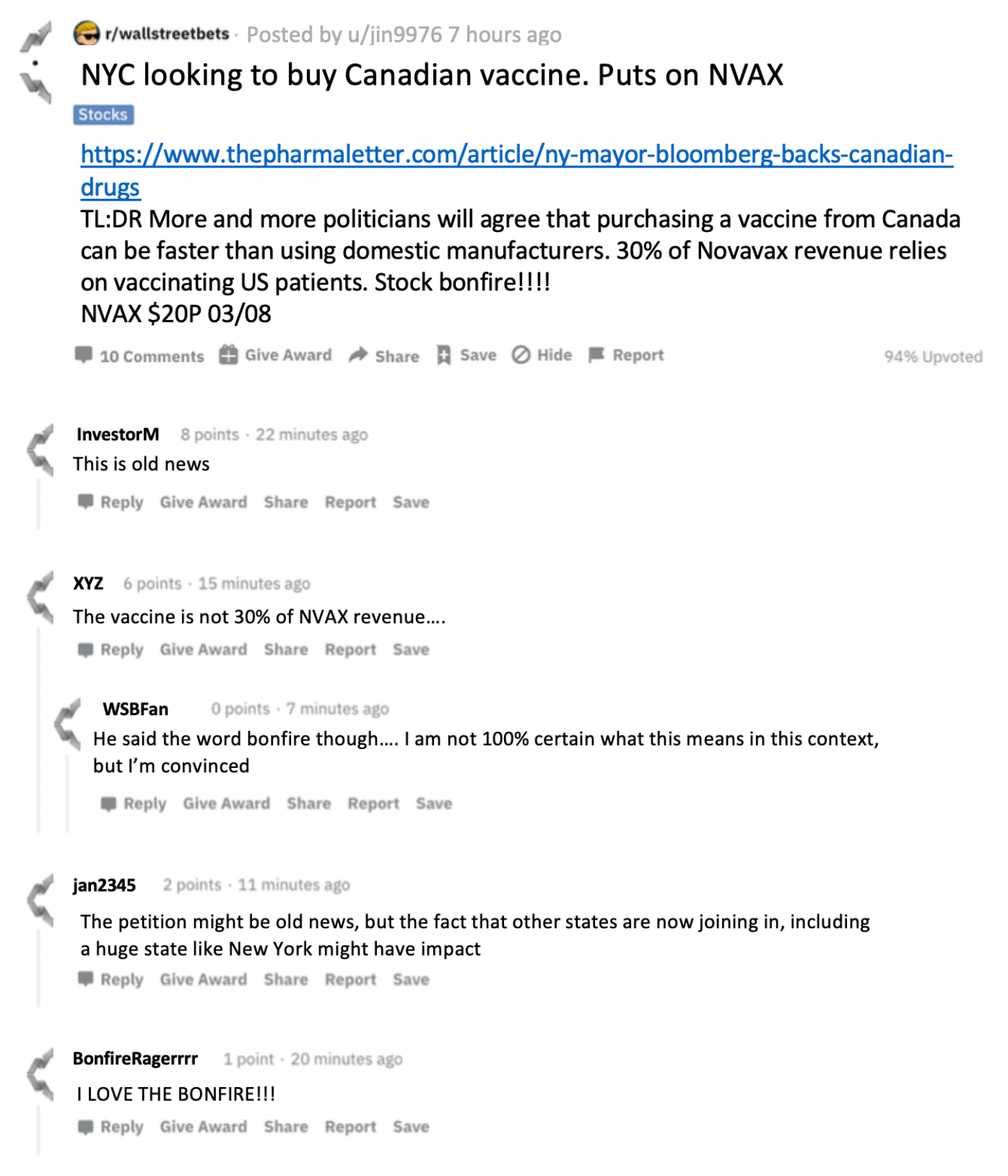}}
                \caption{A typical discussion on WSB} 
                \label{fig:sample_conversation}
            \end{subfigure}
            \hfill
            \begin{subfigure}{.46\textwidth}
                \centering
                \frame{\includegraphics[width=\linewidth]{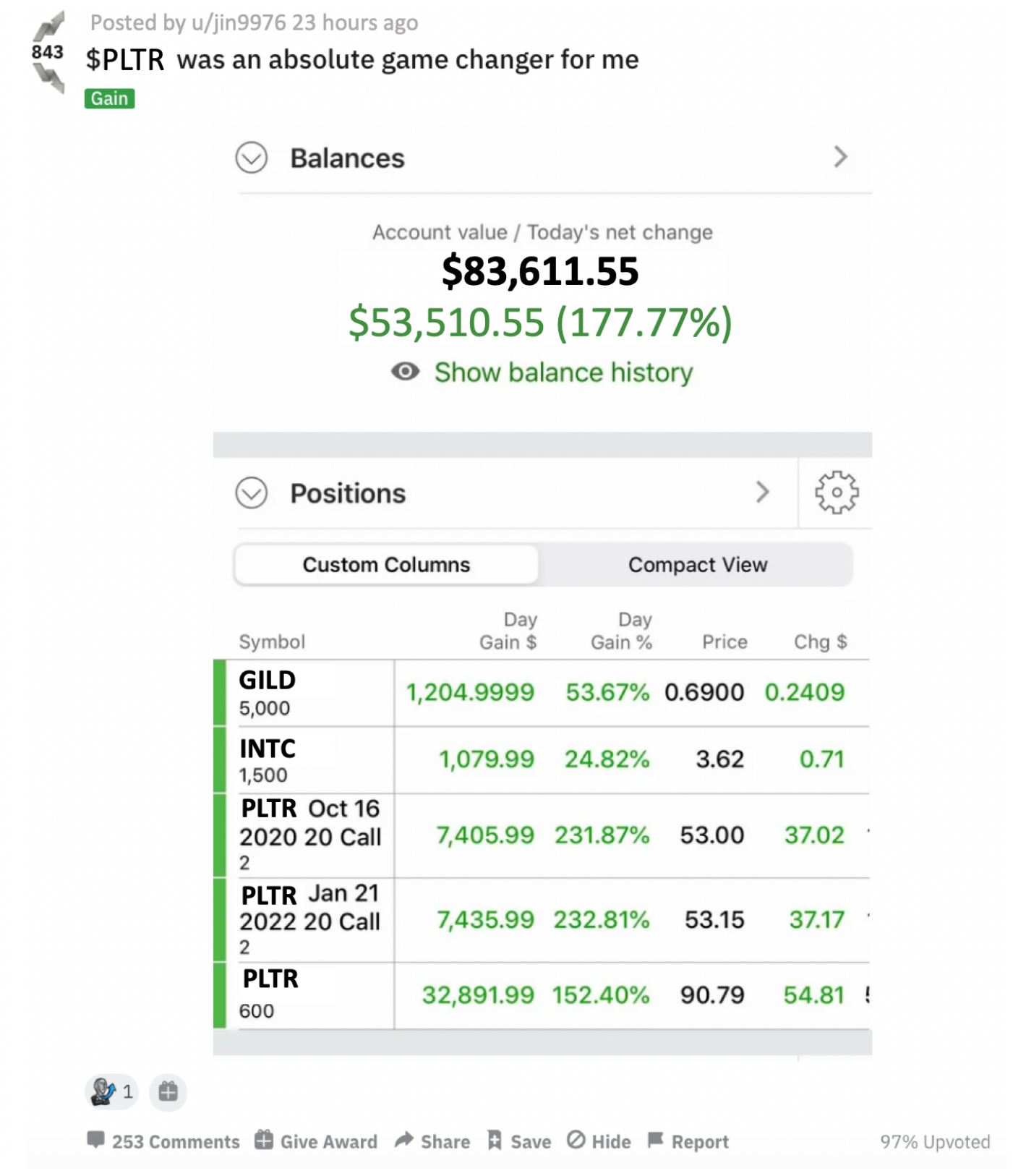}}
                \caption{A sample screenshot of user profits} 
                \label{fig:sample_gains}
            \end{subfigure}
            \caption{\footnotesize{\textbf{What does WSB look like?} These snapshots display typical discussions on WSB. The exact text, usernames, and conversation details have been modified to protect user identities.}}
            \vspace{-0.5em}
            \end{figure} 

        \begin{figure}[ht]
            \begin{center}
             \includegraphics{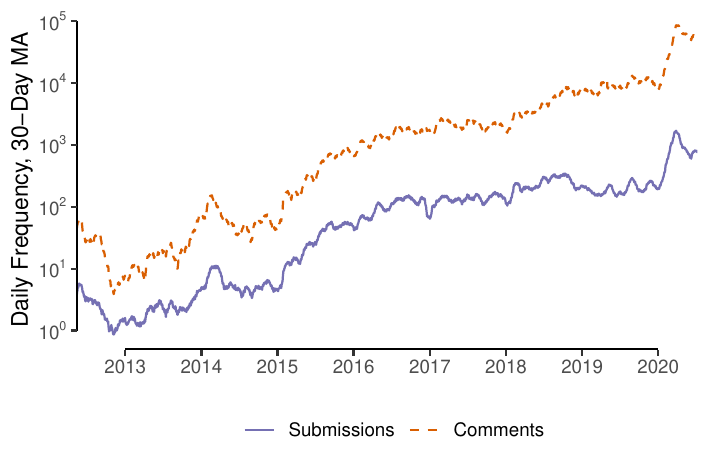}
             \caption{\footnotesize{\textbf{Daily activity on WSB plotted on a logarithmic scale}; the daily submission and comment counts, averaged over 30 days, demonstrate a persistent exponential increase in activity on the WSB forum from 2015 to 2020, with a substantial jump in early 2020.}}
             \label{fig:monthly_post_hist}
            \end{center}
        \vspace{-1.8em}
        \end{figure}
        
Figure \ref{fig:sample_conversation} displays a typical exchange on the WSB forum: individuals discuss stock-related news and their sentiments on whether this will affect stock prices in the future. In addition to market discussions, there is ample evidence of users pursuing the investment strategies encouraged in WSB conversations. Users post screenshots of their investment gains and losses, which subreddit moderators are encouraged to verify, as illustrated in Figure \ref{fig:sample_gains}. 

Figure \ref{fig:monthly_post_hist} displays the evolution of WSB over time. Two jumps are notable: a smaller, seemingly idiosyncratic rise in early 2018, and a sharp spike during the COVID-19 pandemic.
    
\paragraph{Reddit user content presentation}

    Our empirical identification strategy in the \textit{Frequent Posters} approach rests on the premises that users are exposed to random variation in peer sentiments. In this section, we discuss the details of how users are presented with content on Reddit.
    
    Upon logging into Reddit, users are presented with a `home feed'. Historically, the home feed has contained the `top posts' from the subreddits to which a user has subscribed. More recently, Reddit has implemented an algorithm to try and match users to content based on a machine learning algorithm, via the home feed.\footnote{\url{https://reddithelp.com/hc/en-us/articles/4402284777364-What-are-home-feed-recommendations-}} However, this change has only taken effect recently.\footnote{\url{https://www.reddit.com/r/help/comments/rrkptm/home_feed_has_changed_drastically/}} Subreddit top posts are not individually tailored to the specific user. Users have several sort options based on whether they prefer to see most recent or most highly rated content.\footnote{\url{https://www.reddit.com/r/help/comments/7l7686/order_of_posts/}} 
    
    Reddit has, relatively recently, introduced the option to follow individual users, however, following a user means getting exposed to what they post directly to \textit{their own page}, similarly to following an additional subreddit dedicated exclusively to this user. The experience is described as:
    \begin{quote}
        Following is just like subscribing to a subreddit, except the subreddit is your profile page. Reddit recently added the ability to post directly to your profile instead of to a specific subreddit. Posts you post to your profile will be seen on a user’s front page feed if they follow you. Outside of that, following does nothing else except your username will be listed in their subscribed subreddits list.\footnote{\url{reddit.com/r/NoStupidQuestions/comments/9dzp9y/what_does_following_someone_on_reddit_do/}}
    \end{quote} 
    The content viewed on WSB would, therefore, remain consistent for all users regardless of their followership, with random temporal variation, across users. Furthermore, given the fact that the anonymity of the forum is of great appeal, followership ties are rare.\footnote{\url{https://www.reddit.com/r/NoStupidQuestions/comments/7lkwqs/do_reddit_users_actually_follow_other_people/}} We consider how pervasive followership relationships are on Reddit by studying which users actually post to their own profiles (the only way to target content directly at followers). We look through all 42,036 authors who create posts about individual tickers within our sample and observe that less than 3\% of users create content on their own individual user profile pages prior to our data cutoff time, demonstrating the relative lack of content generated for followers and the insignificance of followership relationships on Reddit. Furthermore, we observe that content posted to WSB user's own profile (which we retrieve) is generally unrelated to investment -- investment advice is typically shared on investment-related forums to reach a targeted audience. We test the sensitivity of our results to the users that post to their own profiles remaining in our sample by removing them and rerunning the \textit{Frequency Posters} estimation procedure: the results remain unchanged when these users are removed. 
    
    We conclude that individual users were exposed to WSB content based on the content on the forum that was most recent and popular at the time of their logging on, rather than based on personal preference. This, in turn, allows us to assert that users are exposed to random, temporal variation in peer sentiment. 

\subsection{Tickers mentioned on WSB}
\label{app:most_frequent_tickers}
    
\begin{table}[ht]
\begin{center}
\caption{Most frequent ticker mentions} 
\label{tab:most_frequent_ticker}

\begin{tabular}{llrrr}
  \hline
    Ticker & Name & Comments & Submissions & Sum \\  
    \\[-1.8ex]
    
    SPY & S\&P 500 Index & 291,279 & 9,408 & 300,687 \\ 
    AMD & Advanced Micro Devices, Inc. & 124,685 & 5,721 & 130,406 \\ 
    TSLA & Tesla, Inc. & 124,222 & 5,910 & 130,132 \\ 
    MU & Micron Technology, Inc. & 86,611 & 3,941 & 90,552 \\ 
    AAPL & Apple Inc. & 48,345 & 1,880 & 50,225 \\ 
    AMZN & Amazon.com, Inc. & 44,426 & 1,534 & 45,960 \\ 
    MSFT & Microsoft Corporation & 41,152 & 1,799 & 42,951 \\ 
    SNAP & Snap Inc. & 40,766 & 2,043 & 42,809 \\ 
    NVDA & NVIDIA Corporation & 38,012 & 1,556 & 39,568 \\ 
    SPCE & Virgin Galactic Holdings, Inc. & 30,758 & 1,640 & 32,398 \\  
    FB & Facebook, Inc. & 26,143 & 1,446 & 27,589 \\ 
    DIS & The Walt Disney Company & 25,611 & 1,088 & 26,699 \\ 
    BYND & Beyond Meat, Inc.& 23,299 & 906 & 24,205 \\ 
    NFLX & Netflix, Inc. & 20,800 & 936 & 21,736 \\ 
    JNUG & Direxion Daily Jr Gld Mnrs Bull 3X ETF & 15,761 & 1,095 & 16,856 \\ 
    GE & General Electric Company & 15,730 & 929 & 16,659 \\ 
    RAD & Rite Aid Corporation & 14,781 & 839 & 15,620 \\ 
    SQ & Square, Inc. & 14,003 & 824 & 14,827 \\ 
    ATVI & Activision Blizzard, Inc. & 13,076 & 674 & 13,750 \\ 
    USO & United States Oil & 12,949 & 667 & 13,616 \\ 
   \hline
\end{tabular}
\end{center}
\footnotesize{\textit{Notes}: this table lists the 20 most mentioned assets on WSB, observed by submissions which uniquely mention the related ticker. `Comments' is the number of comments posted on these submissions, `Submissions' counts submissions, and `Total' is the sum of the two. The name of the asset corresponding to the identified ticker is retrieved from \textit{Yahoo Finance}.}
\end{table}

    \begin{figure}[ht]
        \begin{center} 
            \includegraphics{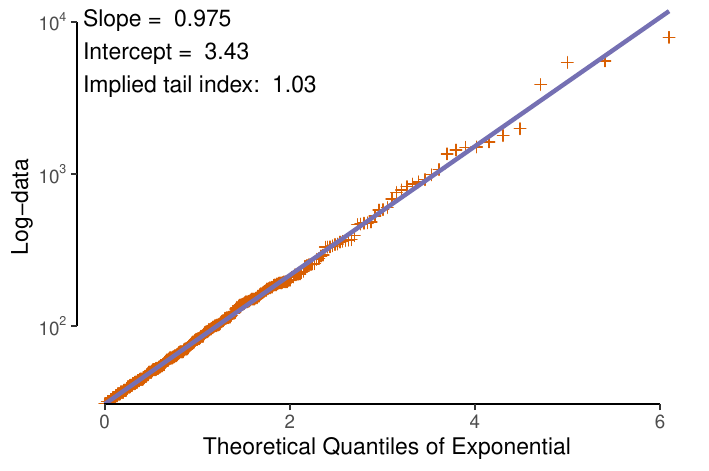} 
            \caption{\footnotesize{\textbf{QQ Plot of the tail in ticker mentions on WSB}; the number of submissions for each ticker (on a log-scale) is plotted against the theoretical quantiles of an exponential distribution. Quantiles are calculated as $q(i) = -\log(1-i/(N+1)$, where $N$ is the number of observations, and $i$ the order of the statistic, from 1 to $N$. The linear fit suggests that the data follows a Pareto distribution, with the tail index equal to the inverse of the slope. The threshold for a ticker to be part of the `tail' is 31 mentions; note the intercept, at $\text{exp}(3.43) \approx 31$.}}
            \label{fig:tail_ticker_mention}
        \end{center}
        \vspace{-2em}
    \end{figure} 
    
    Conventionally, submissions or comments that mention a ticker will spell it using uppercase letters, or following a dollar sign. However, a challenge is that not all uppercase words are valid tickers. 
    
    We first match words in WSB submissions to assets by identifying any succession of two to five capital letters. Subsequently, we used a pre-determined list of tickers from CRSP to check whether a match is indeed present in the available financial data. Some abbreviations or capitalised words which are not valid tickers might still show up, such as `USD' (\textit{ProShares Ultra Semiconductors}), `CEO' (\textit{CNOOC Limited}), and `ALL' (\textit{The Allstate Corporation}). Single characters also appear, such as `A' (\textit{Agilent Technologies, Inc.}). We manually created a list of such tickers, and removed matches featured in WSB submissions, to build a preliminary list of candidate ticker mentions. We refined a second list of candidates by checking whether a collection of one to five letters, lower or uppercase, is preceded by a dollar sign. Any mentions of `\$CEO' or `\$a' count as the tickers `CEO' and `A', respectively. These extracts are, again, checked against the list of available tickers.
    
    A small fraction of the 4,650 tickers we extract dominate the discourse on WSB. 90\% of tickers are mentioned fewer than 31 times, and more than 60\% are mentioned fewer than five times. The~frequency distribution of tail of ticker mentions demonstrates this point, for which Figure \ref{fig:tail_ticker_mention} displays a QQ-plot. We arbitrarily selected tickers with the number of mentions in the top 10\textsuperscript{th} percentile. Even though threshold of mentions for this top decile is 30 submissions, the most popular, SPY, features in almost 8,000 submissions. The orange crosses in Figure \ref{fig:tail_ticker_mention} locate the empirical densities, on a log scale, which are plotted against the theoretical quantiles of an exponential distribution on the x-axis. Under the assumption that ticker mentions are heavy-tailed (similarly to vocabulary distributions), the logarithm of the mentions follows an exponential distribution, with the intercept at the threshold, and the slope equal to the inverse of the tail index. Indeed, the linear fit in Figure \ref{fig:tail_ticker_mention} is close to perfect, supporting the assumption that the popularity of assets in WSB is heavy-tailed, with an estimated tail exponent of approximately $1.03$. In what follows, we used submissions for which we identified a single ticker, unless otherwise specified, forming a dataset of 103,205 submissions with unique ticker mentions by our cutoff date.
    

\newpage
\subsection{Sentiment modeling in WSB posts}
\label{app:bert}

    \begin{table}[ht]
\centering
{\begin{tabular}{c r| c c c}
\multicolumn{2}{c}{}  &  \multicolumn{3}{c}{Predicted Label} \\[2ex]
\multirow{4}{*}{\rotatebox{90}{True Label}}
     &&  \textbf{-} & \textbf{0} & \textbf{+} \\ \cline{2-5}
& \textbf{-} &  64\% & 28\% & 7\% \\
& \textbf{0} &  6\% & 77\% & 17\% \\
& \textbf{+} &  6\% & 27\% & 67\% \\
\end{tabular}}
\caption{\footnotesize{\textbf{Fine-tuned FinBERT confusion matrix:} We use 10\% of our hand-labeled data to test the performance of FinBERT on out-of-sample sentiment prediction. The results highlight the model's ability to predict sentiment with reasonably high accuracy.}}
\label{tab:finbert}
\end{table}
\vspace{-0.7em}
    
        
        
        Our goal, with regards to the text data in WSB, is to gauge whether discussions on certain assets express an expectation for their future price to rise, the `bullish' case, to fall, the `bearish' case, or to remain unpredictable, the `neutral' case. Among other alternatives, we pursued a supervised-learning approach to identify the sentiment expressed about an asset within a WSB submission. This required a training dataset, for which we manually labelled 4,932 random submissions with unique ticker mentions as either `bullish', `bearish' or `neutral' with respect to the authors' expressed expectations for the future price. We used the FinBERT algorithm for labeling \citep{araci2019finbert} - a financially oriented modification of Google's Bidirectional Encoder Representations from Transformers (BERT) algorithm \citep{devlin2018bert}. Work not shown here implements an alternative regression-based approach as a robustness check, but FinBERT performs better out-of-sample.
        
        We trained FinBERT on 75\% of the labelled data, and used the remaining 25\% for validation and the test set. Table \ref{tab:finbert} plots the out-of-sample confusion matrix. For the out-of-sample test, we train FinBERT on 75\% of the available data and use 15\% for validation; we then compute what the algorithm predicts for the remaining 10\% of data. We achieve 70\% accuracy on the test set. This is better than a LASSO regression's accuracy, which was implemented separately and is not cover here.
        
        \begin{figure}[ht!]
            \begin{subfigure}{0.4\textwidth}
                \centering
                \includegraphics[width=\linewidth]{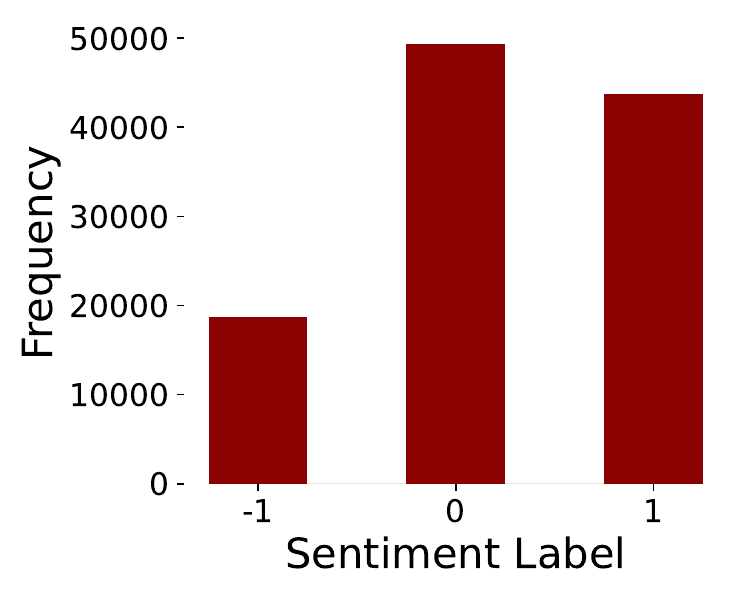}
                \caption{Distribution of Sentiment Labels - Output from FinBERT Classifier} 
                \label{fig:app_sentiment_dist}
            \end{subfigure}
            \hfill
            \begin{subfigure}{0.4\textwidth}
                \centering \includegraphics[width=\linewidth]{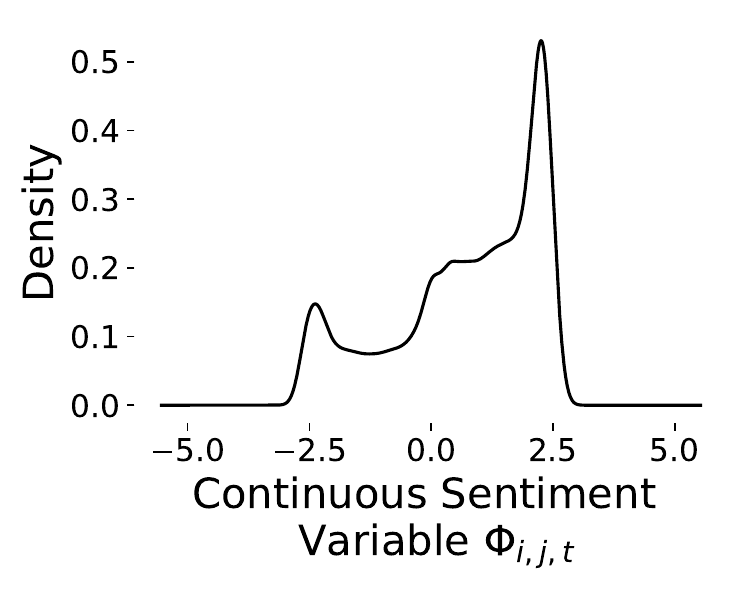}
                \caption{Continuous Sentiment Variable Distribution} 
                \label{fig:app_density_con}
            \end{subfigure}
            \caption{\footnotesize{\textbf{Distribution of Expressed Sentiments on WSB}; We present the density plot of the labeled posts on WSB, and our key continuous variable of log-odds of positive over negative sentiment $\Phi_{i,j,t}$.}}
            \vspace{-0.9em}
            \label{fig:app_density1}
            \vspace{-1.2em}
    \end{figure} 
    
        \paragraph{Data Description} Our final samples contains 111,765 submissions that have a mention of a single, identifiable asset. Figure \ref{fig:app_sentiment_dist} shows that the sample is slightly unbalanced - more posts are labeled as neutral than the other two categories, and more posts appear bullish than bearish. The distribution of our continuous variable is shown in Figure \ref{fig:app_density_con}.
\subsection{Market variables}
\label{app:market_data}

We include a set of market return and volatility control variables. The data source for these variables are the daily stock files issued by the Center for Research in Security Prices (CRSP), accessed through Wharton Research Data Services.

\paragraph{Market variables in Sections \ref{subsec:wsb_characteristics}\&\ref{subsec:consensus}} 

The following market variables serve as controls.

$r_{j,t}$: the log return for asset $j$ on trading day $t$. From CRSP, we calculate it using their `RET' variable: $r_{j,t} = \log(RET_{j,t} - 1)$, which automatically corrects the percentage change in closing prices for share splits and dividend distributions.

$\Bar{r}_{j,t}$: the average log returns for asset $j$ in the five days prior to $t$ (the log return on day $t$ is not included). A minimum of three daily log-return observations is required, otherwise the observation is set as missing. 

$\sigma^2_{j,t}$: the variance of log returns for asset $j$ in the five days prior to $t$ (the log return on day $t$ is not included). A minimum of three daily log-return observations is required, otherwise the observation is set as missing.

\paragraph{Matching submission timings to trade timings}

If a post occurs before 16:00:00 EST on day $t$, we match it with the log-return on the same day $t$. If a post occurs after 16:00:00 EST on a given day, we match it with market data for the next trading day, $t+1$. This is done to capture the fact that many news announcements occur after hours and someone posting after the market close may be exposed to these after-hour moves. Instance in which submissions are made on weekends, or holidays, are matched to the next possible trading day. For example, a submission made at 5pm on Friday is paired to the observed log return for the following Monday. 





\subsection{How prevalent are hype traders?}
\label{app:data_hype_traders}

    \begin{figure}[ht!]
    \begin{center}
    
        \begin{subfigure}[t]{.45\textwidth}
            \includegraphics[width=\textwidth]{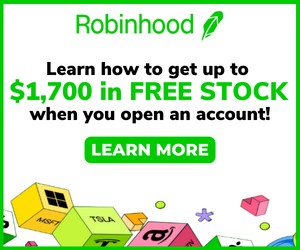}
            \subcaption{Trading platform ad - example 1}
            \label{fig:trade_platform_ad1}
        \end{subfigure}
        \hfill
        \begin{subfigure}[t]{.45\textwidth}
            \includegraphics[width = \textwidth]{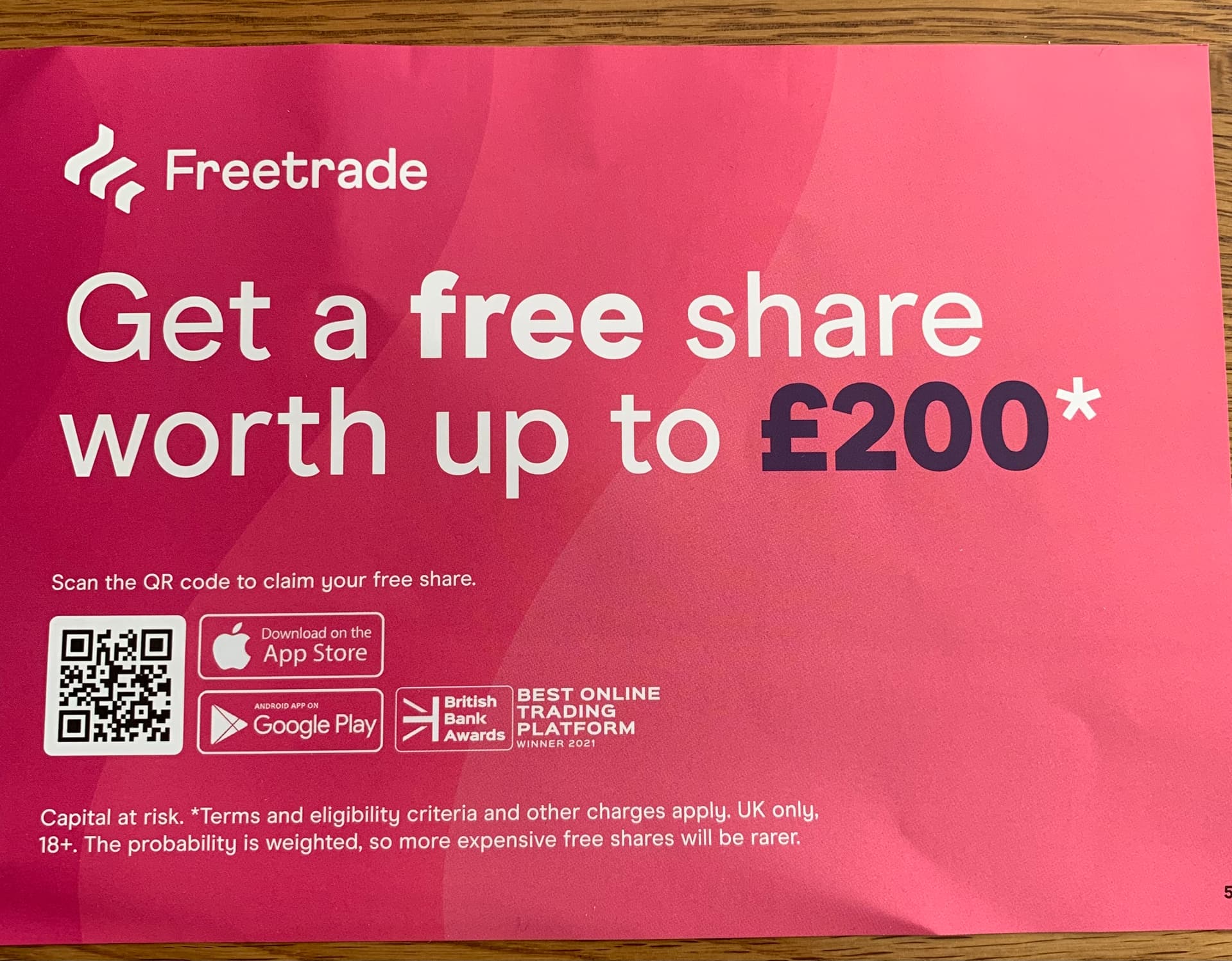}
            \subcaption{Trading platform ad - example 2}
            \label{fig:trade_platform_ad2}
        \end{subfigure}  
        \begin{subfigure}[t]{.45\textwidth}
            \includegraphics[width=\textwidth]{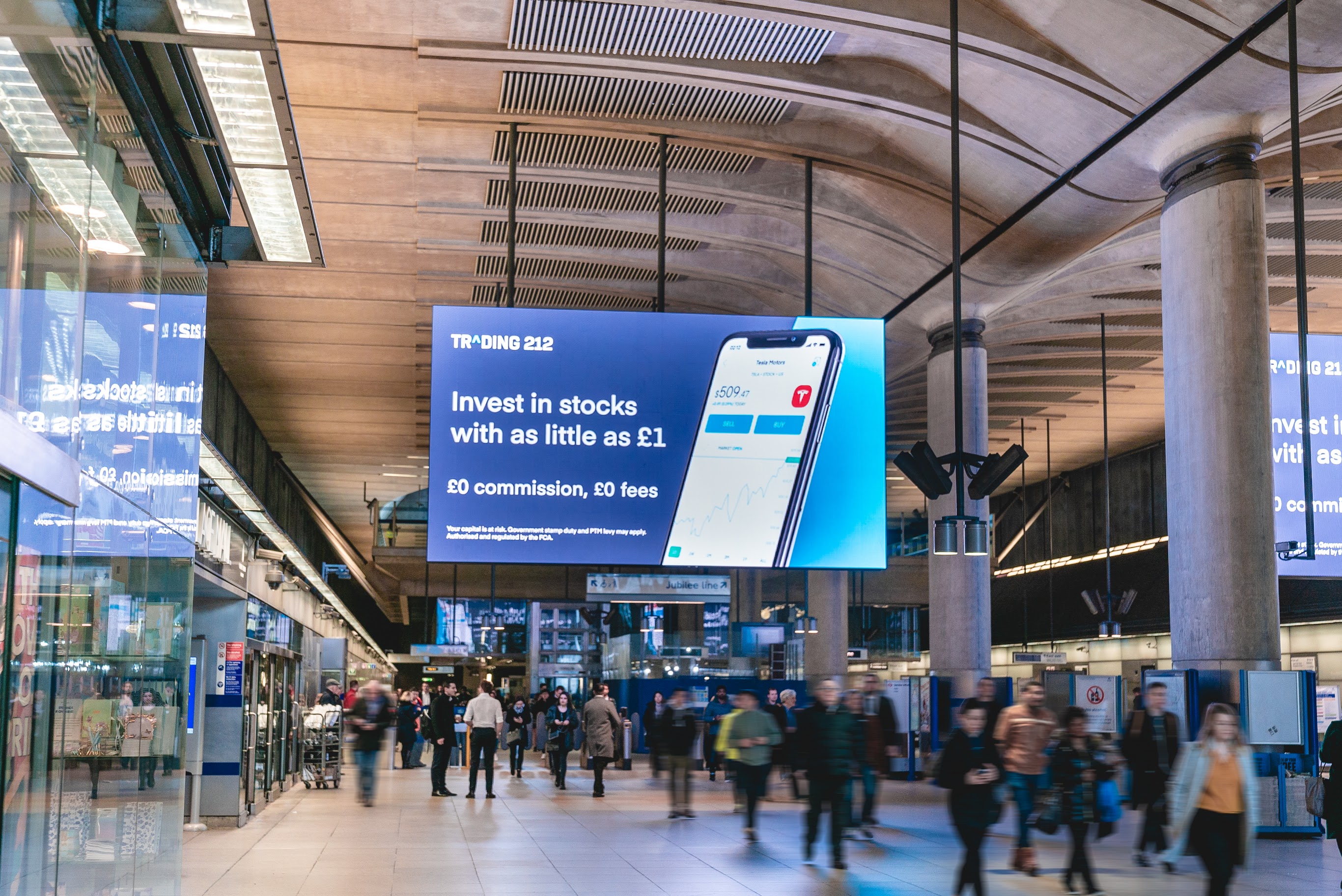}
            \subcaption{Trading platform ad - example 3}
            \label{fig:trade_platform_ad3}
        \end{subfigure}  
        \hfill
        \begin{subfigure}[t]{.45\textwidth}
            \includegraphics[width=\textwidth]{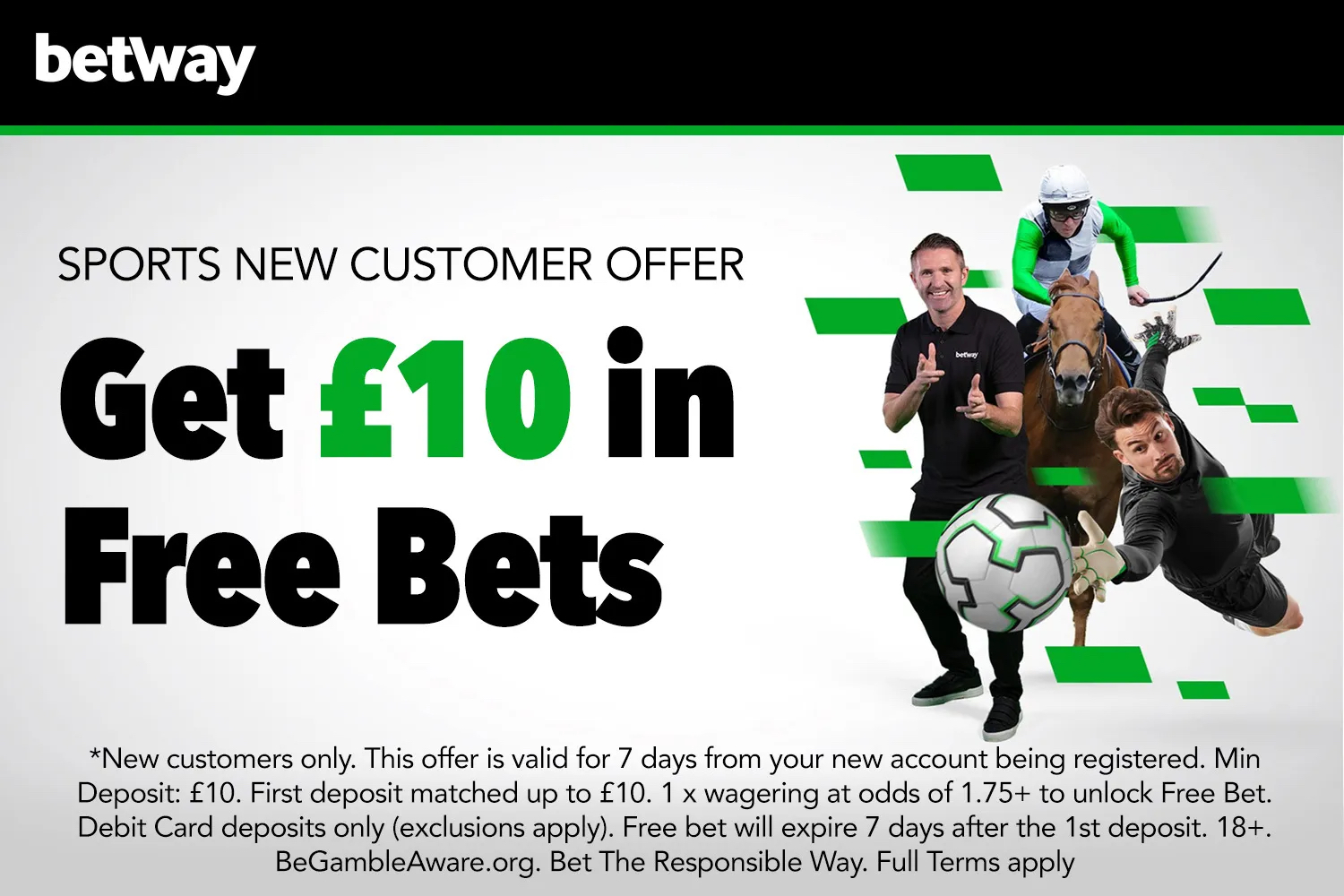}
            \subcaption{Gambling ad}
            \label{fig:gambling_ad}
        \end{subfigure} 
        \vspace{-0.2em}
        \caption{\footnotesize{\textbf{Sample trading and gambling advertisements}; We note the similarity in the style in certain trading advertisement to that of a gambling advertisement - encouraging the retail trader to get `hooked' with an initial free share offering, or to get in with virtually no money in their account.}}
        \label{fig:ads}
        \end{center}
        \vspace{-1em}
    \end{figure}

This study focuses primarily on the WSB discussion forum. However, a potential outstanding question is whether our findings extend to the broader trading population. We present several facts to support the broader relevance of our findings. Firstly, we note that anonymous stock market related forums have skyrocketed in their popularity. WSB, as we had noted previously, has experienced exponential growth and currently boasts fourteen million followers -- putting these numbers into perspective, The Times newspaper recently boasted 7.5 million subscriptions.\footnote{\url{https://www.nytimes.com/2021/02/04/business/media/new-york-times-earnings.html }} However, retail trader appetite for the hype is not satiated, as new forums are gaining popularity: \url{r/StockMarketLeakz} is currently one of the top-growing subreddit forums, while \url{r/personalfinance}, \url{r/CryptoCurrency}, \url{r/bitcoin}, \url{r/stocks} are all among the top 100 forums by number of subscribers. 

The rise of hype traders has not gone unnoticed as a prospective business opportunity with a tremendous rise in the number of retail trading platforms. Even though many cite long-term investment as a key reason to join, others use an advertisement approach with clear parallels to the gambling industry: offering a free trade or a free stock upon taking up the platform, or to trade with virtually no initial money in their account, as shown in Figure \ref{fig:ads}. Other trading platforms, in turn, provide investors the opportunity to seamlessly execute upon the social and psychological biases studied within this text: `CopyTrader', for example, on the platform eToro allow investors to automatically copy the trades of others. A recent report by the UK Parliament discusses several important anecdotes pointing to the future relevance of this study: i) UK’s largest investment platform, Hargreaves Lansdown, reported a 40\% jump in net new business in the final six months of 2020, ii) the average age of platform users has dropped from 54 in 2012, to 47, reflecting a rise in younger investors (which are more prevalent on discussion forums, such as WSB), iii)  Trading 212 (a different platform) announced on 2 February 2021 that it would temporarily pause new account openings due to huge demand.\footnote{\url{https://commonslibrary.parliament.uk/the-rise-of-armchair-retail-trading-risks-and-regulation/}} Even though investor discussion forums have had an influence on markets in the past, we argue that the rise of WSB constitutes the prevalence of a new type of retail trader - the `hype' trader - which has permanently altered the composition and behaviour of retail traders, has changed the dynamics of financial markets, and whose behaviours are likely to increase in importance.

\section{Model}
\label{app:model}
\subsection{Static model with investment complementarities}
\label{app:model_complementarity}

We study the role of complementary investment decisions. To that end, the model operates in two stages. In the first stage, investors build their expectation for the asset's value, using observed signals from their peers and their expectation of the market-clearing price as a function of the expected, and as of yet hidden, supply shock. In the second stage, the asset supply shock is revealed, and investors execute their trades according to their demand curve. 

Investor $i$ who expects value $\mathbb{E}_i(v)$ and understands the price setting mechanism. Investor $i$'s maximised payoff from Eq. \ref{eq:mean_variance_tradeoff} is
\begin{align}
    \mathcal{L}(\phi_i^\ast) &= \frac{\mathbb{E}_i^2(v-p)}{2\gamma \mathbb{E}_i(v-p)^2} \label{eq:optimal_payoff} \\
    &= \frac{1}{2} \mathbb{E}_i(v - p) \phi^\ast_i  \\
    &= \frac{1}{2} [\mathbb{E}_i(v) - \mathbb{E}(v)]\phi^\ast_i + \frac{\gamma \sigma^2}{2}  \left( \frac{1}{N} \sum_j^N \phi_j^\ast \right) \phi^\ast_i, \label{eq:payoff_complementarity}
\end{align}
where $\mathbb{E}(v) = {1}/{N} \sum \mathbb{E}_i(v)$ as before. Here, investors base their price expectations on the simple equilibrium in Eq. \ref{eq:simple_equilibrium}, but use their personal expectations and constant uncertainty $\sigma^2$ to forecast price in the second stage. Eq. \ref{eq:payoff_complementarity} demonstrates that the investor's payoff depends on their peers in two regards. First, payoffs increase to the degree that the investor in question expects to outperform others, in terms of the value they realise in the asset. This is seen in the first component, by which buying(selling) the asset increases the payoff to the extent that $i$'s expected value $\mathbb{E}_i(v)$ is higher(lower) than that of their peers. Second, the payoff increase by the average optimal asset demand of all investors in the economy. 

The asset demand model predicts that social interactions -- knowledge of other's asset purchases -- is a significant component of investors' welfare \textit{in expectation}. Eq. \ref{eq:payoff_complementarity} is a well-known formulation for strategic interactions between agents acting under quadratic loss \citep{zenou2016key}. Deriving Eq. \ref{eq:payoff_complementarity} with respect to two investors' demands reveals their strategic complementarity:
\begin{align}
    \frac{d^2\mathcal{L}(\phi_i^\ast)}{d\phi_j^\ast d\phi^\ast_i} = \frac{\gamma \sigma^2}{2N} > 0. \label{eq:strategic_complementarity}
\end{align}
The emergence of strategic complementarities is due to a crowding effect that investors have on price. The higher asset demand by other investors, the higher the realised price will turn out to be. Before the value of the asset is revealed, investors are motivated to gauge demand by others to better estimate what the price will be, in excess of their personal valuation.

The unweighted average of peer sentiment in Eq. \ref{eq:payoff_complementarity} emerges because we did not provide a specific mechanism by which information about asset demand is transmitted. The acquisition of information under some cost to the investor is an interesting extension, although already studied by \citet{hellwig2009knowing}. Their study offers more rigorous insight into the manifestation of strategic complementarities, as well as the emergence of multiple equilibria, when investors seek to learn about the underlying price from a set of possible signals. 

\subsection{Discussion on Assumptions \ref{assumption:complementarity}, \ref{assumption:mech_extrapolation}}

\paragraph{Assumption \ref{assumption:complementarity}} Assumption \ref{assumption:complementarity} is in line with the finding that our asset demand model produces strategic complementarities in investor asset demands in Eq. \ref{eq:strategic_complementarity}. Several extensions of the simple formulation are possible to account for greater complexities in social interactions. It can, for example, be extended to take into account \textit{key players} \citep{zenou2016key} through changing the way that people weight the demand of others $\phi_{i,t-1}$ in the sum $1/N \sum_i \phi_{i,t-1}$ to $\sum_i s_i \phi_{i,t-1}$, where $s_i$ captures the influence of player $i$ and $\sum_i s_i = 1$. 

In more complex settings, we can consider the unique complementarities between connected individuals as described in \cite{zenou2016key}, clusters of investors \citep{bouchaud2003theory}, or alternative information spreading / individual targeting models, which have received attention in the recent literature \citep{galeotti2020targeting}. The added complexity would affect aggregate demand through the expectations of other's demand $\phi_{t-1}$.

\paragraph{Assumption \ref{assumption:mech_extrapolation}} 

Mechanical extrapolation is our preferred way to introduce a relationship between prices and demand \citep{barberis2018extrapolation}. A model with mechanical extrapolation has several shortcomings, one of which is the inability to relate expectation updates to psychological underpinnings. However, our assumption is justified by our empirical work in Section \ref{sec:consensus_results} which demonstrates that individuals update their outlook based on recent asset returns. 

\subsection{Persistent fluctuations}

The reversal in returns is an important feature that emerges from social contagion in investors' price expectations. If large enough, these can produce bubbles in asset prices: initial momentum from positive news creates a price run-up, before an absence of news creates a drought of new asset demand. The subsequent price crash carries on its own momentum. We can treat demand as a latent variable to see these oscillations manifest in return data. Substituting lagged demand into the equation for returns, and iterating infinitely yields
\begin{align}
\label{eq:return_time_series}
    r_t = -\sum_{T = 1}^\infty \left(\frac{\alpha}{\gamma \sigma^2}\right)^T r_{t-T} + \frac{S_t\gamma \sigma^2}{\beta N}
\end{align}
as long as $\alpha / \gamma \sigma^2 < 1$, so that the contribution of demand fluctuations to returns converges to zero over time. This is an autoregressive model with infinite lags, where the coefficients decrease exponentially with lag size $T$. Without any knowledge of asset demand, the second term encapsulates an unobservable error term, which the model links to exogenous changes in the asset's supply. Eq. \ref{eq:return_time_series} demonstrates that an exogenous increase in returns at time $t$ is followed by a smaller decrease in $t+1$. This oscillation persists indefinitely, and would converge to zero rapidly if the social signal $\alpha / \gamma \sigma^2$ is sufficiently small.

\subsection{Bubble dynamics}
\label{app:bubble_model}

In addition to the equilibrium setting, we demonstrate how peer effects are relevant in modeling bubble dynamics through incorporating them in an extension of \cite{barberis2018extrapolation}. We highlight the relevant model details below, however, direct the reader to the original paper for the full model setup. In the original model, extrapolators determine their demand from a `fundamental signal' with weight $w_i$, as well as an extrapolation signal with weight $(1-w_i)$, and trade with fundamental trades in the market. The demand function for extrapolators with non-varying temporal weights is: 
\begin{align}
    w_i \frac{F_t}{\gamma \sigma^2} + (1-w_i) \frac{M_t}{\gamma \sigma^2}.
\end{align}
\cite{barberis2018extrapolation} define $M_t$ ($X_t$ in the original text) as:
\begin{align*}
    M_t &= (1-\theta_E) \sum_k^{t-1} \theta^{k-1} (P_{t-k} - P_{t-k-1}) + \theta^{t-1} X_1,\\
    &= (1-\theta_E) (P_{t-1} - P_{t-2}) + \theta_E M_{t-1}. 
\end{align*}
where $\theta_E$ is the weight placed on recent versus older price changes, and is between zero and one. 

We propose to modify extrapolator signal to incorporate a social component, $M_t^s$:
\begin{align}
    M_t^s = (1 - \theta_E) \phi_t + \theta_E M_{t-1}^s,
\end{align}
where $\phi_t$ is the average sentiment determined from past price returns (extrapolation), as well as a past expressed sentiments (persistent demand driven by peer effects),
\begin{align}
    \phi_t = \beta^* (P_{t-1} - P_{t-2}) + \alpha^* \phi_{t-1},
\end{align}
where $\beta^*$ and $\alpha^* $ sum to one. 

    \begin{figure}[ht]
        \begin{center} 
            \includegraphics[width = \textwidth]{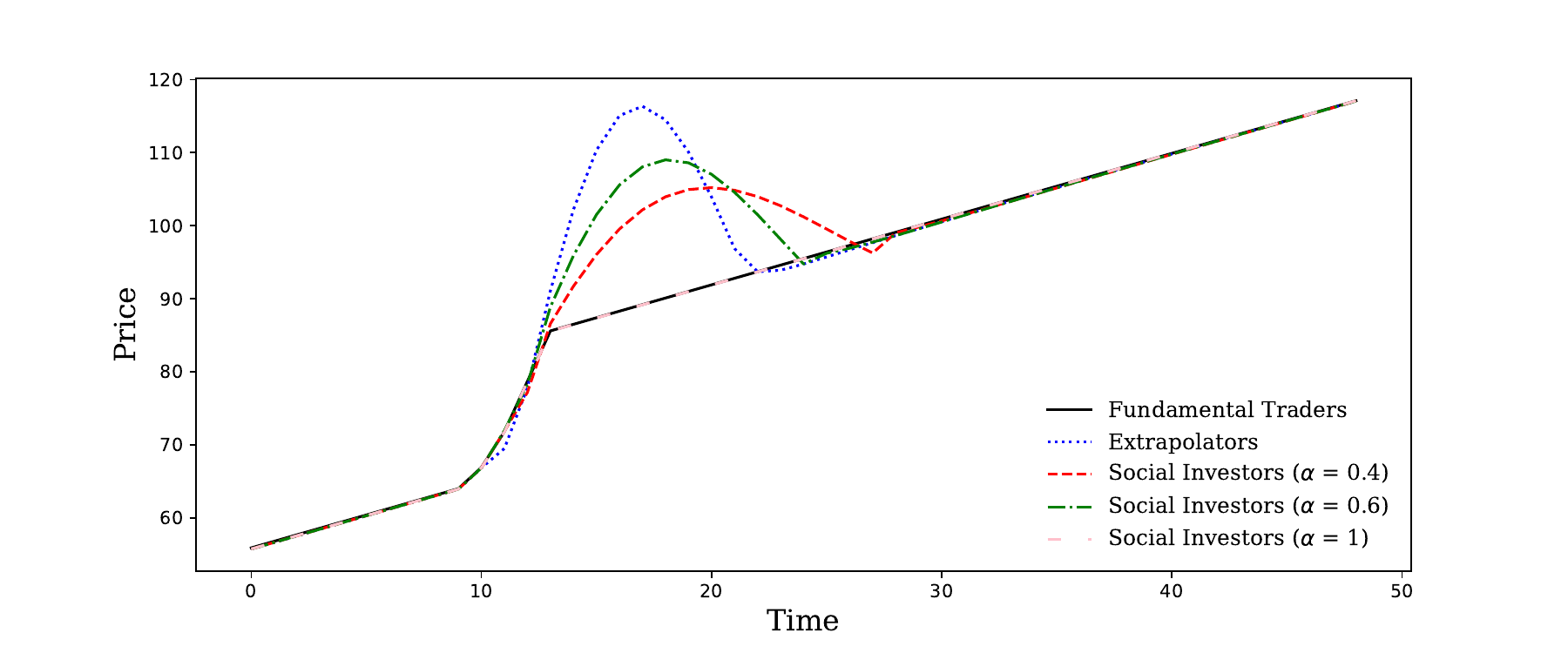} 
            \caption{\footnotesize{\textbf{Bubbles with Different Parameter Values for $\alpha^*$, $\beta^*$}; We choose initial parameters similar to those from Figure (1) in \cite{barberis2018extrapolation}: $w_i= 0.1$, $\sigma^2=3$, $\gamma = 0.1$, $\theta_E = 0.1$; extrapolators / social investors make up 70\% of investors, the remainder are fundamental traders; the quantity of the asset is set to 1. In Figure \ref{fig:barberis_bubble1}, we choose that the fundamental value of the asset remains unchanged except for periods 11-14 when information is revealed resulting in an increase in the future dividend of the asset by 2,4,6,6 (respectively).}}
            \label{fig:barberis_bubbles_appendix}
        \end{center}
    \vspace{-1em}
    \end{figure} 

We use our estimates for $\beta$ and $\alpha$ to compare resulting bubble dynamics in the presence of peer effects to the original findings in \cite{barberis2018extrapolation} -- our estimates demonstrate that social investors on WSB press a relative weight $\alpha^*$ of 0.6 on the sentiments of peers and a relative weight $\beta^*$ of 0.4 on recent returns. Figure \ref{fig:barberis_bubbles_appendix} demonstrates that the values for $\alpha^*$, $\beta^*$ control how long of a memory investors have. As $\alpha^*$ increases, we observe that the bubble takes a longer period of time to form and dissipate.

\section{Extra results for peer effects}
\label{app:consensus}
\subsection{Target independent variable}
\label{app:utility_framework}

    We build a discrete-choice empirical strategy to suit our model. Under the assumption that $u_{i,t}$ is drawn from a standard type-I Extreme Value Distribution, we model the log-odds of expressed investor sentiments $\phi_i$ by a standard multivariate logistic function,
    \begin{align}
        \label{eq:consensus1}
        \log \left( \frac{P(\phi_{i,t} = +1)}{P(\phi_{i,t} = 0)}\right) &= g(b_{i,t}) + f(\Bar{\phi}_{-i,(t-1,t)}) - \theta \sigma^2_{i,t} + u_{i,t}^+, \\
        \log \left( \frac{P(\phi_{i,t} = -1)}{P(\phi_{i,t} = 0)}\right) &= -g(b_{i,t}) - f(\Bar{\phi}_{-i,(t-1,t)}) - \theta \sigma^2_{i,t} + u_{i,t}^-,
    \end{align}
    where $t$ denotes time, and $(t-1,t)$ an interval preceding $t$. The goal of this paper, in light of Prediction 1, is to test  empirically whether $f(\cdot)$ is increasing. To that end, we aggregate bullish and bearish sentiments into one continuous variable, $\Phi_{i,t}$:
    \begin{align}
        \Phi_{i,t} = \frac{1}{2} \log \left( \frac{P(\phi_{i,t} = +1)}{P(\phi_{i,t} = -1)}\right) &= g(b_{i,t}) + f(\Bar{\phi}_{-i,(t-1,t)}) + \frac{u_{i,t}^+ - u_{i,t}^-}{2}.
    \end{align}
    In the main body, the error term is expressed as $\epsilon_{i,t}$. Under the assumption that $u_{i,t}^+$ and $u_{i,t}^-$ are independent and identically distributed, $u_{i,t}^+ - u_{i,t}^-$ will follow a logistic distribution with finite variance. 

\subsection{Full regression estimates}
\label{app:consensus_full_reg}
Tables \ref{tab:consensus_full_reg} and \ref{tab:reg_network_full} present our full regression estimates. Table \ref{tab:network_FS} presents our First Stage estimates for our \textit{Commenter Network} approach, which has multiple IVs.


\begin{table}[ht!]
\centering
\caption{Peer influence: \textit{Frequent Posters} -- full regression estimates}
\begin{tabular}{ll@{\hskip 0.2in}c@{\hskip 0.2in}c@{\hskip 0.2in}c}
\toprule \toprule
 && \multicolumn{3}{c}{\textit{Dependent Variable: $\Phi_{i,j,t}$}}\\
 && Reduced Form & Full Second Stage & Random Peers \\
 && (1) & (2) & (3) \\ [2ex]
\toprule
\multirow{6}{*}{\rotatebox{90}{\textit{Independent}}}
\multirow{6}{*}{\rotatebox{90}{\textit{Variables}}}
& $\Phi_{i,j,t-1}$            &    0.154 (0.010) *** &   0.129 (0.011) *** &  0.158 (0.010) ***\\
& $\Bar{\Phi}_{-i,j,(t-1,t)}$ &   0.055 (0.011) *** &    0.036 (0.010) *** &   0.005 (0.009)\\
& $r_{j,t}$                   &  0.022 (0.004) *** &   0.025 (0.005) *** &   0.023 (0.004) ***\\
& $\Bar{r}_{j,t}$             &      0.007 (0.004) &     0.007 (0.005) &     0.006 (0.004) \\
& $\sigma^2_{j,t}$           &      -0.003 (0.004) &     0.003 (0.008)  &      -0.003 (0.004)\\
&Ticker Fixed Effects & Yes & Yes & Yes \\
\toprule
& No. Observations: & 14,396 & 11,122 & 14,371\\
& $R^2$: & 0.12 & 0.08 & 0.11\\
& $R^2_{adj}$: & 0.08 & 0.06 & 0.08\\
\bottomrule
\end{tabular} \\
\bigskip
\vspace{-.7em}
\begin{tabular}{cccc}
\multicolumn{4}{l}{%
  \begin{minipage}{17cm}%
    \footnotesize{ \textit{Notes}: The dependent variable is individual investor sentiment about an asset, scaled continuously between $(-\infty, \infty)$, is estimated by the individual's previously expressed sentiment about the same asset ($\Phi_{i,j,t-1}$) and a set of market control variables ($r_{j,t},\Bar{r}_{j,t},\sigma^2_{j,t}$), using OLS. The sentiment of peers ($\Bar{\Phi}_{-i,j,(t-1,t)}$) is estimated in several ways. In Column (1), we use observed, average sentiment of peers between an author's two posts. In Column (2), we estimate the sentiment of peers using an IV. In Column (3), we select a random cohort to estimate peer sentiment. Robust standard errors, clustered at the ticker level, are presented in parentheses. Observations with incomplete market data are dropped.\\
*** Significant at 1\% level
** Significant at 5\% level
* Significant at 10\% level}
  \end{minipage}}
\end{tabular}
\vspace{-0.5em}
\label{tab:consensus_full_reg}
\end{table}

\begin{table}[!ht]
\begin{center}
  \caption{First Stage estimates for \textit{Commenter Network} approach} 
  \label{tab:network_FS} 
\begin{tabular}{@{\extracolsep{5pt}}lcccc} 
\toprule \\[-1.8ex] 
& $\phi_{i,j,t-1}^{-1}$  & $\phi_{i,j,t-1}^{0}$ & $\phi_{i,j,t-1}^{+1}$ & $\Bar{\Phi}_{-i,j,t-1}$ \\
\cline{2-5} \\[-1.8ex] 
\textit{Dependent variable:} \\ 
Sentiment of Peers & -0.30 (0.04) *** & 0.12 (0.03) *** & 0.25 (0.03) ***  & 0.14 (0.01) *** \\
\bottomrule \\[-1.8ex] 
\end{tabular} 
\end{center}
\vspace{-.9em}
\footnotesize{ \textit{Notes}: The dependent variable is individual investor sentiment about an asset expressed in a single submission, scaled continuously between $(-\infty, \infty)$, modeled using IVs. We estimate it using the individual's previously expressed sentiment about the same asset ($\phi_{i,j,t-1}$) as a categorical variable, with the author not having posted previously ($\phi_{i,j,t-1}^{NA}$) as the baseline, as well as the average sentiment of posts that the author commented on previously ($\Bar{\Phi}_{-i,j,t-1}$). We user the timing of IVs to control for common shocks, as discussed in the main text. Our regression has 24,013 observations and an F-statistic of 118.

*** Significant at 1\% level ** Significant at 5\% level * Significant at 10\% level}
\vspace{-0.9em}
\end{table}


\begin{table}[ht!]
\centering
\caption{Peer influence: \textit{Commenter Network} -- full regression estimates}
\begin{tabular}{ll@{\hskip 0.4in}c@{\hskip 0.4in}c@{\hskip 0.4in}c}
\toprule \toprule
 && \multicolumn{3}{c}{\textit{Dependent Variable -- $\Phi_{i,j,t}$}}\\
 && Reduced Form & Full Second Stage & Random Network \\
 && (1) & (2) & (3) \\ [2ex]
\toprule
\multirow{8}{*}{\rotatebox{90}{\textit{Independent Variables}}}
& $\phi_{i,j,t-1}^{-1}$            &   -0.226 (0.026) ***  & -0.215 (0.021) *** &  -0.342 (0.040) ***\\
& $\phi_{i,j,t-1}^{0}$            &  0.047 (0.023) ** & 0.034 (0.022)  & 0.073 (0.036) **\\
& $\phi_{i,j,t-1}^{+1}$            &   0.160 (0.028) *** &  0.141 (0.031) *** &    0.244 (0.042) ***\\
& $\Bar{\Phi}_{-i,j,t-1}$ &   0.041 (0.009) *** &   0.022 (0.009) **  &  0.009 (0.009)\\
& $r_{j,t}$                   &   0.020 (0.003) *** &   0.023 (0.006) *** &   0.031 (0.005) ***\\
& $\Bar{r}_{j,t}$             &      0.005 (0.003)  &   0.006 (0.006)  &      0.008 (0.005)\\
& $\sigma^2_{j,t}$           &      0.044 (0.289) &     0.512 (0.508)  &      0.078 (0.449)\\
&Ticker Fixed Effects & Yes & Yes & Yes \\
\toprule
& No. Observations: & 24,902 & 16,514 & 25,220\\
& $R^2$: & 0.09 & 0.07 & 0.09\\
& $R^2_{adj}$: & 0.06 & 0.06 & 0.06\\
\bottomrule
\end{tabular} \\
\bigskip
\vspace{-.7em}
\begin{tabular}{cccc}
\multicolumn{4}{l}{%
  \begin{minipage}{17cm}%
    \footnotesize{ \textit{Notes}: The dependent variable is individual investor sentiment about an asset expressed in a single submission, scaled continuously between $(-\infty, \infty)$. We estimate it using the individual's previously expressed sentiment about the same asset ($\phi_{i,j,t-1}$) as a categorical variable, with the author not having posted previously ($\phi_{i,j,t-1}^{NA}$) as the baseline. We control for a set of market control variables ($r_{j,t},\Bar{r}_{j,t},\sigma^2_{j,t}$). The sentiment of posts that the author commented on previously ($\Bar{\Phi}_{-i,j,t-1}$) is estimated several ways. In column (1), we present the estimate using the sentiment of posts the author previously commented on. In column (2), we use an IV to predict the sentiment of posts the author comments on. In column (3), we randomly rewire the network, connecting the author to a random set of posts about the same ticker. Robust standard errors, clustered at the ticker level, are presented in parentheses. Observations with incomplete market data are dropped.\\
*** Significant at 1\% level
** Significant at 5\% level
* Significant at 10\% level}
  \end{minipage}}
\end{tabular}
\vspace{-0.9em}
\label{tab:reg_network_full}
\end{table}

\subsection{Evidence of identification strategy}
\label{app:consensus_identification}
\begin{figure}[ht!]
\begin{center}
    \begin{subfigure}{0.475\textwidth}
        \includegraphics[width=\linewidth]{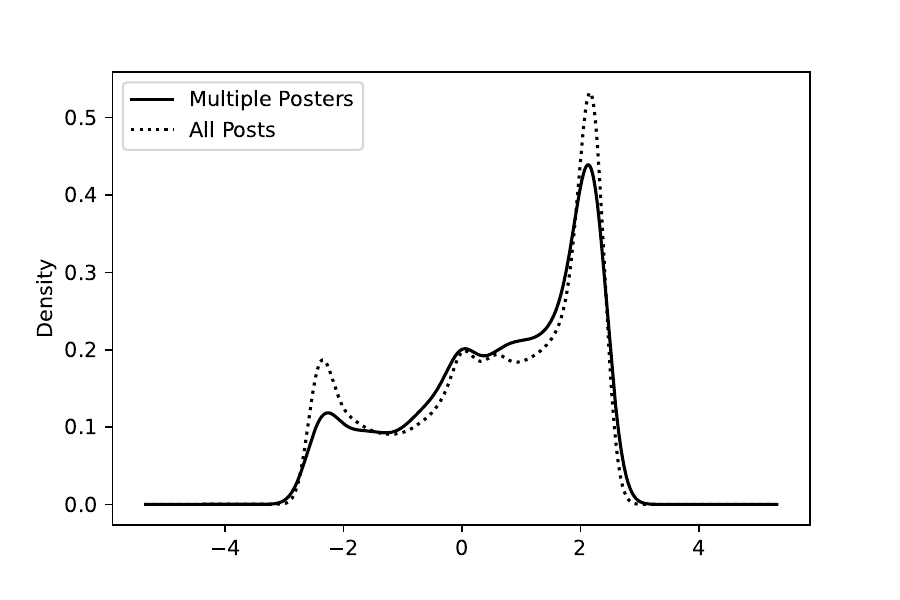}
        \caption{Frequent Posters Sentiment PDF} 
        \label{fig:density_freq}
    \end{subfigure}
    \hspace{0.5cm}
    \begin{subfigure}{0.475\textwidth}
        \includegraphics[width=\linewidth]{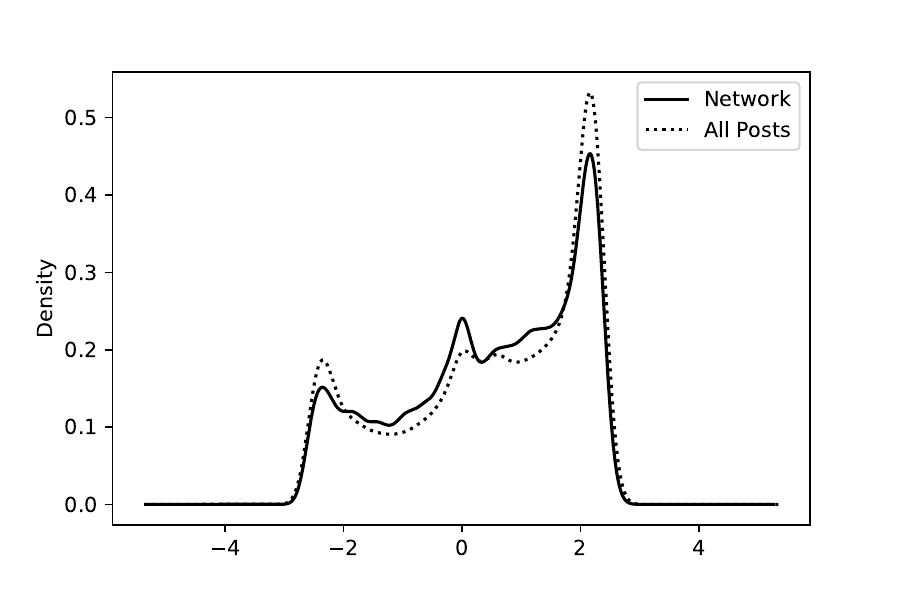}
        \caption{Commenters Sentiment PDF} 
        \label{fig:density_com}
    \end{subfigure}
    \end{center}
    \caption{\footnotesize{\textbf{Density Plot of Sentiments Expressed on WSB}; We present the density plot of the sentiments expressed by users on WSB who post multiple times, labeled as \textit{Multiple Posters}, those who comment on others' posts, labeled as \textit{Network}, and that of all submissions, labeled as \textit{All Posts}.}}
    \vspace{-0.9em}
    \label{fig:density_multiple_vs_all_posters}
\end{figure} 
A potential concern with our approach is whether the sentiments expressed by individuals who post multiple times or are part of the commenters network follow the same distribution as all submissions on the forum. Figure \ref{fig:density_freq} presents the distribution of sentiments for the second or later post of an author about a ticker and Figure \ref{fig:density_com} presents the distribution of sentiments for those who comment on other's posts. Figure \ref{fig:density_multiple_vs_all_posters} provides evidence that the sentiment distributions are similar to that of other posters on WSB, which supports the hypothesis that our analysis offers insight into how all individuals on WSB form opinions. 

A second concern is whether we effectively control for unobserved ticker characteristics. Similarly to \cite{patacchini2016social}, we run `placebo tests', where we replace the composition of an author's peers with a random cohort of people who post on WSB about the same ticker. The random cohort is chosen as follows. We observe how many peers an individual author has. We then select a random sample of the same number of individuals, without replacement, who do not post between an author's two post but post about the ticker at a different time for the \textit{Frequent Posters} approach (if fewer individuals post before, we select all of those individuals), or through a random network rewiring (we select posts randomly about the same ticker before the current post). The results are presented in Tables \ref{tab:consensus_full_reg} and \ref{tab:reg_network_full}, column (3). We observe that all the coefficients remain close to their original values, except for the peer effect, which becomes insignificant. This lends credibility to our peer identification strategy and shows that  unobserved factors that influence within ticker variation are not confounding our estimates.

We cannot directly calculate the J-statistic for our \textit{Commenter Network} approach, since we estimate our IV using observations on several neighbours. We, therefore, take an average of the neighbours past sentiments (transforming the categorical variable into a continuous one) and the average across their neighbour's neighbours sentiments. We use this to compute a J-Statistic with two degrees of freedom.

\subsection{Normalization procedure and non-normalized coefficient estimates}
\label{app:indentification_coeff_interpretation}
In order to compare the relative impacts across variables, we perform mean / standard deviation normalization on the non-categorical variables within our regression. We normalize market variables with respect to the log-returns of all assets discussed on WSB since the forum's creation in 2012. We normalize sentiment variables with respect to the observations within our regressions. The normalization is performed in order to be able to compare the impact of peer effects and returns on sentiment formation. 

Table \ref{tab:reg_consensus_non_norm} presents the non-normalized coefficient estimates for our second stage. We observe that the coefficient on returns and predicted peer sentiment are both higher, and the coefficient on returns is larger than that of predicted peer sentiment: these changes relate to the standard deviations of the two variables which are $0.037$ for daily returns and $0.271$ for predicted peer sentiment, \textit{Frequent Posters}, and $0.168$ for predicted peer sentiment, \textit{Commenter Network}. The first phenomenon is explained by the fact that the standard deviation for both variables is less than one; the second is explained by the fact that daily returns have a substantially smaller standard deviation than that of peer sentiment. 

\begin{table}[ht!]
\begin{center}
\caption{Peer influence in WSB sentiments}
\label{tab:reg_consensus_non_norm}
\resizebox{\textwidth}{!}{
\begin{tabular}{lcc}
\toprule
&    Frequent Posters  &  Network \\ 
& (1) & (2) \\
\cline{2-3} \\[-1.8ex]
\multicolumn{3}{l}{Second Stage -- peer influence estimated using \textit{predicted} average sentiment of peers \textit{(non-normalized)}} \\\\[-1.8ex]
&  \multicolumn{2}{c}{\textit{Dependent Variable}: Investor Sentiment $(\Phi_{i,j,t})$} \\ \cline{2-3} \\[-1.8ex]
    
    Average peer sentiment, && \\
    $\hat{\Phi}_{-i,j,(t-1)}$ \textit{(predicted)} & 0.198 (0.053) ***  & 0.197 (0.083) ** \\
    $r_{j,t}$ & 0.993 (0.185) ***  & 0.921 (0.251) *** \\ \\[-2ex]
    Author \& asset controls ($X_{i,j,t}$) & Yes  &  Yes \\
\bottomrule
\end{tabular}
}
\end{center}
\vspace{-0.6em}
\footnotesize{ \textit{Notes}: this table presents the non-normalized coefficient estimates for the Second Stage of our \textit{Frequent Posters} and \textit{Commenter Network} regressions. \\
*** Significant at 1\% level
** Significant at 5\% level
* Significant at 10\% level}

\end{table}

    \subsection{Contagion dynamics and the origin of bull runs}
    \label{subsec:contagion}
    
    
    In the WSB context, we would expect awareness about specific assets to spread from one user to another, in line with the observations of \cite{shiller2005irrational,banerjee1993economics} and \cite{banerjee2013diffusion}. The emphasis of this section is not on identifying a causal relationship, but rather understanding the dynamics which govern asset interest among investors. These insights, combined with a mechanism for investors' joint sentiment adoption, allow us to paint a more complete picture of retail investor decision-making and the resultant stock market dynamics. 
    
    
    
    We model the log-odds of an author posting about stock $j$ over a baseline using the following linear model:
    \begin{align}
    \label{eq:estimating_contagion}
    l(a_{j,t}) = \log\left(\frac{a_{j,t}}{s_{t}}\right) =  c a_{j,t-1} (1-a_{j,t-1}) + d a_{j,t-1} + \beta_1 \Bar{r}_{j,t-1} + \beta_2 \sigma^2_{j,t-1} + X_j \beta_4 + \zeta_{j,t},
    \end{align}
    where $t$ denotes time (in weeks), the baseline $s_t$ is the probability of posting about a stock that is not widely discussed within the forum (a stock that is mentioned in fewer than 31 submissions within our sample), $a_{j,t-1}$ is the share of all active investors who post about ticker $j$ at times $t-1$ ($a_{j,t} \in [0,1]$ for all $j$ and $t$), $\Bar{r}_{j,t-1}$ is the average log-return in $t-1$, and $\sigma^2_{j,t-1}$ is the variance of the same log-returns (these variables are mostly consistent with Section \ref{subsec:consensus}, and discussed further in our Online Appendix). $X_j$ is a vector of stock dummies. 
    
    Our framework resembles that of Section \ref{subsec:consensus} and is inspired by \citet{banerjee2013diffusion} -- individuals become interested in an asset due to their peers and thanks to a public signal of the asset's performance. Parameter $c$ captures the rate of independent mixing between investors aware of stock $j$, $a_{j,t-1}$, with unaware investors, $1-a_{j,t-1}$. Parameter $d$ captures the rate at which aware investors lose interest. The latter terms control for the asset's perceived profitability and riskiness. Parameter $\beta_1$ is a `quality of signal' term capturing how well the asset has performed in the past, and $\beta_2$ a `noise of signal' term, measuring the asset's recent volatility. We propose that coefficients $c$ and $\beta_1$ are positive -- implying that these dynamics contribute to increased interest in a stock -- while $d$ and $\beta_2$ are negative.
    
    The choice to aggregate over weeks is done to address the sparsity of submissions, especially pre-2017. In addition, we categorise stocks mentioned fewer than 31 times since January 2012 into an `other stocks' group, which forms our benchmark $s_t$.
    
    We also consider a different formulation where we test for the direct impact of historical peer sentiments and the interactions between historical sentiments and returns / volatility: $\Bar{\phi}_{j,t-2}\Bar{r}_{j,t-1}$, $\Bar{\phi}_{j,t-2}$ and $\Bar{\phi}_{j,t-2}\sigma^2_{j,t-1}$. This formulation allows us to evaluate whether WSB users are more likely to discuss a stock if the predictions of their peers have been correct, and accurate, in the past. 

    \subsubsection{Results}

\begin{table}[!ht]
\begin{center}

\caption{Stocks discussed on WSB} 
\label{tab:wsb_stock_dynamics} 
\begin{tabular}{@{\extracolsep{5pt}}lcccc} 

\toprule

& \multicolumn{4}{c}{\textit{Dependent variable:} $l(a_{j,t})$} \\ 
\\[-1.8ex] & (1) & (2) & (3) & (4)\\ 
\hline \\[-1.8ex] 

    $a_{j,t-1} (1-a_{j,t-1})$ & 83.49$^{***}$ (8.20) & 100.20$^{***}$ (9.15) & 46.30$^{***}$ (5.33) & 57.90$^{***}$ (5.13) \\ 
    $a_{j,t-1}$ & $-$48.01$^{***}$ (7.04) & $-$62.37$^{***}$ (7.83) & $-$24.06$^{***}$ (3.94) & $-$33.73$^{***}$ (3.95) \\ 
    $\Bar{r}_{j,t-1}$ & 1.24$^{***}$ (0.39) &  & 1.36$^{***}$ (0.42) &  \\ 
    $\sigma^2_{j,t-1}$ & $-$2.15$^{***}$ (0.60) &  & $-$0.96$^{*}$ (0.54) &  \\ 
    $\Bar{\phi}_{j,t-2} \Bar{r}_{j,t-1}$ &  & 0.56 (1.09) &  & 1.59 (1.10) \\ 
    $\Bar{\phi}_{j,t-2}^2\sigma^2_{j,t-1}$ &  & $-$5.14$^{**}$ (2.19) &  & $-$1.71 (1.53) \\ 
    Constant & $-$3.89$^{***}$ (0.01) & $-$3.88$^{***}$ (0.02) &  &  \\ 
    \hline \\[-1.8ex] 
    Ticker FE & No & No & Yes & Yes \\
    Number of obs. & 13,184 & 6,429 & 13,184 & 6,429 \\ 
    Adjusted R$^{2}$ & 0.28 & 0.36 & 0.10 & 0.14 \\ 
    F-statistic & 1,294 & 920 & 429 & 318 \\ 
    
\bottomrule \\[-1.8ex] 

\end{tabular} 
\end{center}

\footnotesize{ \textit{Notes}: this table presents OLS estimates for the log-odds of users discussing stock $j$ in week $t$, over a collection of stocks that are mentioned fewer than 31 times. Explanatory variables include: the lag in the share of authors discussing $j$, $a_{j,t-1}$, the interaction with the share of authors not discussing $j$, $a_{j,t-1} (1-a_{j,t-1})$, as well as the lag in stock $j$'s weekly average log-return, $\Bar{r}_{j,t-1}$, and variance, $\sigma^2_{j,t-1}$. In columns (2) and (4), the average log-return is multiplied by the two period lag in the average sentiment expressed among WSB submissions on stock $j$, $\Bar{\phi}_{j,t-2}$, and the variance in log-returns by the same sentiment's square, $\Bar{\phi}_{j,t-2}^2$. Columns (3) and (4) include stock-specific fixed effects. Accompanying standard errors, displayed in brackets, are clustered at the stock level, and calculated in the manner of \citet{mackinnon1985some}.

*** Significant at 1\% level ** Significant at 5\% level * Significant at 10\% level}

\end{table}

    The OLS estimates of our model in Eq. \ref{eq:estimating_contagion}, presented in Table \ref{tab:wsb_stock_dynamics}, demonstrate that WSB users follow each other in their choice of stocks. There is strong evidence that the homogeneous mixing property partially explains the uptake of new assets: using estimates in column (1), an increase in the share of authors discussing stock $j$ from 0.1 to 0.2 increases the ratio of authors discussing $j$ over `other stocks' in the following week by approximately threefold. This is contrasted by an increase from 0.2 to 0.3, which prompts a decline in the ratio of authors discussing $j$ over `other stocks' by 50\% -- the difference is driven by the large negative coefficient on $a_{j,t-1}$. This is strongly reminiscent of epidemic contagion models, adapted to the spread of narratives \citep{banerjee1993economics,shiller2017narrative}.
    
    
    When we consider the impacts of stock-specific variables in isolation, presented in columns (1), (3) in Table \ref{tab:wsb_stock_dynamics}, volatility and returns appear to be leading factors for authors deciding what asset to discuss. Average, historical returns are statistically significant at the 1\% level in columns (1) and (3), indicating that discussion sizes are stimulated by large, notably positive, returns. Examining the coefficient in column (3), a stock that experienced a 5\% greater return in one week is the subject of about 7\% more submissions than usual. Volatility appears to play a greater role in our formulation without ticker-specific effects, with its significance declining from column (1) to (3) -- a factor perhaps explained by the choice of hype investors to overlook recent volatility in certain assets, but not others. Our alternative formulation presented in columns (2) and (4), estimating the effect of the correctness and consistency of past WSB predictions in an asset, appears to have limited significance in explaining asset interest.

\section{Extra results for market impact}
\subsection{Portfolio}

\begin{figure}[!ht]
    \centering
    \includegraphics{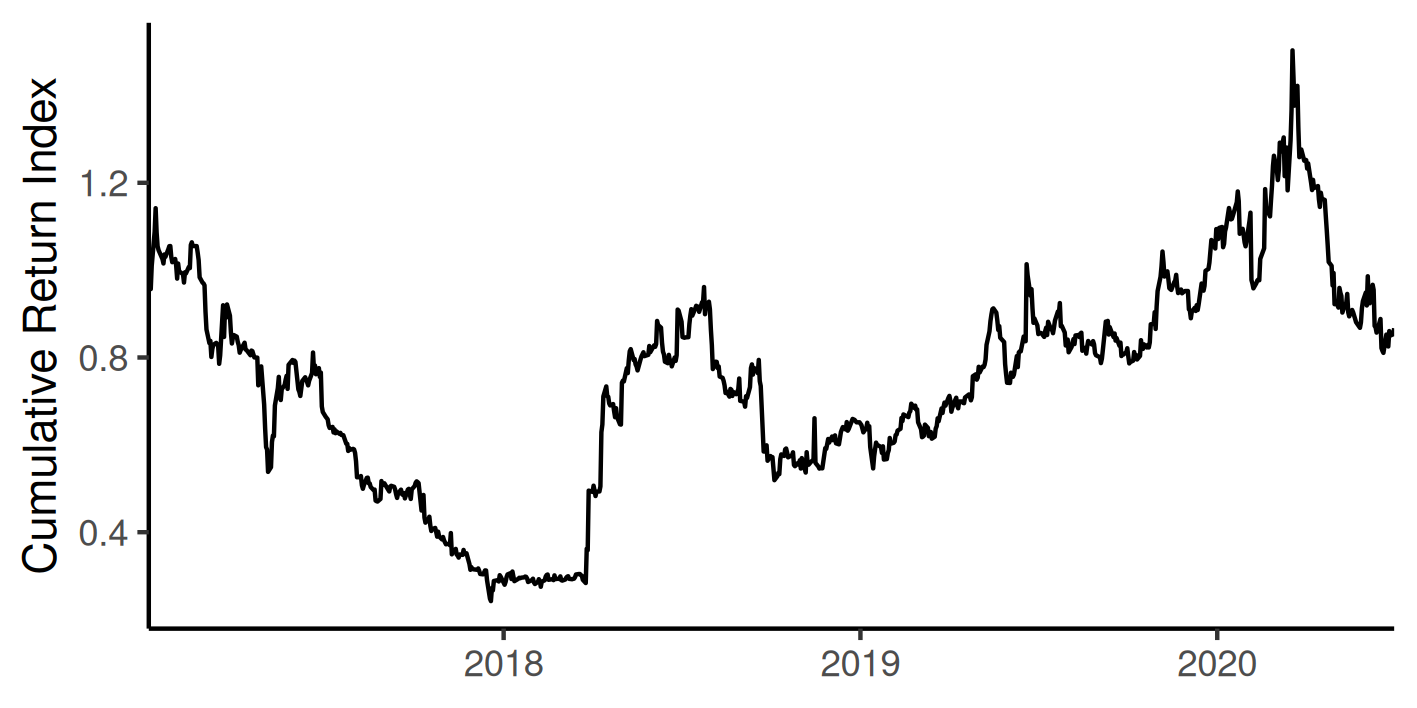}
    \caption{\textbf{Cumulative returns of a portfolio built on WSB sentiment.}}
    \label{fig:portfolio_returns}
\end{figure}

The performance of a portfolio that buys stocks according to WSB sentiments is questionable. Among alternative transformations, we assign weights for stocks with net positive sentiment, $w+_t$, and negate sentiment $w_t^-$. These proxy for long and short positions, with the size of the position by stock measure by the ratio of the stocks' lagged sentiment over the sum of sentiments:
\begin{align}
    w^+_{i,t} = \frac{\Phi_{i,t-1}^+}{\sum_i^N\Phi^+_{i,t-1}}, \quad w^-_{i,t} = \frac{\Phi_{i,t-1}^-}{\sum_i^N\Phi^-_{i,t-1}}.
\end{align}
The cumulative return of the combined long/short portfolio is displayed in Figure \ref{fig:portfolio_returns}. The indexed return from the start of 2017, when WSB data turned frequent enough to trade on daily, to July 2020 displays losses amounting to \%13.3. These are punctuated by various periods of consistent gains or losses, plus certain days of large outsized returns. The information content on WSB therefore does not appear inherently valuable, although this is admittedly not the stated intent; the volatile outcomes speak to the users' aspirations for large, one-off gambles.
\subsection{GIV}
\label{app:giv}

\paragraph{Extracting idiosyncratic social shocks}

In Table \ref{tab:giv_idiosyncratic_shocks}, we present the coefficients from Eq. \ref{eq:giv_social_shocks}. We observe that, consistently with our previous findings and the proposed structure in Figure \ref{fig:frequent_posters_dag}, returns and past user sentiments all impact the expressed sentiment of user $i$ about asset $j$ at time $t$.

\begin{table}[!ht]
\begin{center}
  \caption{Idiosyncratic social shocks estimation} 
  \label{tab:giv_idiosyncratic_shocks} 
\begin{tabular}{@{\extracolsep{5pt}}lc} 
\toprule \\[-1.8ex] 

& \textit{Dependent variable:} $\Phi_{i,j,t}$\\ 
\cline{1-2} \\[-1.8ex]
    $r_{j,t}$ & 0.317*** (0.088) \\ 
    $r_{j,t_w}$ & 0.384*** (0.096)  \\ 
    $\Bar{\Phi}_{j,t_w}$ & 0.202*** (0.041)  \\ 
    \hline \\[-1.8ex] 
    Ticker FE & Yes \\
    Number of obs. & 45,135 \\ 
    R$^{2}$ & 0.077 \\
    F-statistic & 7.85 \\ 
    
\bottomrule 
\end{tabular} 
\end{center}
\vspace{-0.6em}
\footnotesize{ \textit{Notes}: The dependent variable is the log-odds of a given submission by author $i$ at time $t$ on stock $j$ to express bullish over bearish sentiment. Explanatory variables include: the log return on day $t$, $r_{j,t}$, the log-return in week including day $t$ denoted as $t_w$, $r_{j,t_w}$, and the average past sentiment of peers, $\Bar{\Phi}_{i,j,t_w-1}$. Accompanying standard errors, displayed in brackets, are clustered at the stock level.

*** Significant at 1\% level ** Significant at 5\% level * Significant at 10\% level}
\end{table}

\paragraph{GIV  - First Stage} In Table \ref{tab:giv_first_stage}, we present our First Stage regression, where we regress the popularity-weighted idiosyncratic sentiment, as our dependent variable, on the difference between the popularity-weighted and regular average idiosyncratic sentiment. We observe that our GIV is highly predictive of the popularity-weighted sentiment, $\Bar{e}_{j,t_w}$, however, explains only part of the variation. In this way, we are able to extract the element of the popularity-weighted sentiment measure driven by social preferences, versus those driven by other factors. 

\begin{table}[!ht]
\begin{center}
  \caption{GIV - First Stage} 
  \label{tab:giv_first_stage} 
\begin{tabular}{@{\extracolsep{5pt}}lc} 
\toprule \\[-1.8ex] 

& \textit{Dependent variable:} $\Bar{e}_{j,t_w}$\\ 
\cline{1-2} \\[-1.8ex]
    GIV & 0.940*** (0.020) \\ 
    \hline \\[-1.8ex] 
    Number of obs. & 2,441 \\ 
    R$^{2}$ & 0.478 \\
    F-statistic & 2234 \\ 
    
\bottomrule 
\end{tabular} 
\end{center}
\vspace{-0.6em}
\footnotesize{ \textit{Notes}: The dependent variable is the popularity-weighted average idiosyncratic sentiment expressed about asset $j$ in week $t_w$, $\Bar{e}_{j,t_w}$. The explanatory variable is the difference between the popularity-weighted and raw average of idiosyncratic sentiments expressed about asset $j$ in week $t_w$. 

*** Significant at 1\% level ** Significant at 5\% level * Significant at 10\% level}
\vspace{-1.2em}
\end{table}

We illustrate the effect of the GIV through the following two scenarios. In \textit{Scenario One}, a very popular and an unpopular post both express a positive, idiosyncratic sentiment.  In \textit{Scenario Two}, a very popular expresses the positive sentiment, while the unpopular posts expresses no idiosyncratic sentiment. Without employing the GIV, our original popularity-weighted average idiosyncratic sentiment measure $\Bar{e}_{j,t_w}$ would be very similar in both scenarios. However, by using the GIV, we would \textit{predict} zero idiosyncratic sentiment in \textit{Scenario One}, but a large positive idiosyncratic sentiment in \textit{Scenario Two}. In this way, our GIV allows us to distinguish the signal coming from popularity.



\end{document}